\documentclass[twocolumn,superscriptaddress,showpacs,prb]{revtex4-2}
\usepackage{amsmath,bm,amsfonts,dsfont}%
\usepackage[dvipdfm]{color,graphicx}

\def\ov#1{\overline{#1}{}}
\def\un#1{\underline{#1}{}}

\def\ti#1{\tilde{#1}{}}
\def\wti#1{\widetilde{#1}{}}

\def\bg{{\bm g}}\def\bk{{\bm k}}
\def\bw{{\bm w}}

\def\bC{{\bm C}}\def\bD{{\bm D}}\def\bF{{\bm F}}
\def\bG{{\bm G}}\def\bH{{\bm H}}\def\bI{{\bm I}}\def\bK{{\bm K}}
\def\bL{{\bm L}}\def\bP{{\bm P}}
\def\bS{{\bm S}}\def\bT{{\bm T}}\def\bZ{{\bm Z}}

\def\cS{{\cal S}}

\def\pE{\mathbb{E}}\def\pH{\mathbb{H}}\def\pU{\mathbb{U}}\def\pV{\mathbb{V}}

\def\h0{\hat{0}}

\def\ga{\alpha}\def\gb{\beta}\def\gd{\delta}\def\gee{\epsilon}\def\ggg{\gamma}\def\gk{\kappa}\def\gs{\sigma}\def\gt{\theta}
\def\gD{\Delta}\def\gG{\Gamma}

\def\NN{\mathds{N}}\def\RR{\mathds{R}}\def\ZZ{\mathds{Z}}

\def\dH{\mathds{H}}\def\dU{\mathds{U}}\def\dV{\mathds{V}}
\def\one{\mathds{1}}

\def\re{\mathrm{e}}\def\ri{\mathrm{i}}

\def\kost#1#2{(~\!#1~\!|\!~#2~\!)}
\def\ket#1{\mid~\!\!\!{#1}~\!\!\rangle} 

\def\abs#1{\vert#1\vert}

\def\ACB{A\rightarrow C\leftarrow B}
\def\Lexc={\buildrel\leftarrow\over =}
\def\Rexc={\buildrel\rightarrow\over =}
\def\exc={\buildrel \leftrightarrow\over =}
\def\ABCTR{A\buildrel C\over\longleftrightarrow B}
\def\tripletr#1#2#3{#1\buildrel #3\over\longleftrightarrow #2}

\def\d={\buildrel \rm def \over =}
\def\eqmod{\buildrel .\over =}

\def\eqmod3{\buildrel \mathrm{3}\over =}

\def\smatD#1#2{\left(\begin{smallmatrix}
    #1\\ #2\end{smallmatrix}\right)}

\begin{document}

\preprint{XXX}

\title[Q1D OALs]{Topologically constrained obstructed atomic limits in quasi-one-dimensional systems}

\author{Milan Damnjanovi\'c\email[]{yqoq@rcub.bg.ac.rs}}
\affiliation {Serbian Academy of Sciences and Arts,  Knez Mihajlova 35, 11000 Belgrade, Serbia}
\affiliation {NanoLab, Faculty of Physics,
University of Belgrade, P.O. Box 44, 11001 Belgrade, Serbia}

\author{Ivanka Milo\v{s}evi\'c\email[]{ivag@rcub.bg.ac.rs}}
\affiliation {NanoLab, Faculty of Physics,
University of Belgrade, P.O. Box 44, 11001 Belgrade, Serbia}

\begin{abstract}
Possible forms of obstructed atomic limits in quasi-one-dimensional systems are studied using line group symmetry. This is accomplished by revisiting the standard theory with
an emphasis on its group-theoretical background, synthesizing the insights into a theorem that effectively identifies potential cases. The framework is then applied across the classes of quasi-one-dimensional systems, where the obstructed atomic limit serves as the primary criterion for topological characterization. The results are systematically organized and displayed, complemented by several illustrative examples.
\end{abstract}

\pacs{}

\maketitle

\section{Introduction}

Obstructed atomic limits (OALs) emerge when localized Wannier functions cannot align with atomic positions, reflecting a topologically nontrivial system. In the simplest cases, topologically constrained OALs are characterized by the nontrivial Berry phase~\cite{Berry-PRSA-84} of the electronic band. However, while this differential geometry-based approach is highly reliable, it is not readily applicable and must be performed on a case-by-case basis.
For more complex scenarios involving groups of degenerate bands, a more advanced framework is required, the non-Abelian generalization of the Berry phase. Specifically, the Wilson loop operator becomes essential, capturing the holonomy or total Berry phase accumulated as quantum states are parallel transported around a closed loop in momentum space. To determine the positions of Wannier charge centers within a unit cell, the eigenspectrum of the Wilson loop operator must be calculated, and the $\pi$-normalized phase factors of its eigenvalues extracted~\cite{Yang-PRB23}. Again, these calculations necessitate a case-by-case approach (although remaining applicable to any specific situation). Despite these methods are precise and reliable, they are cumbersome and fail to provide a generic understanding of the occurrence of topologically constrained OALs.

In 1D crystals, OALs can give rise to topologically protected edge states due to the mismatch between bulk electron localization and atomic sites. The bulk-boundary correspondence principle ensures that the nontrivial topology of the bulk (related to OALs) induces observable edge states at the boundaries~\cite{Vanderbilt-PRB93}  that emerge as zero-energy in-gap states and are protected by topological invariants; e.g. the Zak phase~\cite{Zak-PRL-89} based $Z_2$-classification of inversion-symmetric 1D-crystals. A topological phase transition, from the obstructed to the trivial atomic limit, typically goes through a gradual change of  the coupling parameters and at the critical point bulk band gap closes and reopens, the  edge states disappear, and the material behaves as a conventional (trivial)  insulator. The OALs together with  the resulting edge states are often protected by the crystal’s symmetries.

In this paper, OALs of quasi-one-dimensional (Q1D) systems are studied in the context of topology within tight-binding framework. The goal is to identify the classes of Q1D systems where the positions of the Wannier charge centers (within a unit cell) do not coincide with the atomic positions.
The paper aims to comprehensively elucidate the symmetry-driven foundations underlying the nontrivial topology and emergence of edge states in Q1D systems having \emph{line group} (LG) symmetry~\cite{YILG}. The analysis begins with an outline of the group-theoretical conditions required to enable this effect. Principles from topological quantum chemistry (TQC)~\cite{BernevigTQC2} are then applied to simplify the problem, reducing it to elementary band representations (EBRs)~\cite{ZakMichelPR,Cano-PRB-18}. Building on these results, we employ group theory to identify the relevant objects, i.e. associated orbits and orbital configurations (models), that qualify as candidates for OALs as detailed in Section~\ref{SMETHOD}. After a brief overview on the LGs (provided in Section~\ref{SLGs}), we give precise formulation and group theoretical conditions enabling OALs in Section~\ref{STheorems}.  This analysis is applied to Q1D systems in Section~\ref{SRES}, followed by an in-depth discussion and various physical (numerical) examples (Section~\ref{SDIS}).

Finally, it should be emphasized that although the LG symmetry-based method introduced here is more efficient and offers broader insight into topologically constrained OALs in Q1D systems than the differential geometry-based approaches, it is not exhaustive. Namely, the exceptions arise where bands with identical (band) representations exhibit different topological properties~\cite{Bacry-Michle-Zak-PRL88,Cano-PRB22}. In such cases, calculating the Berry phase or the Wilson loop operator is necessary to determine the existence of topologically protected OALs.

\section{Group theoretical foundation}\label{SMETHOD}

The following considerations apply to a fully occupied, well-defined set of bands isolated from the rest of the band structure. This assumption is made without loss of generality, as the concepts of obstructed \textsl{vs.} trivial atomic limits are inherently defined within the framework of topological (trivial) insulators.

Let $\bG$ be a symmetry group of a $\wp$ ($\wp=1,2,3$) periodical crystal. Hence, it has Abelian (translational) invariant subgroup $\bT$ with $\wp$ generators, and the finite factor group (isogonal point group) $\bP=\bG/\bT$. Arbitrary group element $g=\kost{G}{\bg}$ (Koster-Seits notation) consists of orthogonal operator $G$ acting in $\RR^3$ and vector $\bg$ from $\RR^3$.

The action on Euclidean space point $\bw$ is given by $g\bw=\kost{G}{\bg}\bw=G\bw+\bg$.  From point $\bw$ group generates infinite \emph{orbit}, with the finite \emph{stabilizer} (little group) $\bF^\bw$. All connected orbits with the mutually conjugated stabilizers $\bF^W$ form a \emph{stratum}, $W$, and different strata form partitions of the space~\footnote{If $\bF^w$ is the stabilizer of $w$, then the stabilizer of the point $gp$ (on the same orbit) is conjugate, i.e. $\bF^{gp}=g\bF^w g^{-1}$. Therefore, the entire  class of mutually conjugated stabilizers characterizes the orbit, and the same irreducible representations (IRs) can be denoted by the symbol $\bF^W$. In addition to these, there may exist isomorphic subgroups that are not conjugate, but their orbits remain geometrically similar.}.
The natural partial order within the set of strata is defined by the subgroup-supergroup relationship: $V\rightarrow W$ means that $\bF^W<\bF^V$. Consequently, stratum $V$  has a lower dimension than stratum $W$, and lies on the boundary of $W$.
An oriented \emph{connectivity graph} is associated with each group, with strata represented as nodes and the partial order defining directional arrows. Selecting one representative point from each orbit yields the \emph{fundamental domain}.

Being Abelian, the translational group $\bT$ has one-dimensional IRs only. The unitary IRs $\gD^{(\bk)}(\bg)=\re^{\ri\bk\cdot\bg}$ are parameterized by the wave vectors $\bk$ forming the Brillouin zone, which is torus $T^\wp$. The whole group acts on it by polar-vector representation: $g\bk=G\bk$. The stabilizers (which include all translations) are $\wp$-periodic groups, with finite orbits; again, the set of \emph{stars} (orbits) with mutually conjugate stabilizers form stratum, $K$, and stabilizer $\bF_K$ represents the class of the conjugated groups. The \emph{irreducible domain} is obtained by selecting one point from each orbit. The partial order among strata is  defined analogously to that in direct space. Allowed irreducible (generally projective) representations  of the stabilizers $\gd^{(\bK\gk\bk)}(\bF_K)$ induce $\bk$-series irreducible representations $D^{(\bK\gk\bk)}(\bG)=\gd^{(\bK\gk\bk)}(\bF_K)\uparrow\bG$, characterized by matrices of the same form along each stratum, with $\bk$-dependent elements arising from the factor system.
The quantum numbers  $\gk$ originate from the stabilizer $\bF_K$ (axial point groups, with IRs given in Tab.~\ref{TAPGIRs}); all the nonequivalent unitary IRs of the group are obtained in this way using only the irreducible domain. Each representation can be decomposed into IRs, $D(\bG)=\sum_{K\gk\bk} f^{K\gk\bk}(D)D^{(K\gk\bk)}(\bG)$, with unique \emph{frequency numbers} $f^{K\gk\bk}(D)$.

Analogously, in the Euclidean space, any representation $d(\bF^W)$ of a stabilizer induces representation of the group, $D(\bG)=d(\bF^W)\uparrow\bG$. Its restriction to the translational subgroup has fixed ($\bk$-independent) frequency numbers: $D(\bG)\downarrow\bT=f(D)\sum_{\bk\in\mathrm{ID}} \gD^{(\bk)}(\bT)$ (i.e. each IR of $\bT$ appears  $f(D)$-times). Representations with this property are called \emph{band representations} (BRs)~\cite{Zak0,ConnectBand,Cano-PRB-18}: the irreducible subspaces of $\bT$ are Bloch spaces $\cS_\bk$, all having the same dimension $\abs{\cS_\bk}=\abs{\cS}$, thus making a vector bundle over the Brillouin zone, and the spectrum of the Hamiltonian (commuting with the translational group) is structured into bands. A direct consequence is that the full decomposition of the band representation is
\begin{subequations}\label{EBand}
\begin{equation}\label{EBandSeries}
D(\bG)=\sum_{K\gk} f^{K\gk}(D)\sum_\bk D^{(K\gk\bk)}(\bG),
\end{equation}
i.e. the frequencies of the IRs of $\bG$ depend solely on the strata and isogonal group quantum numbers, being $\bk$-independent within each stratum.
Thus, the band representations are defined by a finite set of non-negative integers $f^{K\gk}$, corresponding to the finite number of strata and series of associated IRs (indexed by $K$ and $\gk$); this also yields the dimension equality:
\begin{equation}\label{EBandDim}
\abs{\cS}=\sum_{\gk} f^{K\gk}(D)\abs{D^{(K\gk\bk)}(\bG)}/\abs{\bk^*},
\end{equation}\end{subequations}
where $\abs{\bk^*}$ denotes the order of the star of $\bk$.
The requirements \eqref{EBand} provide a general definition of a band representation: in addition to induced representations, included are all positive integer combinations of them.

For certain groups, such as LGs, having a trivial symmetry indicator group~\cite{WatanabeSI2}, it is possible to select a finite set of EBRs, $E_i$ ($i=1,\dots,N$) such that any BR can be decomposed over them~\footnote{For some groups, there exist band representations that cannot be expressed as combinations of the induced representations, resulting in a to non-trivial \emph{symmetry indicator group}.}: $D(\bG)=\sum_if_i(\bG)E_i(\bG)$.
Indeed, set of elementary band representations consists of the induced representations derived from the IRs of the maximal stabilizers $\bF^M$:
\begin{equation}\label{EEbands}
E^{[M\mu]}(\bG)=d^{(\mu)}(\bF^M)\uparrow\bG.
\end{equation}
However, there are \emph{exceptional} cases where this procedure yields an EBR that is equivalent to another one induced from the IR of other maximal stabilizer (type 1,
$d^{(\mu)}(\bF^M)\uparrow\bG=d^{({\mu'})}(\bF^{M'})\uparrow\bG$), or that can be decomposed into other EBRs (type 2,
$d^{(\mu)}(\bF^M)\uparrow\bG=\sum_{M'\mu'}d^{({\mu'})}(\bF^{M'})\uparrow\bG$). In general, the decomposition of band representations into EBRs is \emph{not unique}.

In tight-binding models, the positions of atoms form one or more orbits of the system's symmetry group. Each atom contributes to the (infinite-dimensional) state space $\cS$  by some orbitals that span a finite-dimensional site space $\cS^{Ww}$. Here, $W$ indexes the orbits included in the system, while $w$ refers to atoms positioned on orbit $W$. The site space $\cS^{Ww}$ remains invariant under the site stabilizer $\bF^W$, meaning that $\cS^{Ww}$ carries a representation $d(\bF^W)$.
The same site space (and corresponding orbitals) is associated to each of the atoms of the same orbit. Thus, the total state space of the system is given by $\cS=\sum_{Ww}\cS^{Ww}$.
For a fully occupied band (or set of bands), Wannier functions yield the positions of the centers of the charge. By performing homothetic (symmetry and topology preserving) stretching, the inter-ionic distances can be gradually increased  beyond the interaction range, resulting in flat bands. In this limit, the charge centers reach what is known as the atomic limit, where they are expected to align with the positions of the ions. While this alignment typically holds, there are notable exceptions, known as \emph{obstructed atomic limits}, where the Wannier centers do not coincide with the ionic positions.

To gain insight from a purely group-theoretical perspective, we begin by examining the state space. Let us assume that all ions in the system occupy the Wyckoff position $W$, potentially comprising several orbits of this type.
As a result, the state space $\cS$ carries a band representation $D^W(\bG)=\sum_{W}d(\bF^W)\uparrow\bG$.
Then we define \emph{$W$-elementary band representations} ($W$-EBRs), which are induced from the IRs of $\bF^W$: $D^{W\mu}(\bG)=d^{\mu}(\bF^W)\uparrow\bG$.
Since the site representation  $d(\bF^W)$ can be decomposed into the IRs $d^{\mu}(\bF^W)$ of $\bF^W$ (that is, $d(\bF^W)=\sum_\mu f^\mu_W d^{(\mu)}(\bF^W)$), the induced band representation is: $D^W(\bG)=\sum_\mu f^\mu_W D^{W\mu}(\bG)$.
In the standard tight-binding framework, the electronic orbitals forming bonds are assumed to be centered at the ionic sites. However, an obstruction implies that the electronic orbitals are instead centered at a different stratum, $V$, as verified by the Wannier functions. For now, we assume the stabilizer orders are equal, $\abs{\bF^W}=\abs{\bF^V}$. Consequently, the same band representation $D^W$ can also be induced from stratum $V$: $D^W(\bG)=D^V(\bG)=\sum_\nu f^\nu_V D^{W\nu}(\bG)$. Given the clear analogy with exceptional EBRs, we refer to these BRs as \emph{exceptional band representations} (XBRS). In fact, an exceptional type 1 EBR is simply a specific case of this general concept.

Generalizations of this approach are straightforward. It is only necessary that at least a portion of the state space is filled by such interchangeable subrepresentations. Additionally, restriction to stabilizers of the same order is not required; it is not needed that the electronic orbitals involved in forming OALs share the same Wyckoff positions or the same types of strata.

In conclusion,  potential OALs are indicated when two Wyckoff positions, $A$ and $B$, share a common BR. Specifically, the representations $d^A(\bF^A)$ and $d^B(\bF^{B})$, induce the same BR. EBRs decompositions at $A$ and $B$ may be distinct, as decomposition of BR into EBRs need not be unique.
In addition, for any path within stratum $C$ connecting $A$ and $B$ (thus, the strata involved form a subgraph $\ACB$), there must be a representation $\ggg(\bC)$ that induces the same BR. Such intermediate stratum always exists (at least, a generic stratum adjacent to all other strata) providing a continuous path between two positions. Whenever two endpoints are distinct phases, with different EBRs components (although with the same sum), this path includes a singularity --- a point at which the energy gap closes; this phase transition is denoted as $\ABCTR$.

The succeeding step is to identify these triplets of strata and explore the conditions that lead to differences in the EBR decompositions corresponding to the Wyckoff positions $A$ and $B$. Before that, a brief overview of LGs is presented.

\section{Line groups}\label{SLGs}

The symmetry of systems that are periodic in a single direction (conveniently taken along the z-axis) is described by LGs. These groups are products of the form $\bL=\bZ(f)\bP_n$, where, $\bZ(f)$ represents one of the generalized translations: $\bT(f)$, a pure translational group (generated by $\kost{\one}{f}$, indicating a translation for $f$  along the $z$-axis), a helical group $\bT_Q(f)$ (generated by $\kost{C_Q}{f}$, where $C_Q$ is a rotation by $2\pi/Q$ around the $z$-axis, with $Q\ge1$ being a real number), a glide plane group $\bT'(f)$ (generated by $\kost{\gs_\mathrm{v}}{f}$, where $\gs_\mathrm{v}$ is a vertical mirror plane). The group $\bP_n$ is one of the seven axial point groups: $\bC_n$, $\bS_n$, $\bC_{n\mathrm{h}}$, $\bC_{n\mathrm{v}}$, $\bD_n$,  $\bD_{n\mathrm{d}}$,  $\bD_{n\mathrm{h}}$. There are a total of 13 \emph{families of LGs} among these products, differing by $n$, and also by $Q$ in the 1st and 5th LG-families.

\begin{figure*}[hbt]
\includegraphics[width=12cm]{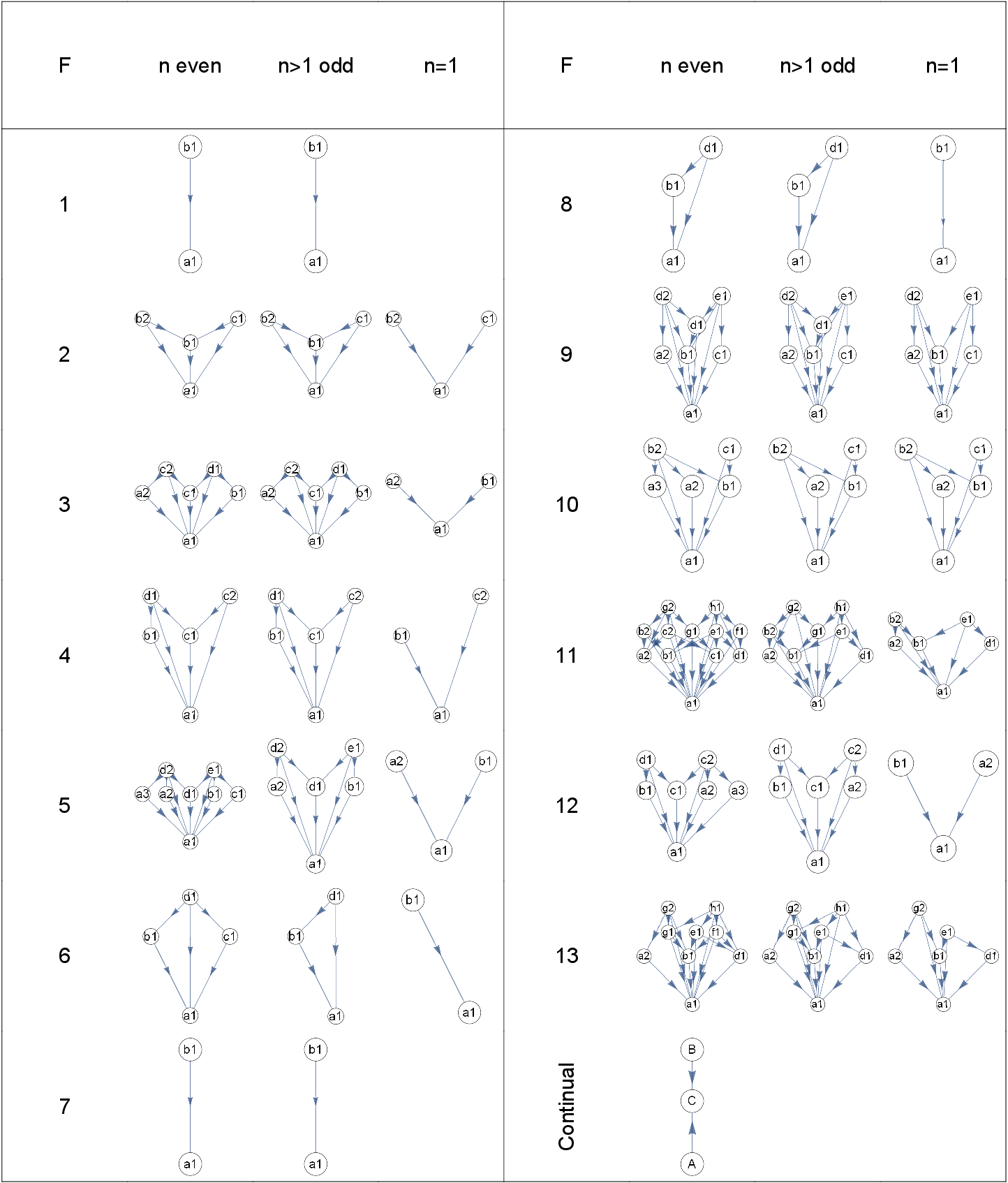}
\caption{\label{F1LGConGr} Connectivity graphs of LGs. For each LG-family (Column F), the graphs for even and odd
$n$ are given in separate columns, along with the case for $n=1$ (degeneration of the general case).
The strata (Wyckoff positions) are labeled as defined in Refs.~\cite{IYJT,YILG}.
The final graph illustrates a triplet of discrete strata corresponding to all continuous  LGs having two maxima.}
\end{figure*}

Subsequently, a complete stratification (orbits and stabilizers) is carried out as described in \cite{IYJT,YILG}. For each LG-family, the strata are listed, with certain distinctions for odd and even $n$, and also for specific exceptions for $n=1$. Knowing the stabilizers, the connectivity graphs (Fig.~\ref{F1LGConGr}) can be constructed.

\emph{Positive groups}, those without elements reverzing $z$-axis,  have a single maximal stratum along the $z$-axis, while the remaining, \emph{negative groups}, possess two maximal strata with representatives at $z=0$ and $z=\tfrac{a}2$), each fixed by $\bP_n$ (the latter is the isomorphic group but with negative symmetry elements, $\kost{\gs_h}{a}$, $\kost{U}{a}$),  positioned at $z=\tfrac{a}2$ and connected by the intermediate stratum on the $z$-axis, with stabilizer being positive subgroup of $\bP_n$.
From the points of maximal strata action of the translational group generates simple chains, while from each point of intermediate stratum generated is double chain, with two points separated by a distance different from $a/2$ per period.

The IRs of LGs have been found and tabulated~\cite{LGIR1,LGIR2,YILG}. The Brillouin zone is a circle; for
the families of \emph{positive groups} (those without elements that reverse the $z$-axis) 1, 6, 7, and 8, this circle is also irreducible domain, consisting the generic stratum $G$, only, while the irreducible domain of the \emph{negative groups}, LG-families 2-5 and 9-13, includes besides the generic stratum $G'=(0,\pi)$, the point strata $\gG=\{0\}$ and $\Pi=\{\pi\}$.
In addition to the wave vector $\bk=k$ (the 1D Brillouin zone), there are the quantum numbers $\gk$ which originate in the isogonal group $\bI=\bG/\bT$. Since $\bI$ is an axial point group, these quantum numbers include $m$, arising from rotations around the $z$-axis, and the parities $\dU$, $\dH$ and $\dV$ associated with the $U$-axis, horizontal and vertical mirror planes, respectively.

\squeezetable
\begin{table}[h]
\caption{\label{T2E} Groups $\bL^{(2)}=\bT\bS_{2n}$. $\bI=\bS_{2n}$. The strata $a_1$, $b_1$, $c_1$ and $b_2$ have stabilizers  $\bC_1$, $\bC_n$,  $\bS_{2n}$ and   $\bS_{2n}$ with respective orders  1, $n$, $2n$ and $2n$. For $n=1$ the $b_1$ stratum does not exist (it merges with $a_1$).}
\begin{tabular}{r|ll|llll}
\hline
No&$ABC$&Orders&XBRs decomposition&R&D&T\\\hline
$1_a$&$b_2c_1b_1$&$2n,2n,n$&$\sum_hD^{[b_2mh]}=\sum_hD^{[c_1mh]}$&
    $I$&$2,2,1$&$X$\\\hline
\end{tabular}\end{table}

\squeezetable
\begin{table}[h]
\caption{\label{T3E} Groups $\bL^{(3)}=\bT\bC_{nh}$. $\bI=\bC_{nh}$. The strata $a_1$, $a_2$, $b_1$, $c_1$, $c_2$ and $d_1$ have stabilizers  $\bC_1$, $\bC_{1h}$,  $\bC_{1h}$, $\bC_n$, $\bC_{nh}$, and   $\bC_{nh}$, with respective orders  1, 2, 2, $n$, $2n$ and $2n$. For $n=1$ the strata $c_1$, $c_2$ and $d_1$ do not exist (they merge with $a_1$, $a_2$ and $b_1$, with $a_2$ and $b_1$ maximal).}
\begin{tabular}{r|ll|llll}
\hline
No&$ABC$&Orders&XBRs decomposition&R&D&T\\\hline
$1_a$&$c_2d_1c_1$&$2n,2n,n$&$\sum_hD^{[c_2mh]}=\sum_hD^{[d_1mh]}$&$I$&$2,2,1$&$X$\\\hline
\end{tabular}\end{table}

For the 1st and 5th LG-families, it is often more convenient (and in incommensurate cases, necessary) to use the helical $\bT_Q$ instead of the translational group. This approach  yields a helical Brillouin zone, a helical isogonal group $\bI_H=\bG/\bT_Q$, and helical momenta quantum numbers.

Elementary band representations for LGs, for both spinless and spinful systems, with and without time reversal symmetry, have been published recently~\cite{LGEBR,YILG}.

The class of groups preserving  $z$-axis includes five \emph{continuous LGs}\cite{YILG}. These groups can be expressed as products $\bG=\bT\bP_\infty$, where $\bT$ is the discrete translational subgroup and $\bP_\infty$ is the subgroup of O(3) that preserves the $z$-axis, containing SO(2)=$\bC_\infty$. The positive groups are  $\bT \bC_{\infty}$ and $\bT \bC_{\infty v}$, while the negative ones are $\bT \bC_{\infty h}$, $\bT \bD_{\infty}$ and $\bT \bD_{\infty h}$. The characteristic of their IRs is that the quantum number of angular momentum, $m$, is any integer or half-integer in positive groups, and a natural or half-natural number in negative groups. For these groups maximal strata we denote by $A$ and $B$, and the intermediate one by $C$. Their orbits, the same as for the discrete groups, are described above.

\squeezetable
\begin{table}
\caption{\label{T4E} Groups $\bL^{(4)}=\bT_{2n}\bC_{nh}$. $\bI=\bC_{2nh}$. The strata $a_1$, $b_1$, $c_1$, $c_2$, $d_1$ and $c_1$ have stabilizers  $\bC_1$, $\bC_{1h}$, $\bC_n$, $\bS_{2n}$, and  $\bC_{nh}$ with respective orders  1, 2, $n$, $2n$ and $2n$. For $n=1$ strata $c_1$ and $d_1$ do not exist (they merge with $a_1$ and $b_1$; maximal $c_2$ and $b_1$).}
\begin{tabular}{r|ll|llll}
\hline
No&$ABC$&Orders&XBRs decomposition&R&D&T\\\hline
$1_a$&$c_2d_1c_1$&$2n,2n,n$&$\sum_hD^{[c_2mh]}=\sum_hD^{[d_1mh]}$&$I$&$2,2,1$&$X$\\\hline
\end{tabular}\end{table}

\squeezetable
\begin{table}
\caption{\label{T5E} Groups $\bL^{(5)}=\bT_{Q}\bD_{n}$. $\bI_H=\bD_{n}$. The strata $a_1$, $a_2$, $a_3$, $b_1$, $c_1$, $d_1$, $d_2$ and $e_1$ have stabilizers  $\bC_1$, $\bD_{1}$, $\bD_{1}$, $\bD_{1}$, $\bD_{1}$, $\bC_n$, $\bD_{n}$, and  $\bD_{n}$ with respective orders  1, 2, 2, 2, 2,  $n$, $2n$ and $2n$.
Strata $a_3$ and $c_1$ appear only in the $n$-even groups.
Additionally, for $n=1$ there are no strata $d_1$, $d_2$ and $e_1$ (they merge with $a_1$, $a_2$ and $b_1$; maximal $a_2$ and $b_1$).}
\begin{tabular}{r|ll|llll}
\hline
No&$ABC$&Orders&XBRs decomposition&R&D&T\\\hline
$1_a$&$d_2e_1d_1$&$2n,2n,n$&$\sum_uD^{[e_10h]}=\sum_uD^{[d_20h]}$&$0$&$2,2,1$&$1_i$\\
  &&&$\sum_hD^{[e_1\frac{n}2h]}=\sum_hD^{[d_2\frac{n}2h]}$&$\tfrac{n}2$&$2,2,1$&$1^e$\\
  &&&$D^{[e_1m0]}\exc= D^{[d_2m0]}$&$I^0$&$2,2,1$&$Y$\\ \hline
\end{tabular}\end{table}

\section{Centres of charge}\label{STheorems}

The initial step in the algebraic classification of atomic limits is to identify all possible subgraphs $\ACB$ within the connectivity graph of the group. For each of these subgraphs, the IRs $d^{(\mu)}(\bF^W)$ of $\bF^W$ (where $W=A,B,C$) are induced to the band representations $D^{W\mu}(\bG)$.
Next, the equation $\sum_\ga a_\ga D^{A\ga}=\sum_\gb b_\gb D^{B\gb}=\sum_\ggg c_\ggg D^{C\ggg}$ is solved for the coefficients $a_\ga$, $b_\gb$, and $c_\ggg$, allowing the same band representation (BR) to be expressed as a combination of the $A$-EBRs, $B$-EBRs and $C$-EBRs, respectively:
\begin{equation}\label{EOALBRs}
B^i_{ABC}=\sum_\ga a^i_\ga D^{A\ga}=\sum_\gb b^i_\gb D^{B\gb}=\sum_\ggg c^i_\ggg D^{C\ggg},
\end{equation}
where $i=1,\dots N$ counts the independent solutions in nonnegative integers\footnote{Note that for each $B^i_{ABC}$,  the solutions for $a^i$, $b^i_\gb$ and $c^i_\ggg$ are unique. In fact,  by assumption, $B^i_{ABC}$ is induced, and for each $W=A,B,C$ there exists a $d^i( \bF^W)$ that induces it. The coefficients are simply frequencies with which $d^i( \bF^W)$  decomposes onto the IRs of $\bF^W$.}. The obtained $N$ independent BRs  are decomposable into EBRs, which is achieved by explicitly solving the equations
\begin{equation}\label{EOALEBRs}
B^i_{ABC}=\sum_j e^i_{js} E^j,
\end{equation}
in terms of the coefficients $e^i_{js}$. Different solutions (for the same $i$) are indexed by $s$.

This algorithm extracts the basis $B^i$ for BRs (non-negative integer combinations of $B^i$) that can be generated by any of the strata $A$, $B$ and $C$: physically, the stabilizer IRs ($d^{(\ga)}(\bF^A)$, $d^{(\gb)}(\bF^B)$ and $d^{(\ggg)}(\bF^C)$, respectively) represent possible orbital models (local spaces for tight-binding) that enable electronic orbitals at each of the three Wyckoff positions. In other words, this kinematic model allows charge centers to occupy any of these positions, while the calculation of Wannier functions for the electronic bands ultimately determines their exact locations.

Note that for each pair of strata, $A$ and $B$, there is at least one additional stratum, $C$ (typically the generic stratum), that completes the target subgraph.
For each subgraph, the corresponding solutions $B^i$ arise naturally from the transitivity of induction.
Starting with the central stratum and its IR $d^{(\ggg)}(\bF^C)$, one directly induces $D^{C\ggg}(\bG)$. By the transitivity of induction, this gives
$D^{C\ggg}(\bG)=(d^{(\ggg)}(\bF^C)\uparrow\bF^A)\uparrow\bG$, which implies that the decomposition of $d^{(\ggg)}(\bF^C)\uparrow\bF^A$ into the IRs of $\bF^A$
directly provides the decomposition of $D^{C\ggg}(\bG)$ into $D^{A\ga}(\bG)$. An important conclusion follows: since there exists a maximal stratum
$M_A\rightarrow A$ (with $M_A=A$ if $A$ is already maximal), the same approach can be used to find the decomposition into EBRs: $D^{C\ggg}(\bG)=\sum_{\mu_A}f^{M_A\mu_A}E^{[M_A\mu_A]}$. As the same arguments apply also to $B$, the following scenarios arise:

\begin{enumerate}
\item\emph{Homo-nested} $ABC$ ($M_A=M_B=M_C=M$) results in a shared decomposition of the induced representations along both directions, containing only $M$-BBRs, and excluding those from other maximal Wyckoff positions:
\begin{subequations}\label{EDec}\begin{equation}\label{EDecM}
B_{ABC}=\sum_{\mu}f^{M\mu}D^{[M\mu]}.
\end{equation}
Of course, if any BR from \eqref{EDecM} is a type 2 exception, $D^{[M\mu]}=D^{\{M\mu\}}$, it is replaced by its decomposition into EBRs from other sites.
\item Otherwise, there is \emph{hetero-nested triplet}. In this case, the total BR decomposes into EBRs from different maximal Wyckoff positions $M_1$ and $M_2$:
\begin{equation}\label{EDecMAMB}
B_{ABC}=\sum_{\mu_1}f^{M_1\mu_1}D^{[M_1\mu_1]}=\sum_{\mu_2}f^{M_2\mu_2}D^{[M_2\mu_2]}.
\end{equation}\end{subequations}
If only one summand exists on either the $M_1$ or $M_2$ side, it represents a type 1 exceptional EBR. If only one side has a single EBR, then the other side corresponds to a type 2 exception. More complex cases arise due to linear dependencies among EBRs, allowing the inclusion of normal EBRs as well. If either stratum
$A$ or $B$ is subordinated to two maximal strata, then the decomposition of common BRs includes EBRs from both maximal strata.
\end{enumerate}

\squeezetable
\begin{table*}
\caption{\label{T9E} Groups $\bL^{(9)}=\bT\bD_{nd}$. $\bI=\bD_{nd}$; strata $a_1$, $a_2$, $b_1$, $c_1$, $d_1$, $d_2$ and $e_1$ have stabilizers  $\bC_1$, $\bC_{1v}$, $\bD_{1}$, $\bC_{1v}$, $\bC_{nv}$, $\bD_{nd}$, and  $\bD_{nd}$ with orders  1, 2, 2, 2, $2n$, $2n$, $4n$ and $4n$, respectively.
For $n=1$ there is no stratum  $d_1$ (becomes $b_1$). In lines 1 and 2 we use:
$X^v_{d_2}=\begin{cases}D^{[d_2\frac{n}200]}& n\text{ even}\\ \sum_{h}D^{[d_2\frac{n}2vh]}& n\text{ odd}\end{cases}$,
$X^v_{e_1}=\begin{cases}D^{[e_1\frac{n}200]}& n\text{ even}\\ \sum_{h}D^{[e_1\frac{n}2hv]}& n\text{ odd}\end{cases}$
(parity $v$ only for $n$ odd).}
\begin{tabular}{r|ll|llll}
\hline
No&$ABC$&Orders&$D^{M_A}=D^{M_B}$&R&D&T\\\hline
$1_e$&$d_2d^0_1b^0_1$&$4n,2n,2$&$\sum_{mh}D^{[d_2m0h]}+\sum_{h}D^{[d_20vh]}+X^v_{d_2}=
                                 \sum_{mh}D^{[e_1m0h]}+\sum_{h}D^{[e_10hv]}+X^v_{e_1}$&
     $I^0,v$&$2n,n,1$&$Z$\\
$2_e$&$d^0_1e_1b^0_1$&$2n,4n,2$&$\sum_{mh}D^{[d_2m0h]}+\sum_{h}D^{[d_20vh]}+X^v_{d_2}=
                                   \sum_{mh}D^{[e_1m0h]}+\sum_{h}D^{[e_10hv]}+X^v_{e_1}$&
     $I^0,v$&$n,2n,1$&$Z$\\
$3_a$&$d_2e_1d^0_1$&$4n,4n,2n$&$\sum_hD^{[d_2m0h]}=\sum_uD^{[e_1m0h]}$&
              $I^0$&4,4,2&$2Y$\\
 &           &&$\sum_hD^{[d_20vh]}=\sum_vD^{[e_10hv]}$&$v$&$2,2,1$&$2_i$\\
 &           &&$\begin{cases}D^{[d_2\frac{n}200]}\exc=D^{[e_1\frac{n}200]}&n\text{ even}\\
                \sum_hD^{[d_2\frac{n}2vh]}=\sum_hD^{[e_1\frac{n}2h\ov{v}]}&n\text{ odd}\end{cases}$&${ \atop v}$&$2,2,1$&${1^e_i\atop2^o_h}$\\\hline
\end{tabular}\end{table*}

In the case with a single maximal stratum, the decomposition onto EBRs is identical for both $A$ and $B$, as indicated by \eqref{EDecM}. This implies that Hamiltonians describing these configurations as the optimal ones are topologically equivalent, the both configurations pertain to the same topological phase, and there is no gap closure in the course of  transition along $C$. Specifically, as the connectivity graph of the positive LGs has a single maximum, so such systems do not have topologically constrained OALs exhibiting distinct band topologies.

Furthermore, in the spinless case, if $C$ is a generic stratum, $C=a_1$, with a trivial stabilizer, $\bF^C=\{e\}$, having only the unit IR $\one(\bC)$, then the only
$C$-EBR  is $D^{C1}$, which is the regular representation $R$ of $\bG$.  Additionally, intermediate induction to any subgroup, $\bF^A$ and $\bF^B$ in particular, also yields the regular representation of the subgroup. Therefore, all common BRs induced from $a_1$, are simply multiples of $R$, regardless of the intermediate groups $A$ or $B$. When spin is taken into account, the double groups are relevant, and the stabilizer of the generic stratum is $\ti{\bC}_1$. It contains the identity and a $2\pi$-rotation, with IRs $m=0$ (unit IR) and $m=\tfrac12$, respectively (Appendix~\ref{SNotation}). When these representations are induced to $\wti{\bG}$, they yield $D^{C0}$ and $D^{C\un1}$, being integer and half-integer regular representations, respectively, each containing only integer and half-integer IRs, with frequencies matching their dimensions.
This conclusion applies equally to all cases (not only to generic stratum, when it is inherently required) where the solutions of equation \eqref{EOALBRs} include (a multiple of) the regular representation of $\bF^C$.

\squeezetable
\begin{table}[hbt]
\caption{\label{T10E} Groups $\bL^{(10)}=\bT_c\bS_{2n}$. $\bI=\bD_{nd}$. The strata $a_1$, $a_2$, $a_3$, $b_1$, $b_2$ and $c_1$ have stabilizers  $\bC_1$, $\bD_{1}$, $\bD_{1}$, $\bC_{n}$, $\bD_{n}$ and  $\bS_{2n}$ with respective orders  1, 2, 2, $n$, $2n$ and  $2n$.
Stratum $a_3$ appears only in the $n$-even groups. Strata $b_1$ and $b_2$ for $n=1$ become $a_1$ and $a_2$ (maximal $c_1$ and $a_2$).}
\begin{tabular}{r|ll|llll}
\hline
No&$ABC$&Orders&$D^{M_A}=D^{M_B}$&R&D&T\\\hline
$1^1_a$&$b_2c_1b_1$&$2n,2n,n$&$\sum_hD^{[b_20h]}=\sum_hD^{[c_10h]}$&$0$&$2,2,1$&$1_i$\\
 &           &&$\sum_hD^{[b_2\frac{n}2h]}=\sum_hD^{[c_1\frac{n}2h]}$&$\tfrac{n}2$&$2,2,1$&$1^e$\\
 &           &&$D^{\{b_2m0\}}\Lexc= \sum_hD^{[c_1mh]}$&$I^0$&$2,2,1$&$Y$\\\hline
\end{tabular}\end{table}

\squeezetable
\begin{table*}[hbt]
\caption{\label{T11E} Groups $\bL^{(11)}=\bT\bD_{nh}$. $\bI=\bD_{nh}$; strata $a_1$, $a_2$, $b_1$, $b_2$, $c_1$, $c_2$, $d_1$, $e_1$, $f_1$, $g_1$, $g_2$, $h_1$; stabilizers
 $\bC_1$, $\bC_{1h}$, $\bC_{1v}$, $\bD_{1h}$, $\bD_{1}$, $\bD_{1h}$, $\bC_{1h}$, $\bD_{1h}$,
 $\bD_{1h}$,      $\bC_{nv}$,   $\bD_{nh}$, $\bD_{nh}$; orders 1, 2, 2, 4, 2, 4, 2, 4, 4, $2n$, $4n$,  $4n$.
 Strata $c_1$, $c_2$ and $f_1$ appear only in the $n$-even groups;
 for $n=1$ also strata $g_1$, $g_2$ and $h_1$ become $b_1$, $b_2$ and $e_1$ (maximal $b_2$ and $e_1$).}

\begin{tabular}{l|ll|llll}
\hline
No&$ABC$&$D^{M_A}=D^{M_B}$&R&D&T\\\hline
$1^e$&$b^{g_2}_2c^{g_2}_2a^{g_2}_2$
       &$\sum_{mh}D^{[g_2m0h]}=\sum_{mh}D^{[h_1m0h]}$
       &$I^0_h$&$2,2,1$&$1_h$\\
  &$4,4,2$
       &$2\sum_{mh}D^{[g_2m0h]}+\sum_{h}D^{[g_20v(vh)]}+\sum_{h}D^{[g_2\tfrac{n}2v(vh)]}=$NO ($h_1$)
       &$I^0_i,v$&$2,2,1$&$2_i$\\

$2^1_g$&$b^{g_2}_2e^{h_1}_1b^0_1$
       &$\sum_{mh}D^{[g_2m0h]}+D^{[g_2\tfrac{n}200]}=\sum_{mh}D^{[h_1m0h]}+D^{[h_1\tfrac{n}200]}$
       &$I^0_h$&$2,2,1$&$1_h$\\
    &$4,4,2$
       &$\sum_{mh}D^{[g_2m0h]}+\sum_h(D^{[g_20vh]}+D^{[g_2\tfrac{n}2vh]})=
         \sum_{mh}D^{[h_1m0h]}+\sum_h(D^{[h_10hv]}+D^{[h_1\tfrac{n}2hv]})$
       &$I^0_i,v$&$2,2,1$&$2_i$\\

$3_f$&$b^{g_2}_2g^0_1b^0_1$
       &$\sum_{mh}D^{[g_2m0h]}+D^{[g_2\tfrac{n}200]}=\sum_{mh}D^{[h_1m0h]}+D^{[h_1\tfrac{n}200]}$
       &$I^0_h$&$2,n,1$&$1_h$\\
  &$4,2n,2$
       &$\sum_{mh}D^{[g_2m0h]}+\sum_h(D^{[g_20vh]}+D^{[g_2\tfrac{n}2vh]})=
         \sum_{mh}D^{[h_1m0h]}+\sum_h(D^{[h_10hv]}+D^{[h_1\tfrac{n}2hv]})$
       &$I^0_i,v$&$2,n,1$&$2_i$\\

$4$&$b^{g_2}_2g_2a^{g_2}_2$
       &$\sum_{mh}D^{[g_2m0h]}=\sum_{mh}D^{[h_1m0h]}$
       &$I^0_h$&$2,2n,1$&$1_h$\\
   &$4,4n,2$
       &$2\sum_{mh}D^{[g_2m0h]}+\sum_{h}D^{[g_20v(vh)]}+\sum_{h}D^{[g_2\tfrac{n}2v(vh)]}=$NO ($h_1$)
       &$I^0_i,v$&$2,2n,1$&$2_i$\\

$5^e_g$&$c^{g_2}_2e^{h_1}_1c^0_1$
       &$\sum_{mh}D^{[g_2m0h]}=\sum_{mh}D^{[h_1m0h]}$
       &$I^0_h$&$2,2,1$&$1_h$\\
    &$4,4,2$
       &$2\sum_{mh}D^{[g_2m0h]}+\sum_{vh}(D^{[g_20vh]+D^{[g_2\tfrac{n}2vh]}})=
         2\sum_{mh}D^{[h_1m0h]}+\sum_{vh}(D^{[h_10vh]+D^{[h_1\tfrac{n}2vh]}})$
       &$I^0_i,v$&$4,4,2$&$2_i$\\

$6^e_f$&$c^{g_2}_2g^0_1c^0_1$
       &$\sum_{mh}D^{[g_2m0h]}+D^{[g_2\tfrac{n}200]}=\sum_{mh}D^{[h_1m0h]}+D^{[h_1\tfrac{n}200]}$
       &$I^0_h$&$2,n,1$&$1_h$\\
  &$4,2n,2$
       &$\sum_{mh}D^{[g_2m0h]}+\sum_h(D^{[g_20vh]}+D^{[g_2\tfrac{n}2vh]})=
         \sum_{mh}D^{[h_1m0h]}+\sum_h(D^{[h_10hv]}+D^{[h_1\tfrac{n}2hv]})$
       &$I^0_i,v$&$2,n,1$&$2_i$\\

$7^e$&$c^{g_2}_2g_2a^{g_2}_2$
       &$\sum_{mh}D^{[g_2m0h]}=\sum_{mh}D^{[h_1m0h]}$
       &$I^0_h$&$2,2n,1$&$1_h$\\
   &$4,4n,2$
       &$2\sum_{mh}D^{[g_2m0h]}+\sum_{h}D^{[g_20v(vh)]}+\sum_{h}D^{[g_2\tfrac{n}2v(vh)]}=$NO ($h_1$)
       &$I^0_i,v$&$2,2n,1$&$2_i$\\

$8^e$&$e^{h_1}_1f^{h_1}_1d^{h_1}_1$
       &$\sum_{mh}D^{[g_2m0h]}=\sum_{mh}D^{[h_1m0h]}$
       &$I^0_h$&$2,2,1$&$1_h$\\
   &$4,4,2$
       &NO ($g_2$) $=2\sum_{mh}D^{[h_1m0h]}+\sum_{h}(D^{[h_10v(vh)]}+D^{[h_1\tfrac{n}2v(vh)]})$
       &$I^0_i,v$&$2,2,1$&$2_i$\\

$9_f$&$e^0_1g^0_1b_1$
       &$\sum_{mh}D^{[g_2m0h]}+D^{[g_2\tfrac{n}200]}=\sum_{mh}D^{[h_1m0h]}+D^{[h_1\tfrac{n}200]}$
       &$I^0_h$&$2,n,1$&$1_h$\\
   &$4,2n,2$
       &$\sum_{mh}D^{[g_2m0h]}+\sum_h(D^{[g_20vh]}+D^{[g_2\tfrac{n}2vh]})=
         \sum_{mh}D^{[h_1m0h]}+\sum_h(D^{[h_10hv]}+D^{[h_1\tfrac{n}2hv]})$
       &$I^0_i$&$2,n,1$&$2_i$\\

$10^e_f$&$e^{h_1}_1g^0_1c^0_1$
       &$\sum_{mh}D^{[g_2m0h]}=\sum_{mh}D^{[h_1m0h]}$
       &$I^0_h$&$2,2,1$&$1_h$\\
  &$4,2n,2$
       &$2\sum_{mh}D^{[g_2m0h]}+\sum_{vh}(D^{[g_20vh]+D^{[g_2\tfrac{n}2vh]}})=
         2\sum_{mh}D^{[h_1m0h]}+\sum_{vh}(D^{[h_10vh]+D^{[h_1\tfrac{n}2vh]}})$
       &$I^0_i,v$&$2,2,1$&$2_i$\\

11&$e^{h_1}_1h_1d^{h_1}_1$
       &$\sum_{mh}D^{[g_2m0h]}=\sum_{mh}D^{[h_1m0h]}$
       &$I^0_h$&$2,2n,1$&$1_h$\\
  &$4,4n,2$
       &NO ($g_2$) $=2\sum_{mh}D^{[h_1m0h]}+\sum_{h}(D^{[h_10v(vh)]}+D^{[h_1\tfrac{n}2v(vh)]})$
       &$I^0_i,v$&$2,2n,1$&$2_i$\\

$12^e$&$f^{h_1}_1h_1d^{h_1}_1$
       &$\sum_{m}D^{[g_2m0h]}=\sum_{mh}D^{[h_1m0h]}$
       &$I^0_h$&$2,2n,1$&$1_h$\\
   &$4,4n,2$
        &NO ($g_2$) $=2\sum_{mh}D^{[h_1m0h]}+\sum_{h}(D^{[h_10v(vh)]}+D^{[h_1\tfrac{n}2v(vh)]})$
        &$I^0_i,v$&$2,2n,1$&$2_i$\\

$13_a$&$g_2h_1g^0_1$
        &$\begin{cases}\sum_hD^{[g_2\frac{n}2vh]}=\sum_hD^{[h_1\frac{n}2hv]}&n\text{ even }\\
                       D^{[g_2\frac{n}200]}=D^{[h_1\frac{n}200]}&n\text{ odd }\end{cases} $
        &${\tfrac{n}2,v\atop\tfrac{n}2}$&$2,2,1$&${2^e_i\atop 1^o_h}$\\
    &$4n,4n,2n$
        &$\sum_hD^{[g_20vh]}=\sum_hD^{[h_10hv]}$
        &$0,v$&$2,2,1$&$2_i$\\
       &&$\sum_hD^{[g_2m0h]}=\sum_hD^{[h_1m0h]}$
        &$I^0$&$4,4,2$&$2Y$\\\hline

\end{tabular}\end{table*}

\squeezetable
\begin{table}[hbt]
\caption{\label{T12E} Groups $\bL^{(12)}=\bT_c\bC_{nh}$. $\bI=\bD_{nh}$. The strata $a_1$, $a_2$, $a_3$, $b_1$, $c_1$, $c_2$ and $d_1$ have stabilizers
 $\bC_1$, $\bD_{1}$, $\bD_{1}$, $\bC_{1h}$,  $\bC_{n}$, $\bD_{n}$  and  $\bC_{nh}$ with respective orders 1, 2, 2, 2, $n$, $2n$  and  $2n$.
 Stratum $a_3$ appears only in the $n$-even groups.
 In addition, for $n=1$ strata $c_1$, $c_2$ and $d_1$ become $a_1$, $a_2$ and $b_1$ (maximal: $a_2$ and $b_1$).}
\begin{tabular}{r|ll|llll}
\hline
No&$ABC$&Orders&$D^{M_A}=D^{M_B}$&R&D&T\\\hline
$1_a$&$d_1c_2c^0_1$&$2n,2n,n$&
        $\sum_hD^{[d_1\frac{n}2h]}=\sum_hD^{[c_2\frac{n}2h]}$&$\frac{n}2,h$&$2,2,1$&$1$\\
  &   &&$\sum_hD^{[d_10h]}=\sum_hD^{[c_20h]}$&$0,h$&$2,2,1$&$1_i$\\
  &   &&$\sum_hD^{[d_1](\pm m)h]}\Lexc=D^{[c_2m]}$&$I^0,h$&$2,2,1$&$2Y$\\\hline
\end{tabular}\end{table}

\begin{table*}[hbt]
\caption{\label{T13E} Groups $\bL^{(13)}=\bT_{2n}\bD_{nh}$. $\bI=\bD_{2nh}$; strata $a_1$, $a_2$, $b_1$, $d_1$, $e_1$, $f_1$, $g_1$, $g_2$ and $h_1$ have stabilizers
 $\bC_1$, $\bD_{1}$, $\bC_{1v}$, $\bC_{1h}$,  $\bD_{1h}$, $\bD_{1h}$, $\bC_{nv}$, $\bD_{nd}$  and  $\bD_{nh}$ with orders 1, 2, 2, 2, 4, 4, $2n$, $4n$  and  $4n$, respectively.
 Stratum $f_1$ appears only in the $n$-even groups, hence the triples of strata with it are only in such groups;
 in $n=1$ case, strata $g_1$ and $h_1$ become $b_1$ and $e_1$ (maximal $g_2$ and $e_1$).}
\begin{tabular}{r|l|llll}
\hline
No&$ABC$&$D^{M_A}=D^{M_B}$&R&D&T\\\hline

$1^e_h$&$e^{h_1}_1f^{h_1}_1b^0_1$
     &$\sum_{mh}D^{[g_2m0h]}=\sum_{mh}D^{[h_1m0h]}$
     &$I^0_h,h$&$2,2,1$&$1_h$\\
  &$4,4,2$
     &$\sum_{mh}D^{[g_2m0h]}+\sum_{h}D^{[g_20hv]}+D^{[g_2\frac{n}200]}=
       \sum_{mh}D^{[h_1m0h]}+\sum_{h}D^{[h_10hv]}+D^{[h_1\frac{n}2hv]}$
     &$I^0_i,v,h$&$2,2,1$&$2_i$\\
$2^e$&$e^{h_1}_1f^{h_1}_1d^{h_1}_1$
     &$\sum_{mh}D^{[g_2m0h]}=\sum_{mh}D^{[h_1m0h]}$
     &$I^0_h,h$&$2,2,1$&$1_h$\\
  &$4,4,2$
     &NO ($g_2$)$=2\sum_{mh}D^{[h_1m0h]}+\sum_{h}D^{[h_10h(vh)]}+D^{[h_1\frac{n}2h(vh)]}$&
           $I^0_i,v,h$&$2,2,1$&$2_i$\\
$3_f$&$e^{h}_1g^0_1b^0_1$
     &$\sum_{mh}D^{[g_2m0h]}=\sum_{mh}D^{[h_1m0h]}$
     &$I^0_h,h$&$2,n,1$&$1_h$\\
  &$4,2n,2$
     &$\sum_{mh}D^{[g_2m0h]}+\sum_{h}D^{[g_20vh]}+D^{[g_2\frac{n}200]}=
       \sum_{mh}D^{[h_1m0h]}+\sum_{h}D^{[h_10hv]}+D^{[h_1\frac{n}2hv]}$
     &$I^0_i,v,h$&$2,n,1$&$2_i$\\
$4^1_d$&$e^{h_1}_1g_2b^0_1$
     &$\sum_{mh}D^{[g_2m0h]}+D^{[g_2\tfrac{n}2vh]}=\sum_{mh}D^{[h_1m0h]}+$
     &$I^0_h,v,h$&$2,2n,1$&$1_h$\\
 &$4,4n,2$
     &$\sum_{mh}D^{[g_2m0h]}+\sum_{h}D^{[g_20hv]}+D^{[g_2\frac{n}200]}=
       \sum_{mh}D^{[h_1m0h]}+\sum_{h}D^{[h_10hv]}+D^{[h_1\frac{n}2hv]}$
     &$I^0_i,v,h$&$2,2n,1$&$2_i$\\
$5$&$e^{h_1}_1h_1d^{h_1}_1$
     &$\sum_{mh}D^{[g_2m0h]}=\sum_{mh}D^{[h_1m0h]}$
     &$I^0_i,v,h$&$2,2n,1$&$1_h$\\
  &$4,4n,2$
      &NO ($g_2$)$=2\sum_{mh}D^{[h_1m0h]}+\sum_{h}D^{[h_10h(vh)]}+D^{[h_1\frac{n}2h(vh)]}$
      &$I^0_h,h$&$2,2n,1$&$2_i$\\
$6^e_f$&$f^{h_1}_1g^0_1b^0_1$
       &$\sum_{mh}D^{[g_2m0h]}=\sum_{mh}D^{[h_1m0h]}$
       &$I^0_h,v,h$&$2,n,1$&$1_h$\\
   &$4,2n,2$
       &$\sum_{mh}D^{[g_2m0h]}+\sum_{h}D^{[g_20hv]}+D^{[g_2\frac{n}200]}=
         \sum_{mh}D^{[h_1m0h]}+\sum_{h}D^{[h_10hv]}+D^{[h_1\frac{n}2hv]}$
       &$I^0_i,v,h$&$2,n,1$&$2_i$\\
$7^e_d$&$f^{h_1}_1g_2b^0_1$
       &$\sum_{mh}D^{[g_2m0h]}=\sum_{mh}D^{[h_1m0h]}$
       &$I^0_i,v,h$&$2,2n,1$&$1_h$\\
  &$4,4n,2$
       &$\sum_{mh}D^{[g_2m0h]}+\sum_{h}D^{[g_20hv]}+D^{[g_2\frac{n}200]}=
         \sum_{mh}D^{[h_1m0h]}+\sum_{h}D^{[h_10hv]}+D^{[h_1\frac{n}2hv]}$
       &$I^0_i$&$2,2n,1$&$2_i$\\
$8^e$&$f^{h_1}_1h_1d^{h_1}_1$
       &$\sum_{mh}D^{[g_2m0h]}=\sum_{mh}D^{[h_1m0h]}$
       &$I^0_i,v,h$&$2,2n,1$&$1_h$\\
  &$4,4n,2$
       &NO ($g_2$)$=2\sum_{mh}D^{[h_1m0h]}+\sum_{h}D^{[h_10h(vh)]}+D^{[h_1\frac{n}2h(vh)]}$
       &$I^0_h,h$&$2,2n,1$&$2_i$\\
$9_e$&$g^0_1g_2b^0_1$
       &$\sum_{mh}D^{[g_2m0h]}=\sum_{mh}D^{[h_1m0h]}$&$I^0_h,v,h$&$n,2n,1$&$2_i$\\
  &$2n,4n,2$
       &$\sum_{mh}D^{[g_2m0h]}+\sum_{h}D^{[g_20hv]}+D^{[g_2\frac{n}200]}=
         \sum_{mh}D^{[h_1m0h]}+\sum_{h}D^{[h_10hv]}+D^{[h_1\frac{n}2hv]}$
       &$I^0_h,v,h$&$n,2n,1$&$2_i$\\
$10_a$&$g_2h_1g^0_1$
       &$\sum_hD^{[g_2m0h]}=\sum_hD^{[h_1m0h]}=D^{g_1m0}$
       &$I^0,h$&$4,4,2$&$Y$\\
   &$4n,4n,2n$
       &$\sum_hD^{[g_20vh]}=\sum_hD^{[h_10hv]}$
       &$0,v,h$&$2,2,1$&2\\
      &&$\begin{cases}D^{[g_2\frac{n}200]}\Lexc=\sum_vD^{[h_1\frac{n}2vh]}&n\text{ even }\\
                     \sum_hD^{[g_2\frac{n}2vh]}\Rexc=D^{[h_1\frac{n}200]}&n\text{ odd }\end{cases}$
       &$\tfrac{n}2,v,h$&$2,2,1$&${1^e_i\atop 1^o_h}$\\\hline
\end{tabular}\end{table*}

Note that when all components of the common BR are type 1 exceptional EBRs, there is no topological difference between the $A$ and $B$ bands. This situation is an exception, where no OALs exist despite the distinct maxima of $A$ and $B$.

\begin{figure}[b]
\includegraphics[width=7cm]{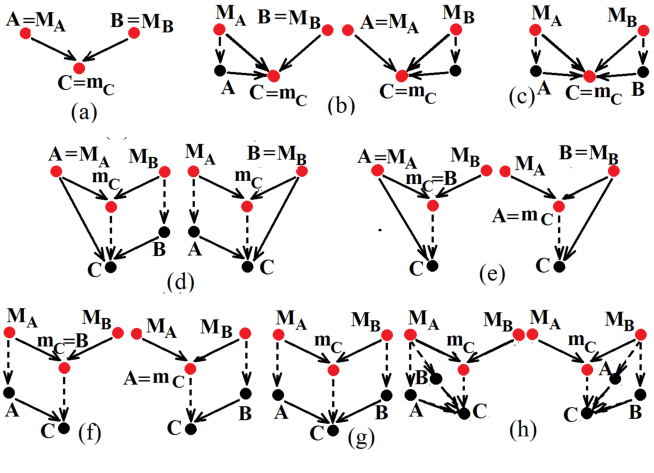}
\caption{\label{FSubH}
Subgraphs $\ACB$  of the connectivity graph of $\bG$ leading to OALs. The graphs are derived from
$M_A\rightarrow m_C\leftarrow M_B$ (where $M_A$ and $M_B$ are maximal strata, and  $m_C$is the maximal stratum connecting them). (a): represented is the case where all three strata remain maximal (this case always exists). (b) The case where one stratum (not $C$) is lower than in (a). (c) and (d): two strata are lowered. (g): all strata lowered. (e) and (f): special cases of (d) and (g), where either $A$ or $B$ coincides with $m_C$. (h): nearly homonested, with $C$ forming the bordering stratum.}
\end{figure}

The insights discussed enable the formulation of the following general algorithm (not limited to LGs) for identifying OALs:
\begin{enumerate}
\item[(i)] Firstly, identify the subgraphs $M_A\rightarrow m_C\leftarrow M_B$ of the group connectivity graph, where $M_A$ and $M_B$  are distinct maximal strata, and $m_C$ is the maximal stratum connecting them.
\item[(ii)] Next, the subgraphs $\ACB$ are extracted from the group connectivity graph, being related to those found in (i), as shown in Fig.~\ref{FSubH}.
\item[(iii)] Finally, the BRs allowed by equation \eqref{EOALBRs} for all such graphs should be found (and retained those that include at least one BR induced from non-regular representation of $\bF^C$).
\end{enumerate}

\section{Results}\label{SRES}

All results were obtained using the POLSym code, which performs algebraic and numerical group-theoretical algorithms~\cite{MGPT} tailored for (solid-state) physics applications. As there is infinitely many line groups, here we give analytical expressions for the families; to enlighten the geometry for each of the found triplet, we give explicit results and figure within Supplemental data~\ref{SLG6}. The triplets $ABC$ with non-regular XBRs are listed in column 2 ($ABC$) for each LG-family of negative groups (Tables~\ref{T2E}--\ref{T13E}), with their enumeration provided in the first column ("No")
The next column, "Orders", lists the orders of the stabilizers for single groups (one half of the double groups value). In Tables \ref{T11E} and \ref{T13E}, these data appear in the first column (below the triplet).
Following the section of the tables related to the strata, the basis of the XBRs is provided in the "XBRs" column as decompositions into EBRs of the maximal strata
$M_A$ and $M_B$, which majorize $A$ and $B$, respectively. The quantum numbers used for labeling the EBRs correspond to those of the isogonal axial point groups (Appendix~\ref{SAPG}).

The range of the counters ($m$ and parities) appears in the next column, "R": besides values $0$ and $\frac{n}2$, explicitly specified for $m$ in certain cases, intervals
$I$ and $I^0$ denote $m\in(-\text{n/2},n/2]$ and $m\in(0,n/2)$ (including both integer and half-integer values),while the notation $I^0_i$ (and $I^0_h$) specifies
$m$ taking only integer (half-integer) values from $I^0$; the indices $v$ and $h$ take the values $\pm1$. Only non-dummy entries are listed  to ensure accurate counting.

Since XBRs follow a general momentum superslection rule, i.e., they never mix integer and half-integer values of $m$ but instead distribute these
quantum numbers across different XBRs. Therefore, if $m$ is non-dummy (no summation), each integer or half-integer value specifies a distinct XBR. In contrast, when summation over $m$ is indicated in the tables, it independently refers to two XBRs: one with $m$ taking integer values, and the other with half-integer values within the interval specified in the column "R". The intervals $I^0$ and $I^0_h$  signify that only integer or only half-integer values of $m$ are allowed.
Column $T$ provides information on the total number of XBRs in the row, as well as the number of integer $T_i$ and half-integer values. Specifically, in the tables,
$X$, $Y$ and $Z$ denote the following:
\begin{subequations}\label{EIntHalfNum}
\begin{eqnarray}\label{Ea}
&X:&\ m\in(-\text{n/2},n/2]\quad X_i=X_h=n,\phantom{X} X=2n;\\ \label{Eb}
&Y:&\ m\in(-0,n/2)\ Y_i=[\tfrac{n-1}2],Y_h=[\tfrac{n}2],\ Y=n;\\ \label{Ec}
&Z:&\text{ if }n\text{ even }\ Z_i=2,\ Z_h=1,\phantom{XXXXx} Z=3,\\\nonumber
&  &\text{ if }n\text{ odd \ }\ Z_i=2,\ Z_h=2,\phantom{XXXXx} Z=4,
\end{eqnarray}\end{subequations}
where $[x]$ represents the integer part of $x$.

Triplets containing only integer and half-integer regular (Appendix~\ref{SNotation})) XBRs are not listed. In fact, the decomposition of such XBRs into $W$-EBRs is straightforward: $R^W=\sum_\mu \abs{\mu}D^{W\mu}$, separately for integer and half-integer IRs; the same applies to the maximal strata, as well, when decomposition on EBRs is obtained. There are many such cases, and they are easily identified: all triplets with $C=a_1$ belong here, meaning that $A$ and $B$ form an arbitrary pair of two non-generic strata, connected only by generic stratum. Tables list pairs connected also non-generic stratum.

Certain strata (specified in the caption) exist only for $n$ even, and triples containing such strata are marked with a superscript
$e$ over the first column counter to indicate it. In other cases, some $W$-EBRs (distinguished by the value
$m=n/2$ of the momentum quantum number) are present exclusively for $n$ even or within the double group.
In spinless cases with $n$ odd, these terms are disregarded. If all terms in a row are of this type, the entire solution is omitted (and should not be counted); this is indicated by a superscript $e$ in the last column.

For $n=1$, the number of strata is further reduced (the missing strata are listed in the caption). The remaining triplets predominantly have $C=a_1$;
in fact, there is only one nontrivial triplet for $n=1$ group of LG-families 9, 10, 11, and 13 (indicated by a superscript on the ordinal in column 1);
all of these triplets include both maximal strata.
Additionally, $m=0$ solutions are integer, as denoted by the superscript $i$ in the last column. Similarly, solutions that appear only when
$n$ is odd are marked with a superscript $o$, while those with a subscript $h$ correspond to half-integer solutions.

The superscript on the stratum denotes the maximal stratum that majorizes it (omitted for maximal strata); a superscript of 0 indicates the border strata, which are subordinated to both maxima.
Thus, homo-nested triplets have the same superscript for all three elements. In the case of a hetero-nested triplet, which indicates an OAL, a letter subscript on the triplet ordinal refers to the corresponding graph in Fig.~\ref{FSubH}.
In hetero-nested cases, when the XBR decomposition in both $M_A$ and $M_B$ contains a single term, it corresponds to a common EBR for both
$M_A$ and $M_B$, and is automatically classified as type 1 exceptional. This is denoted as $D^A\exc=D^B$ (the last XBR in LG-families 5 and 9).
Also, if only $D^{M_A}$ consists of a single term, then on the $B$-side, the $M_A$-EBR is decomposed into $M_B$-EBRs (which are the EBRs for the group). Hence, in this case,
$D^{M_A}$ is a type 2 exceptional, indicated by $D^A\Lexc=D^B$ (analogously $D^A\Rexc=D^B$), which establishes the OAL, as seen in the last XBRs of LG-families 10, 12, and 13.

\squeezetable
\begin{table}
  \caption{\label{TCX}
  XBRs of continual negative line groups. After the group, XBRs decomposed over both maxima are listed. Possible values of $m$ and the index $v$ (when appears) are in the next two columns. Site spaces' dimensionsand total number of the XBRs in the raw are in "D" and "T" ($\abs{\ZZ_2}$  counts the members of the first line, and $\abs{\NN}$ natural numbers.) }
\begin{tabular}{l|lllll}
\hline
$G$&XBRs decompositions&$m$&$v$&D&T\\\hline
$\bT\bC_{n\textrm{h}}$&$\sum_hD^{[c_2mh]}=\sum_hD^{[d_1mh]}$&$0,\pm\tfrac12,\dots$&$ $&$2,2,1$&$\abs{\ZZ_2}$\\
$\bT_{Q}\bD_{\infty}$&$\sum_hD^{[e_10h]}=\sum_uD^{[d_20h]}$&$0$&$ $&$2,2,1$&$1_i$\\
  &$D^{[e_1m0]}\exc= D^{[d_2m0]}$&$\pm\tfrac12,\pm1$& &$2,2,1$&$4\abs{\NN}$\\
$\bT\bD_{nh}$&$\sum_hD^{[g_20vh]}=\sum_hD^{[h_10hv]}$&$0$&$\pm1$&$2,2,1$&$2_i$\\
             &$\sum_hD^{[g_2m0h]}=\sum_hD^{[h_1m0h]}$&$\tfrac12,1,\dots$&$ $&$4,4,2$&$2\abs{\NN}$\\\hline
\end{tabular}\end{table}

\section{Discussions}\label{SDIS}

A symmetry-based framework for identifying conditions that enable obstructed atomic limits (OALs) is presented. While the approach is specifically applied to Q1D systems described by LGs, the underlying group-theoretical technique has universal applicability, extending to layered systems and three-dimensional crystals.  In fact, only  calculations and results, presented in section \ref{SRES}, are related to the LGs.

The central concept focuses on XBRs, which represent physical tight-binding models enabling to connect two Wickoff positions (WPs), $A$ and $B$. In the context of OALs, these WPs are interpreted either as centers of trivial limits (coinciding with ion positions) or as obstructed limits (not coinciding with ion positions). We demonstrate that topologically distinct phases arise exclusively when subgraph $\ACB$ is hetero-nested. This condition is colloquially related to the existence of "inversion symmetry", and we have shown that precise meaning is appearance of at least two maximal strata of the group; in Q1D systems this
phenomenon implies $z$-reversal symmetry (only negative LGs) and a hetero-nested model.
In the simplest cases this is demonstrated by the SSH model~\cite{SSH79,SSH80}, a canonical example of a 1D topological system~\cite{TopInsShort}.

It should be emphasized that the same framework can be extended to describe multiple phases differentiated by positions of the Wannier centers.

As previously noted, XBRs define the orbitals that must be taken into account within the tight-binding model corresponding to the transition $\ABCTR$. Specifically, the notation \eqref{EOALBRs}, $\sum_\mu c^\mu_PD^{W\mu}$, explicitly indicates that each site within the stratum $W$ ($W=A,B,C$) contributes to the state space via the carrier space of the representation $d(\bF^W)=\sum_\mu c^\mu_Pd^{(\mu)}(\bF^W)$. This essentially identifies the tight-binding model underlying the OAL, a topic to be discussed shortly. However, it is worth noting beforehand that this is why the decompositions of each XBR into $W$-EBRs are given in Appendix~\ref{SPEBRs} (while the possible triplets are for $n=6$ line groups illustrated in Supplemental data~\ref{SLG6}).

In the spinless case, the irreducible representations (IRs) of the orthogonal group $O(3)$ are first reduced to the IRs of the factor group. The IRs of $O(3)$ are denoted as $d^{(l\pm)}(O(3))$, where $l=0,1,\dots$ corresponds to the angular momentum, and $\pm$ distinguishes between representations that are even or odd under spatial inversion.
The integer IRs are characterized by atomic orbitals $\ket{lm}$ ($m=-l,\dots,l$), corresponding to the IR $d^{(l(-)^l)}(O(3))$, expressed as spherical harmonics in coordinate representation. These representations are then subduced (restricted) to $\bF^W$, and the resulting forms are decomposed into the IRs of $\bF^W$. Since the spatial stabilizers of LGs are axial point groups, $m$ remains the angular momentum quantum number of the orbital. More specifically, $m$ represents the $z$-component of angular momentum, modulo $n$ of the stabilizer in discrete LGs (while $n=\infty$ in continuous cases). The other parities are determined analogously, based on the resulting action of the second-order symmetry operations on the orbitals:
\begin{subequations}\label{EModels}
\begin{eqnarray}\label{EHY}
&\gs_hY^l_m(\gt,\phi)&\d=Y^{l}_m(\pi-\gt,\phi)=(-1)^{l+m}Y^l_{m}(\gt,\phi);\\\label{EUY}
&UY^l_m(\gt,\phi)&\d=Y^{l}_m(\pi-\gt,-\phi)=(-1)^lY^l_{-m}(\gt,\phi);\\ \label{EVY}
&\gs_vY^l_m(\gt,\phi)&\d=Y^{l}_m(\gt,-\phi)=(-1)^mY^l_{-m}(\gt,\phi).
\end{eqnarray}\end{subequations}
Thus, for LG-families 2 and 3, as well as $\bT\bC_{\infty h}$, each orbital remains invariant. In other cases, the invariant subspaces are those spanned by $Y^l_{m}$ and $Y^l_{-m}$. In these situations, a more convenient basis is provided by the linear combinations $Y^{l\pm}_m=\frac1{\sqrt2}(Y^l_m\pm\ri Y^l_{-m})$ ($m=1,\dots,l$). In both cases, the corresponding (1D or 2D) representation is denoted $d^{(m)}_o$.

In the spinful case, this part of the site state space is augmented by a 2D complex space carrying the natural spin representation~\cite{DLG} $d_s$ of SU(2), defined by the matrices:
$d_s(\gs_h)=\smatD{-i&0}{0&i}$, $d_s(U)=\tfrac{1}{\sqrt{2}}\smatD{-i&-1}{1&i}$ and
$d_s(\gs_v)=\tfrac{1}{\sqrt{2}}\smatD{1&-i}{-i&1}$. Hence, the spinful model becomes $d_o\otimes d_s$, where  $d_o$ incorporates irreducible components determined by $m$ and parities, as described above.

For each triplet $\ACB$, there exists at least one solution: the regular representation, an XBR that can be induced from all the site stabilizers and corresponds to the induced regular representation of the stabilizers. This is evident after realizing that the regular representation is obtained by induction, $R(\bG)=\one(\{e\})\uparrow\bG$, where $\one(\{e\})$ represents the unit representation of the identity subgroup $\{e\}$. Using the transitivity of induction, it follows that for any arbitrary subgroup $\bH$, $\one(\{e\})\uparrow\bH=R(\bH)$ and $R(\bH)\uparrow\bG=(\one(\{e\})\uparrow\bH)\uparrow\bG=R(\bG)$. These cases are not shown in the tables but can be easily derived if needed. In fact, since $a_1$ is a border stratum, all these solutions decompose over both maximal strata EBRs.

In the cases when all half-integer IRs of $\bF^W$ are of even dimension (this is not possible for all integer IRs) and divided into pairs (with an opposite parity $P$) inducing the same representations, the half-regular representation \begin{equation}\label{EHRBR}
\sum_{\mu}\frac{\abs{\mu}}2D^{[M_A\mu]}=\sum_\nu\frac{\abs{\nu}}2D^{[M_B\nu]},
\end{equation}
takes the role of half-integer regular representation. It turns out that it is accompanied by two integer BRs, one for each sign of $P$. Such XBRs are found for all triples besides the last one, in the families 9, 11 and 13. This includes the homo-nested situation, when integer solutions can be found only on the nest-side, while the half-integer decomposable over EBRs of the both maximal strata; in fact, the parity $P$ being geometrical obstacle for decomposition in terms of another maximal stratum for integer XBRs, becomes indistinguishable for half-integer BRs, making geometrical position of the strata irrelevant.

Number of distinct sites within an orbit is determined by the ratio $\abs{\bG}/\abs{\bF^W}$. For periodic groups, this value is infinite, so it is normalized to the elementary cell by dividing by the "order of translations"  $\abs{\bT}$. Consequently, the number of ions per elementary cell is finite, $n_W=\abs{\bI}/\abs{\bF^W}$, where $\bI=\bG/\bT$ is the isogonal group. In the LG-families 1 and 5, the helical group $\bT_Q$ takes the role of $\bT$, and the helical "isogonal" group is $\bI_H=\bG/\bT_Q$.
In this context, $n_W=\abs{\bI_H}/\abs{\bF^W}$  is the number of atoms in orbit $W$ within the monomer, effectively replacing the elementary cell.
 The requirement for dimensional equality of the state spaces constructed from atoms $A$, $B$ and $C$ leads to~\footnote{The induction of the representation $d(\bF^W)$ to $\bG$  involves adding a block of dimension $\abs{d}$ for each point in the orbit to the total matrix. As a result, the total dimension per elementary cell is  $n_W\abs{d}$.}:
\begin{equation}\label{EDimBR}
\abs{\cS_{A}}n_A=\abs{\cS_{B}}n_B=\abs{\cS_{C}}n_C,
\end{equation}
where $\abs{\cS_{A}}=\sum_{\ga}a_{A\ga}\abs{d^{\ga}(\bF^A)}$ (likewise for $B$ and $C$).

Notably, the examples reported thus far in the literature \cite{Benalcazar2017,BernevigTQC2,Cano-PRB22} implicitly  assume that $A$ and $B$ have the same number of atoms per period, $n_A=n_B$, which implies equality of the site spaces, $\abs{\cS_{A}}=\abs{\cS_{B}}$. In contrast, our procedure allows for cases where $n_A$ and $n_B$ differ, enlightening more general scenarios. Often, at least one of the two positions corresponds to centers of charge displaced from the ionic positions. This is evident in several listed examples, as shown in the Tables~\ref{T9E},\ref{T11E} and \ref{T13E} (columns "Orders" and $\abs{D}$).

The presented theory  seamlessly generalizes to multi-orbit systems, where ions and displaced charge centers are distributed across multiple orbits, which may be of the same or different types. Consider a scenario where ions occupy $m_A$ orbits of type $A$, while the centers of charge shift to $m_B$ orbits of type $B$. Since all $A$-type orbits contribute identical  basis functions (and similarly for $B$-type), the expression in \eqref{EDimBR} remains unaffected. The number of atoms per cell stays constant, while
$\abs{\cS_{A}}$  becomes an $m_A$-multiple of the single orbit space dimension (i.e., the frequencies of the stabilizer IRs are scaled by $m_A$). An intriguing scenario arises when $m_A\neq m_B$, potentially leading to effects similar to those described in the preceding paragraph.
Further generalizations along these lines are also possible.

Introducing the stratum $C$, which connects $A$ and $B$, often leads to an increase in the dimensions of the site spaces associated with $A$ and $B$.
 This results from  the induction procedure, where the IR blocks of $\bF^C$ are fewer than those of $\bF^A$ and of $\bF^B$.
  However, the number of blocks in $\bF^C$ is proportionally greater than that in $\bF^A$ and  $\bF^B$, in accordance with the inverse ratio of their orders.
  Consequently, if the initial induction step from the IRs of $\bF^C$ to the stabilizers of $A$ and $B$ does not fully recover the IRs of the latter, the total space expands with the addition of $C$. This suggests that  a tunneling-like effect between $A$ and $B$, without $C$, can be realized with less dimensional models. Notice here that with help model based on regular representations transition between  arbitrary pair of strata can be realized.

As emphasized before, for all $n>1$ LGs, the maximal strata are situated on the $z$-axis, with orbits making a simple chain. In negative LGs, the two such chains have the same configuration, being mutually shifted for a half of the period; they represent well-elaborated example of a 1D chain with two topological phases~\cite{TopInsShort}. This is more appropriately and accurately described using continuous LGs~\cite{YILG}, as the chain is invariant under $\bC_\infty=SO(2)$. The three symmetry classes with $z$-reversal are $\bT \bC_{\infty h}$, $\bT \bD_{\infty}$, and $\bT \bD_{\infty h}$.
As discussed, strata $A$ and $B$ are linked by the line $C$, which is the sole remaining stratum containing discrete orbits; each its orbit is a "double" chain,  with stabilizer $\bF^C$ being either $\bC_{\infty}$ or $\bC_{\infty\textrm{v}}$. The two maximal strata, referred to here as $A$ and $B$ (details provided in Appendix~\ref{SCLG}), represents a well-elaborated example of a 1D chain with two topological phases~\cite{TopInsShort}.  A shared characteristic of all these configurations is that the angular momentum quantum number can take any integer value,  supporting a range of physical models that vary with $m$, as detailed in Table~\ref{TCX}.

\appendix

\section{Notation and techique}\label{SNotation}

IRs of group $\bF$ are denoted as $d^{(\mu)}(\bF)$, its dimension is $\abs{\mu}$; the superscript $\mu$ in parentheses gathers all quantum numbers (QNs), distinguishing various IRs. QN of angular momentum (its $z$-component) takes different integer or half-integer values; when used as index, the notation is shortened by using overline for negative value ($\ov{m}=-m$), and underline (of the index value) denotes half ($\un{m}=\tfrac{m}2$). Similarly, QNs $\pm1$ for parities are shortened as $\pm$.

BR induced from  $d(\bF^W)$ is $D^{Wd}$;  in particular, when $d$ is IR with QNs $\mu$, the induced BR is $D^{W\mu}$. Further, when $W$ is a maximal stratum, induced IR of $\bF^W$ is EBR, which is emphasized as by brackets $D^{[W\mu]}$, unless it is type 2 exceptional, when braces are used $D^{\{W\mu\}}$; recall that type 1 exceptions are still EBRs (although induced also from other stabilizer(s)). $W$-EBR is BR elementary for stratum $W$, i.e. representation induced from an IR of the stabilizer $\bF^W$; it is elementary in the set of the representations induced from  $\bF^W$. XBR is BR which can be induced from each of several different Wyckoff positions, $A$, $B$,\dots; if necessary the relevant Wyckoff position are listed:  $AB\cdots$-XBR.

To obtain the results, we utilized the POLSym code, employing a modified group projector method to decompose BRs into IRs of the group, and subsequently into EBRs.
 In fact, we showed\cite[Eq.(41)]{MGPT} that the reduction can be performed using only the stabilizer and the representation to be induced, relying solely on the generators of the stabilizer group.

The double group of group $\bG$ is denoted as $\wti{\bG}$. Here we more carefully analyze regular representation of double group $\wti{\bG}$. Generally the regular representation of the group $\bG$ is induced from the unit representation of the identity subgroup (stabilizer of generic stratum), and it contains $\abs{\mu}$ copies of each IR $d^{(\mu)}$ of $\bG$. However, for double groups generic stratum is fixed by $\wti{\bC}_1=\bZ_2=\{\ti{e},\ti{e}'\}$ ($\ti{e}$ stands for identity and $\ti{e}'$ for rotation for $2\pi$ around any axis), which is central invariant subgroup. It has two IRs: $\gd^{(\pm)}(\ti{e})=1$ and $\gd^{(\pm)}(\ti{e}')=\pm1$. The frequencies of the IRs of $\bG$ in the representation induced from $\gd^{(\pm)}$
are \cite{MGPT}: $f^\mu_\pm=\frac12({\rm Tr}(d^{(\mu)}(\ti{e})\pm d^{(\mu)}(\ti{e}')))$. It is clear that the representation induced from $\gd^{(+)}$ contains each integer IR (characterized by $d^{(\mu)}(\ti{e}')=\one$) exactly $\abs{\mu}$ times, while for half-integer IRs, when $d^{(\mu)}(\ti{e}')=-\one$, the terms cancel,
i.e. this is regular representation of the (single) group $\bG$ (whose elements are pairs $\{g,\ti{e}'g\}$ of equally represented elements). For the induced representation $\gd^{(-)}(\wti{\bC}_1)\uparrow\wti{\bG}$, situation is just the opposite, and frequency of  half-integer IRs is equal to their dimension, while there is no integer components. In this sense these two of the representations are used as regular ones.

\section{Axial pont groups}\label{SAPG}
There are seven finite families of the axial point groups. $\bC_n$, $\bC_{n\mathrm{h}}$, $\bS_{2n}$, $\bD_n$, $\bC_{n\mathrm{v}}$, $\bD_{n\mathrm{d}}$ and $\bD_{n\mathrm{h}}$ ($n=1,2,\dots$). All of them have $\bC_n$ as the subgroup, giving the basic  classification of IRs by the angular momentum QN $m$; it takes integer and half-integer (for double-groups only) values from the interval $(-\tfrac n2, \tfrac n2]$. Besides, families 2-5 have one, and  families 6 and 7 two additional generators, which square is in $\bC_n$; as the IRs are constructed by induction from the index-two subgroup (in one or two steps), these generators introduce parity QNs, with value 0 when representation are without parity, while otherwise $\pm1$ for integer and $\pm\ri$ for half-integer representations. Summarizing, the QNs are $m$, and one or two parities. There are continual LGs, $\bC_\infty$, $\bC_{\infty\mathrm{h}}$,  $\bD_\infty$, $\bC_{\infty\mathrm{v}}$ and $\bD_{\infty\mathrm{h}}$. The first one is Lie subgroup in the others, with IRs counted by $m=0\pm\tfrac12,\pm1,\dots$; parities are the same as in the corresponding finite groups.

In  literature there are several (e.g. crystallographic) notations for the IRs. Precisely, the main part of the crystallographic symbol ${}^{\pV}X^{\pH}_m$ is letter,  $X=A,E$ (traditionally denoting the one- and two-dimensional IRs). Angular momentum $m$ is (right) subscript. The parity QN introduced by $\gs_v$ (preserves $z$-axis) is given as the left superscript (in fact, in the  most traditional crystallographic labels, parity of $\gs_v$ is denoted by $A$ and $B$, instead of superscript $\pV$), while the right one is reserved for $z$-reversing parities (due to $S_{2n}$, $U$ and $\gs_h$).
However, this (and any similar notation) is cumbersome to be applied at superscript position in Tables of the XBRs. Thus,  we use the minimal symbol, purely the complete set of QNs, $m\pV\pH$, instead (according to bijective correspondence). In Table~\ref{TAPGIRs} matrices of IRs (given for generators; this is sufficient to derive the IRs on the whole group) are associated QNs. So, in the tables of XBRs the IRs of stabilizers are indicated by QNs only in the superscript after the site: $D^{Pm\pV\pH}$  stands for  $W$-EBR ${}^{\pV}X^{\pH}_m(\bF^W)\uparrow\bG$. Clearly, if there is no parity of particular type, it is omitted from the label of from the symbol of IR and $W$-BRs.

Notice the difference between $m=n/2$ for $n$ odd and $n$ even in the last two groups. While in $\bD_{n\mathrm{d}}$ there is single integer IR ($n$ even), for $n$ odd there are 4 half-integer ones differing by both parities; just the opposite situation is in $\bD_{n\mathrm{d}}$.

Further, although $\bS_{2n}$ is generated by $S_{2n}=C_{2n}\gs_{\mathrm{h}}$, also matrix of $C_n=S^2_{2n}$ is given; the same refers to $\bD_{n\mathrm{d}}$, generated by $S_{2n}$ and $\gs_{\mathrm{v}}$.

\squeezetable
\begin{table}
\caption[Reps]{\label{TAPGIRs}Irreducible representation of 7 families of axial point groups. For each group $\bP_n$ (Column 1),
the IRs are specified by the values of $n$ for which the corresponding IR exists, its dimension $D$,
values of the group specific QNs, and the allowed range of angular momentum $m$. The remaining columns are the matrices of the elements: the first one is rotation $R_\phi$ for the angle $\phi$  around $z$-axis; next one is $\gs_{\mathrm{v}}$ if element of  $\bP_n$, while the last one is $z$-reversing generator ($S_{2n}$ in $\bS_{2n}$ and $\bD_{n\mathrm{d}}$), $U$ in $\bD_n$, and $\gs_{\mathrm{h}}$ in $\bC_{n\mathrm{h}}$, $\bD_{n\mathrm{d}}$ and $\bD_{n\mathrm{h}}$.
Five infinite (Lie) groups are indicated by $n=\infty$. Abbreviations:
$e_m=\re^{\ri m\phi}$, $e'_m=\begin{cases}\re^{\ri (m\phi+\pi/4)}&\text{ if $m$ half-integer,}\\
                              e_{m/2}&\text{ if $m$ integer}\end{cases}$
In finite cases $\phi=2\pi/n$ ($R_\phi=C_n$).}
\begin{ruledtabular}\begin{tabular}{ll|ccc|ccc}
$\bP_n$&$n$&$D$&QNs&$m$&$R_\phi/C_n$&$\gs_{\mathrm{v}}$&$S_{2n}/\gs_{\mathrm{h}}/U$\\\hline
$\bC_n$&$1,2\dots;\infty$&1&$m$&$(-\tfrac{n}2,\tfrac{n}2]$&$e_m$&\\\hline
$\bS_{2n}$&1,2\dots&1&$m\pH$&$(-\tfrac{n}2,\tfrac{n}2]$&$e_m$&&$\pH e_{\un{m}}$\\\hline
$\bC_{n\mathrm{h}}$&$1,2\dots;\infty$&1&$m\pH$&$(-\tfrac{n}2,\tfrac{n}2]$&$e_m$&&$\pH$\\\hline
$\bD_n$&$3,4\dots;\infty$&2&$m0$&$(0,\tfrac{n}2)$&$\smatD{e_{\ov{m}}&0}{0&e_m}$&&$\smatD{0&\gee}{1&0}$\\
&$1,2\dots;\infty$&1&$0\pU$&$0$&1&&$\pU$\\
&1,2\dots&1&$\frac{n}2\pU$&$\tfrac{n}2$&$-1$&&$\pU$\\\hline
$\bC_{n\mathrm{v}}$&$3,4\dots;\infty$&2&$m0$&$(0,\tfrac{n}2)$&$\smatD{e_{\ov{m}}&0}{0&e_m}$&$\smatD{0&\gee}{1&0}$&\\
&$1,2\dots;\infty$&1&$0\pV$&$0$&1&$\pV$&\\
&1,2\dots&1&$\frac{n}2\pV$&$\tfrac{n}2$&$-1$&$\pV$&\\\hline

$\bD_{n\mathrm{d}}$&3,4\dots&2&$m0\pH$&$(0,\tfrac{n}2)$&$\smatD{e_{\ov{m}}&0}{0&e_m}$&$\smatD{0&\gee}{1&0}$&$\pH\smatD{0&-e'_{m}}{e'_{\ov{m}}&0}$\\
&1,2\dots&1&$0\pV\pH$&$0$&1&$\pV$&$\pH$\\\cline{2-2}
&1,3,\dots&1&$\frac{n}2\pV\pH$&$\tfrac{n}2$&$-1$&$\pV$&$\pH$\\
&2,4\dots&2&$\frac{n}200$&$ $&$\smatD{-1&0}{0&-1}$&$\smatD{1&0}{0&-1}$&$\smatD{0&1}{1&0}$\\\hline

$\bD_{n\mathrm{h}}$&$3,4\dots;\infty$&2&$m0\pH$&$(0,\tfrac{n}2)$&$\smatD{e_{\ov{m}}&0}{0&e_m}$&$\smatD{0&\gee}{1&0}$&$\pH\smatD{0&1}{1&0}$\\
&$1,2\dots;\infty$&1&$0\pV\pH$&$0$&1&$\pV$&$\pH$\\\cline{2-2}
&1,3\dots&2&$\tfrac{n}200$&$\tfrac{n}2$&$\smatD{-1&0}{0&-1}$&$\smatD{\ri&0}{0&-\ri}$&$\smatD{0&-1}{1&0}$\\
&2,4\dots&1&$\tfrac{n}2\pV\pH$&$ $&$-1$&$\pV$&$\pH$\\
\end{tabular}\end{ruledtabular}\end{table}

\begin{widetext}
\section{Continual line groups}\label{SCLG}

To analyse the chains, the continual LGs are appropriate. There are only sparse data on them~\cite{YILG}, and here we give necessary ones, IRs and EBRs, which are used in OALs calculations presented in Table~\ref{TCX}.

\squeezetable
\begin{table*}
\caption[Reps]{\label{TLGContIRs}Irreducible representation of 5 continual line groups. For each group $\bG$ (Column 1),
the IRs are specified by its dimension $D$,
values of the group specific QNs, and the allowed ranges of momentum $k$ and angular momentum $m$. The remaining columns are the matrices of the elements: the first one is translation $\kost{\one}{1}$ for the period along $z$-axis, then follows rotation $R_\phi$ for the angle $\phi$  around $z$-axis; next one is $\gs_{\mathrm{v}}$ if element of  $\bP_n$, while the last one is $z$-reversing generator ($S_{2n}$ in $\bS_{2n}$ and $\bD_{n\mathrm{d}}$), $U$ in $\bD_n$, and $\gs_{\mathrm{h}}$ in $\bC_{n\mathrm{h}}$, $\bD_{n\mathrm{d}}$ and $\bD_{n\mathrm{h}}$.
Abbreviations: $\pE=(-1)^{2m}$ ($\pm1$ for integer/half-integer IRs),
$e_k=\re^{\ri k}$, $e_m=\re^{\ri m\phi}$, $K=\text{diag}[e_k,e_{k},e_{-k},e_{-k}]$, $M=\text{diag}[e_m,e_{-m},e_{m},e_{-m}]$, $A=\smatD{0&\pE}{1&0}\oplus\smatD{0&1}{\pE&0}$, $B=\smatD{0&\pE}{1&0}\otimes\smatD{1&0}{0&1}$.}
\begin{ruledtabular}\begin{tabular}{ll|lll|cccc}
$\bG$&$D$&QNs&$k$&$m$&$\kost{\one}{0}$&$R_\phi$&$\gs_{\mathrm{v}}$&$\gs_{\mathrm{h}}/U$\\\hline
$\bT\bC_\infty$&1&$km$&$(-\pi,\pi]$&$0,\pm\tfrac12,\pm1,\dots$&$e_k$&$e_m$&\\\hline
$\bT\bC_{\infty\mathrm{h}}$&2&$km0$&$(0,\pi)$&$0,\pm\tfrac12,\pm1,\dots$&$\smatD{e_k&0}{0&e_{-k}}$&$\smatD{e_m&0}{0&e_m}$&&$\smatD{0&\pE}{1&0}$\\
                           &1&$km\pH$&$0,\pi$&$0,\pm\tfrac12,\pm1,\dots$&$e_k$&$e_m$&&$\pH$\\\hline
$\bT\bD_\infty$&2&$km0$&$(0,\pi)$&$0,\pm\tfrac12,\pm1,\dots$&$\smatD{e_k&0}{0&e_{-k}}$&$\smatD{e_m&0}{0&e_{-m}}$&&$\smatD{0&\pE}{1&0}$\\
               &2&$km0$&$0,\pi$&$\tfrac12,1,\dots$&$\smatD{e_k&0}{0&e_{-k}}$&$\smatD{e_m&0}{0&e_{-m}}$&&$\smatD{0&\pE}{1&0}$\\
               &1&$k0\pU$&$0,\pi$&$0$&$e_k$&$1$&&$\pU$\\\hline
$\bC_{\infty\mathrm{v}}$&2&$km0$&$(-\pi,\pi]$&$\tfrac12,1,\dots$&$\smatD{e_k&0}{0&e_k}$&$\smatD{e_m&0}{0&e_{-m}}$&$\smatD{0&\pE}{1&0}$&\\
                        &1&$k0\pV0$&$(-\pi,\pi]$&0&$e_k$&1&$\pV$&\\\hline
$\bD_{\infty\mathrm{h}}$&4&$km00$&$(0,\pi)$&$\tfrac12,1,\dots$&$K$&$M$&$A$&$B$\\
                        &2&$k0\pV0$&$(0,\pi)$&$0$&$\smatD{e_k&0}{0&e_{-k}}$&$\smatD{1&0}{0&1}$&$\pV\smatD{1&0}{0&\pE}$&$\smatD{0&\pE}{1&0}$\\
                        &2&$km0\pH$&$0,\pi$&$\tfrac12,1,\dots$&$\smatD{e_k&0}{0&e_k}$&$\smatD{e_m&0}{0&e_{-m}}$&$\smatD{0&\pE}{1&0}$&$\pH\smatD{1&0}{0&\pE}$\\
                        &1&$k0\pV\pH$&$0,\pi$&$0$&$e_k$&$1$&$\pV$&$\pH$\\
\end{tabular}\end{ruledtabular}\end{table*}

\squeezetable
\begin{table*}
  \caption{\label{TEBRC}
  EBRs of continual line groups. Maximal stabilizer(s) in column $M$ are followed by their stabilizers and label in terms of QNs.
  Range of QNs is in the next column. Finally, there is the decomposition of the EBR over IRs of the LG.
  Here, $G=(-\pi,\pi]$, $H=[0,\pi]$; summation in $k$ over these regions is shorten as $\sum^X_k{_kY}={_XY}$. $\pV$, $\pU$ and $\pH$ take values $\pm1$ for integer, and $\pm\ri$ for half-integer representations.
  The exceptions are only in the group $\bT(a)\bD_{\infty}$, where IRs $E_m$ of both maximal stabilizers induce the same EBR (type 2).}
\begin{tabular}{lllllll}\hline\noalign{\smallskip}
     Group&$M$&$\bF^{M}$&$D$&IR&m&Reduction\\ \noalign{\smallskip}\hline\noalign{\smallskip}

 $\bT(a)\bC_\infty$&$b_1$&$\bC_\infty$&1&$d^{(m)}$&$0,\pm\tfrac12,\dots$&$\sum^G_kD^{(Gkm)}$\\
        \noalign{\smallskip}\noalign{\smallskip}

 $\bT(a)\bC_{\infty\text{h}}$&$c_2$&$\bC_{\infty\text{h}}$&1&$d^{(m\pH)}$&$0,\pm\tfrac12,\dots$&
         $D^{(\Gamma m\pH)}+\sum^H_kD^{(Hkm0)}+D^{(\Pi m\ov{\pH})}$\\
 &$d_1$&$\bC_{\infty\text{h}}$&1&$d^{(m\pH)}$&$0,\pm\tfrac12,\dots$&
         $D^{(\Gamma m\pH)}+\sum^H_kD^{(Hkm0)}+D^{(\Pi m\pH)}$\\
  \noalign{\smallskip}\noalign{\smallskip}

 $\bT(a)\bD_{\infty}$&$d_2$&$\bD_{\infty}$&1&$d^{(0\pU)}$&$0$&$D^{(\Gamma 0\pU)}+\sum^H_kD^{(Hk00)}+D^{(\Pi0\ov{\pU})}$\\
     &$e_1$&$\bD_{\infty}$&1&$d^{(0\pU)}$&$0$&$D^{(\Gamma 0\pU)}+\sum^H_kD^{(Hk00)}+D^{(\Pi0\pU)}$\\
 &$d_2,e_1$&&2&$d^{(m0)}$&$\tfrac12,1,\dots$&
    $D^{(\Gamma m0)}+\sum^H_k(D^{(Hkm0)}+D^{(Hk\ov{m}0)})+D^{(\Pi m0)}$\\
  \noalign{\smallskip}\noalign{\smallskip}

 $\bT(a)\bC_{\infty\text{v}}$&$d_1$&$\bC_{\infty\text{v}}$&1&$d^{(0\pV)}$& $0$&$\sum^G_kD^{(Gk0V)}$\\
 &$d_1$&&1&$d^{(m0)}$&$\tfrac12,1,\dots$&$\sum^G_kD^{(Gkm0)}$\\
  \noalign{\smallskip}\noalign{\smallskip}

 $\bT(a)\bD_{\infty\text{h}}$&$g_2$&$\bD_{\infty\text{h}}$&2&$d^{(m0\pU)}$&$\tfrac12,1,\dots$&
      $D^{(\Gamma m0\pH)}+\sum^H_kD^{(Hkm00)}+D^{(\Pi m0\ov{\pH})}$\\
 &$g_2$&&1&$d^{(0\pV\pH)}$&$0$&
     $D^{(\Gamma0\pV\pH)}+\sum^H_kD^{(Hk0]pV0)}+D^{(\Pi0]pV\ov{\pU})}$\\
 &$h_1$&$\bD_{\infty\text{h}}$&2&$d^{(m0\pU)}$&$\tfrac12,1,\dots$&
     $D^{(\Gamma m0\pH)}+\sum^H_kD^{(Hkm00)}+D^{(\Pi m0\pH)}$\\
 &$h_1$&&1&$d^{(0\pV\pH)}$&$0$&
     $D^{(\Gamma0\pV\pH)}+\sum^H_kD^{(Hk0\pV0)}+D^{(\Pi0\pV\pU)}$\\
  \noalign{\smallskip}\hline\noalign{\smallskip}
\end{tabular}\end{table*}

\section{Decompositions over local $W$-EBRs}\label{SPEBRs}

It is useful to have decompositions of the XBRs over $W$-EBRs dor $W=A,B,C$. In fact, while from the point of view of the OALs classification and results, decomposition over EBRs is sufficient, the physical model is directly seen from the representations of the stabilizers, having the same labels as the $W$-EBRs For this reason for families 9, 11 and 13 we give also this decompoosition, neatly following the ordering in the tables \ref{T9E}, \ref{T11E} and \ref{T13E} (for other groups only maximal strata for $A$ and $B$ are allowed, i.e. these and previous decompositions are the same).

Decompositions of the XBRs over $W$-EBRs can, in some cases, appear to break symmetry, with certain signs of momentum or parity quantum numbers seemingly preferred over their opposites. However, this is merely a reflection of the fact that both IRs of each parity of the stabilizer generate the same BR. For example, the representations $A_m$ and $A_{-m}$  of  $\bC_n$ induce the same 2D representation,  $E_m$ of $\bD_n$. Consequently, the induced representations
$D^{\bC_nm}(\bT_Q\bD_n)=D^{\bC_n(-m)}(\bT_Q\bD_n)$ are equivalent and correspond to $D^{[\bD_nm]}(\bT_Q\bD_n)$. This exemplifies the apparent asymmetry in quantum numbers encountered in various scenarios, such as in the LG-family 12, d1. Similarly, in the LG-family 13, an exceptional case arises where $D^{g_2\frac{n}2}=\sum_vD^{h_1\frac{n}2v+}=\sum_vD^{h_1\frac{n}2v-}$. Here, any choice on the right-hand side seemingly violates $\gs_h$-parity.

\squeezetable
\begin{table*}
\caption{\label{T9} Groups $\bL^{(9)}=\bT\bD_{nd}$. $X^u=\begin{cases}D^{[W\frac{n}200]}& n\text{ even}\\
\sum_{u}D^{[W\frac{n}2uv]}& n\text{ odd}\end{cases}$, for $W=d_2,e_1$ (lines 1 and 2).
$m=n/2$ za $n$ parno ide zajedno sa $m=0$, te polucele nemaju $u$, KOd $n$ odd ima 2 cele i dve polucele}
\begin{tabular}{r|ll|llll}
\hline
No&$ABC$&$D^A=D^B=D^C$&R\\\hline
$1$&$d_2d^0_1b^0_1$&$\sum_{mh}D^{[d_2m0h]}+\sum_{h}D^{[d_20vh]}+X^v_{d_2}=
       \sum_{m}D^{d_1m0}+D^{d_10v}+D^{d_1\frac{n}2v}=D^{b_1v}$&$I^0,v$\\
$2$&$d^0_1e_1b^0_1$&$\sum_{m}D^{d_1m0}+D^{d_10\overline{v}}+D^{d_1\frac{n}2v}=
       \sum_{mh}D^{[e_1m0h]}+\sum_{h}D^{[e_10hv]}+X^v_{e_1}=D^{b_1v}$&$I^0,v$\\
$3$&$d_2e_1d^0_1$&$\sum_hD^{[d_2m0h]}=\sum_hD^{[e_1m0h]}=D^{d_1m0}$&$I^0$\\
        &&$\sum_hD^{[d_20vh]}=\sum_hD^{[e_10hv]}=D^{d_10v}$&$v$\\
        &&$\begin{cases}D^{[d_2\frac{n}200]}\exc=D^{[e_1\frac{n}200]}=D^{d_1\frac{n}21}&n\text{ even}\\
           \sum_hD^{[d_2\frac{n}2vh]}=\sum_hD^{[e_1\frac{n}2hv]}=D^{d_1\frac{n}2v}&n\text{ odd}\end{cases}$&${ \atop v}$\\\hline
\end{tabular}\end{table*}

\squeezetable
\begin{table*}
\caption{\label{T11} Groups $\bL^{(11)}=\bT\bD_{nh}$.
 For the triplet $9^e$ in the Decomposition for $B=h_1$ for $n=4k$ only, in summation (starred) in $m$, for $m=k$ the term is $2D^{[h_1k(-h)]}$ (the opposite $h$  with respect to other terms).assymmetric: 1.2,2.2,$c_2e_1c_1$,$e_1g_1c_1$}
\begin{tabular}{l|ll|llll}
\hline
No&$ABC$&$D^A=D^B=D^C$&R\\\hline
1&$b^{g_2}_2c^{g_2}_2a^{g_2}_2$&$D^{b_2\frac1200}=D^{c_2\frac1200}=D^{a_2\frac121}$&$\frac12$\\
     &             &$\sum_hD^{b_20h(vh)}=\sum_hD^{c_20hv}=D^{a_20v}$&$0,v$\\
2&$b^{g_2}_2e^{h_1}_1b_1$&$D^{b_2\frac1200}=D^{e_1\frac1200}=D^{a_2\frac121}$&$\frac12$\\
     &             &$\sum_hD^{b_20vh}=\sum_vD^{e_10vh}=D^{b_10v}$&$0,v$\\
3&$b^{g_2}_2g^0_1b_1$&$D^{b_2\frac1200}=\sum_mD^{g_1m0}=D^{b_1\frac121}$&$I^0_h$\\
  &&$\sum_hD^{b_20vh}=\sum_mD^{g_1m0}+D^{g_10v}+D^{g_1\frac{n}2v}=D^{b_1v}$&$I^0_i,v$\\
4&$b^{g_2}_2g_2a^{g_2}_2$&$D^{b_2\frac1200}=\sum_{mh}D^{[g_2m0h]}=D^{a_2\frac121}$&$I^0_h,h$\\
     &&$\sum_vD^{b_20v(hv)}=2\sum_mD^{[g_2m0h]}+\sum_v D^{[g_20v(hv)]}+
               \sum_v D^{[g_2\frac{n}2v(hv)]}=D^{a_2h}$&$I^0_i,h$\\
5&$c^{g_2}_2e^{h_1}_1c^0_1$&$D^{c_2\frac1200}=D^{e_1\frac1200}=D^{c_1\frac121}$&$\frac12$\\
     &&$\sum_{vh}D^{c_20vh}=\sum_{vh}D^{e_10vh}=\sum_hD^{c_10h}$&$0$\\
6&$c^{g_2}_2g^0_1c^0_1$&$D^{c_2\frac1200}=\sum_{m}D^{g_1m0}=D^{c_1\frac121}$&$I^0_h$\\
    &&$\sum_hD^{c_20vh}=\sum_mD^{g_1m0}+D^{g_10v}+D^{g_1\frac{n}2\overline{v}}=D^{c_10v}$&$I^0_i,v$\\
7&$c^{g_2}_2g_2a^{g_2}_2$&$D^{c_2\frac1200}=\sum_{mh}D^{[g_2m0h]}=D^{a_2\frac121}$&$I^0_h$\\
   &&$\sum_vD^{c_20vh}=2\sum_mD^{[g_2m0h]}+\sum_vD^{[g_20v(vh)]}+
               \sum_vD^{[g_2\frac{n}2v(vh)]}=D^{a_2h}$&$I^0_i,h$\\
8&$e^{h_1}_1f^{h_1}_1d^{h_1}_1$&$D^{e_1\frac1200}=D^{f_1\frac1200}=D^{d_1\frac121}$&$\frac12$\\
     &&$\sum_vD^{e_10v(hv)}=\sum_vD^{e_10hv}=D^{d_10h}$&$v,h$\\
9&$e^{h_1}_1g^0_1b^0_1$&$D^{e_1\frac1200}=\sum_mD^{g_1m0}=D^{b_1\frac121}$&$I^0_h$\\
     &&$\sum_hD^{e_1vh}=\sum_mD^{g_1m0}+D^{g_10v}+D^{g_1\frac12v}=D^{d_1v}$&$I^0_i,v$\\
10&$e^{h_1}_1g^0_1c^0_1$&$D^{e_1\frac1200}=\sum_mD^{g_1m0}=D^{e_1\frac121}$&$I^0_h$\\
     &&$\sum_{hv}D^{e_1vh}=2\sum_mD^{g_1m0}+\sum_{hv}D^{g_1\frac12vh}=\sum_{h}D^{c_10h}$&$I^0_i$\\
11&$e^{h_1}_1h_1d^{h_1}_1$&$D^{e_1\frac1200}=\sum_{mh}D^{[h_1m0h]}+D^{h_1\frac{n}200}=D^{d_1\frac121}$&$I^0_h$\\
     &&$\sum_vD^{e_10v(vh)}=2\sum^*_mD^{[h_1m0h]}+\sum_vD^{[g_10v(vh)]}+\sum_vD^{[g_1\frac{n}2v(vh)]}=
               D^{g_1h}$&$I^0_i,h$\\
$12$&$f^{h_1}_1h_1d^{h_1}_1$&$D^{f_1\frac1200}=\sum_{mh}D^{[h_1m0h}=D^{d_1\frac121}$&$I^0_h$\\
     &&$\sum_vD^{f_1hv}=2\sum_mD^{[h_1m0h]}+\sum_vD^{[h_10v(vh)]}+\sum_vD^{[h_1\frac{n}2v(vh)]}=D^{d_1h}$&
               $I^0_i,h$\\
13&$g^a_2h^a_1g^0_1$&$
    \begin{cases}\sum_hD^{[g_2\frac{n}2vh]}=\sum_hD^{[h_1\frac{n}2hv]}=D^{g_1\frac{n}2v}&n\text{ even }\\
     D^{[g_2\tfrac{n}200]}=D^{[h_1\frac{n}200]}=D^{g_1\frac{n}21}&n\text{ odd }\end{cases} $&
     ${\tfrac{n}2,v\atop\tfrac{n}2}$\\
  &          &$\sum_hD^{[g_20vh]}=\sum_hD^{[h_10hv]}=D^{g_10v}$&$0,v$\\
  &          &$\sum_hD^{[g_2m0h]}=\sum_hD^{[h_1m0h]}=D^{g_1m0}$&$I^0$\\\hline
\end{tabular}\end{table*}

\begin{table*}
\caption{\label{T13} Groups $\bL^{(13)}=\bT_{2n}\bD_{nh}$.}
\begin{tabular}{r|ll|llll}
\hline
No&$ABC$&$D^A=D^B=D^C$&R\\\hline

1&$e^f_1g^0_1bv_1$&$D^{e_1\frac1200}=\sum_mD^{g_1m0}+D^{g_1\tfrac{n}21}=D^{d_1\frac121}$&$I^0_h$\\
        &&$\sum_hD^{e_1hv}=\sum_mD^{g_1m0}+D^{g_10v}+D^{g_1\frac{n}2v}=D^{b_10v}$&
              $I^0_i,v,h$\\
2&$e_1f^{h_1}_1d^0_1$&$D^{e_1\frac1200}=D^{f_1\frac1200}=D^{d_1\frac12}$&$\frac12$\\
     &         &$\sum_vD^{e_10v(hv)}=\sum_vD^{f_10hv}=D^{d_10h}$&$0,h$\\

3&$e^{h_1}_1g^0_1b^0_1$&$D^{e_1\frac1200}=\sum_{m}D^{g_1m0}=D^{b_1\frac121}$&$I^0_h,v$\\
 &&$\sum_hD^{e_1hv}=\sum_{m}D^{g_1m0}+\sum_hD^{[g_10h]}+D^{g_1\frac{n}21}=D^{b_10v}$&$I^0_i,v$\\

4&$e^d_1g^d_2b^0_1$&$D^{e_1\frac1200}=\sum_{mh}D^{g_2m0h}+\sum_hD^{g_1\tfrac{n}21h}=D^{d_1\frac121}$&$I^0_h,h$\\
 &&$\sum_hD^{e_1hv}=\sum_{mh}D^{[g_2m0h]}+\sum_hD^{[g_20vh]}+D^{[g_2\frac{n}200]}=D^{b_1v}$&$I^0_i,v$\\

5&$e^{h_1}_1h_1d^{h_1}_1$&$D^{e_1\frac1200}=\sum_{mh}D^{[h_1m0h]}+D^{[h_1\frac{n}200]}=D^{d_1\frac12}$&$I^0_h$\\
     &&$\sum_vD^{e_10v(hv)}=2\sum_mD^{[h_1m0h]}+\sum_vD^{[h_10v(hv)]}+\sum_vD^{[h_1\frac{n}2v(hv)]}=D^{d_1h}$&
              $I^0_i,h$\\

6&$g^0_1f^{h_1}_1b^0_1$&$\sum_{m}D^{g_1m0}=D^{[f_1\frac1200]}=D^{b_1\frac121}$&$I^0_h$\\
     &&$\sum_mD^{g_1m0}+D^{g_10v}+D^{g_1\frac{n}21}=\sum_hD^{f_10v(hv)}=D^{b_10v}$&
              $I^0_i,v$\\

7&$g_2f^{h_1}_1b^0_1$&$\sum_{mh}D^{[g_2m0h]}=D^{f_1\frac1200}=D^{b_1\frac121}$&$I^0_h$\\
     &&$\sum_{mh}D^{[g_2m0h]}+\sum_hD^{[g_20vh]}+D^{[g_2\frac{n}200]}=\sum_hD^{f_10h(vh)}=D^{b_10v}$&
              $I^0_i,v$\\

8&$f^{h_1}_1h_1d^0_1$&$D^{e_1\frac1200}=\sum_{mh}D^{[h_1m0h]}=D^{d_1\frac12}$&$I^0_h$\\
     &&$\sum_vD^{f_10hv}=2\sum_mD^{[h_1m0h]}+\sum_vD^{[h_10v(hv)]}+\sum_vD^{[h_1\frac{n}2v(hv)]}=D^{d_1h}$&
              $I^0_i,h$\\
9&$g^0_1g_2b^0_1$&$\sum_{m}D^{g_1m0}=\sum_{mh}D^{g_2m0h}=D^{b_1\frac121}$&$I^0_h$\\
     &&$\sum_mD^{g_1m0}+D^{g_10v}+D^{g_1\frac{n}21}=\sum_{mh}D^{[g_2m0h]}+\sum_hD^{[g_20vh]}+D^{[g_2\frac{n}200]}=
               D^{b_10v}$&$I^0_i,v$\\

10&$g_2h_1g^0_1$&$\sum_hD^{[g_2m0h]}=\sum_hD^{[h_1m0h]}=D^{g_1m0}$&$I^0_i$\\
 &&$\sum_hD^{[g_20v]}=\sum_hD^{[h_10hv]}=D^{g_10v}$&$0,h$\\
 &&$\begin{cases}D^{[g_2\frac{n}200]}\Lexc=\sum_vD^{[h_1\frac{n}2v1]}=D^{g_1\frac{n}21}&n\text{ even }\\
     \sum_hD^{[g_2\frac{n}21h]}\Rexc=D^{[h_1\frac{n}200]}=D^{g_1\frac{n}21}&n\text{ odd }\end{cases}
 $&$\tfrac{n}2$\\\hline
\end{tabular}\end{table*}

\end{widetext}

\acknowledgements{This research is funded by Serbian Ministry of Science (Project ON171035, and SANU-F-93).}

%

\clearpage

\begin{widetext}
\section{Supplemental data: $n=6$ AOLs structures in Q1D}\label{SLG6}

In the supplemental material we list one group from each family. We chose the group with  $n=6$, and in the families 1 and 5 also $Q=2n$ (as it is automatically in the families 4, 8 and 13). Actually, the data refer to the spinless and spinful situations, since the double group IRs and EBRs are considered; the integer representations are for single groups.

Families 1 and 7 are not considered as they have only two strata. For the remaining positive families, 6 and 8, all triples are trivial, containing $a_1$ stratum; they are listed only for the completeness.  The data presented contain two type of information.

The first one is about the group (the first page for each group). It starts with the family number and the structure. Different strata (with singled out those with the maximal stabilizers), and their connectivity graph follow in the picture. The page ends with the elementary band representations. Generally, these are  the IRs of the maximal stabilizers induced to the group; therefore, each EBR, $E^{[i]}$ (Column "EBR") is equal to $D^{[M\mu]}$, where $M$ is one of the maximal strata, while $\mu$ are QNs counting IRs $d^{(\mu)}(\bF^M)$ of its stabilizer. For the compactness sake, only $M$-QNs labels are presented in the column "$M|$QNs". The two types of exceptions are emphasized as follows: for type 1 (two such induced representations are equal) both BRs (corresponding to the same EBR) are given in the Column "WP$\backslash$QNs" (see families 5 and 9). Type 2 exceptions are those  $D^{[M\mu]}$ which are not EBRs (thus they are missing in the Column "$M|$QNs"), as they are decomposed (as integer combination) onto the others; they are listed below the table, together with all possible the decompositions (families 10, 12, 13).

The second group of information is on a separate page for each non-trivial triplet of sites. Firstly, the triplet is given (each stratum by different  color corresponding to the next graphics), together with stabilizers and their order. Also, immediately after ordinal stands the symbol: "1", "E", and "A" if triplet exists for each $n$, only for even $n$, and only for $n>1$ (BOLJE OVO OYNACITI I NAPISATI). Next, scheme of an abstract system of this symmetry with Wannier centers (balls in the colors following those of the strata above) on the orbits belonging to strata $A$ or $B$, and intermediate positions $C$  schematically drown with the atoms represented by the (on the top of the page). The triplet of strata is red-emphasized on the connectivity graph; this is useful to see if the triplet is homo- or hetero-nested.

The first table is decomposition on the IRs of the group  of the XBRs $X^i$; for each IR (QNs in the firs row) below, in the row $X^i$ its the frequency number $f^\mu(X^i)$. Then, follow the decompositions of each $W$-EBR $D^{W\mu}(\bG)=d^{(\mu)}(\bF^W)\uparrow\bG$ ($W=A,B$), in the same format. Next table gives decomposition of the XBRs over $W$-EBRs for $W=A,B,C$, together with the dimension of the corresponding physical model state space given by elementary cell (and monomer for familiy 5) and by each  site of the type $A$, $B$ and $C$.

In the last table there is decomposition of the XBRs onto each $M$-EBRs (including also the exceptions!).
Remarkably, some XBRs of homo-nested triples (11.1, 11.4, 11.7, 11.8, 11.12, 13.2, 13.5, 13.8) cannot be decomposed over the opposite maximal stratum EBRs. Here are counted only triples with all three strata $A$, $B$ and $C$ majorized by the same maximal stratum; when either of the strata is on connected to the other maximum, the triplet is hetero-nested. This difference is clearly manifested by the first two triples $\tripletr{e_1}{f_1}{d_1}$and $\tripletr{e_1}{f_1}{b_1}$ in the family 13: $h_1$ majorizes the (same) edges, as well the intermediate stratum  of the first triplet; however, the other intermediate stratum $b_1$ is on the border between two maxima, which is sufficient to make the system hetero-nested.
Hence, in these situation, as well as in the cases of single maximal stratum, OALs are not possible. In hetero-nested triples the decompositions are mutually different, and  therefore $A$ or $B$ electronic centering is seen by the band structure only.

\begin{figure*}[b]\includegraphics[width=0.94\textwidth]{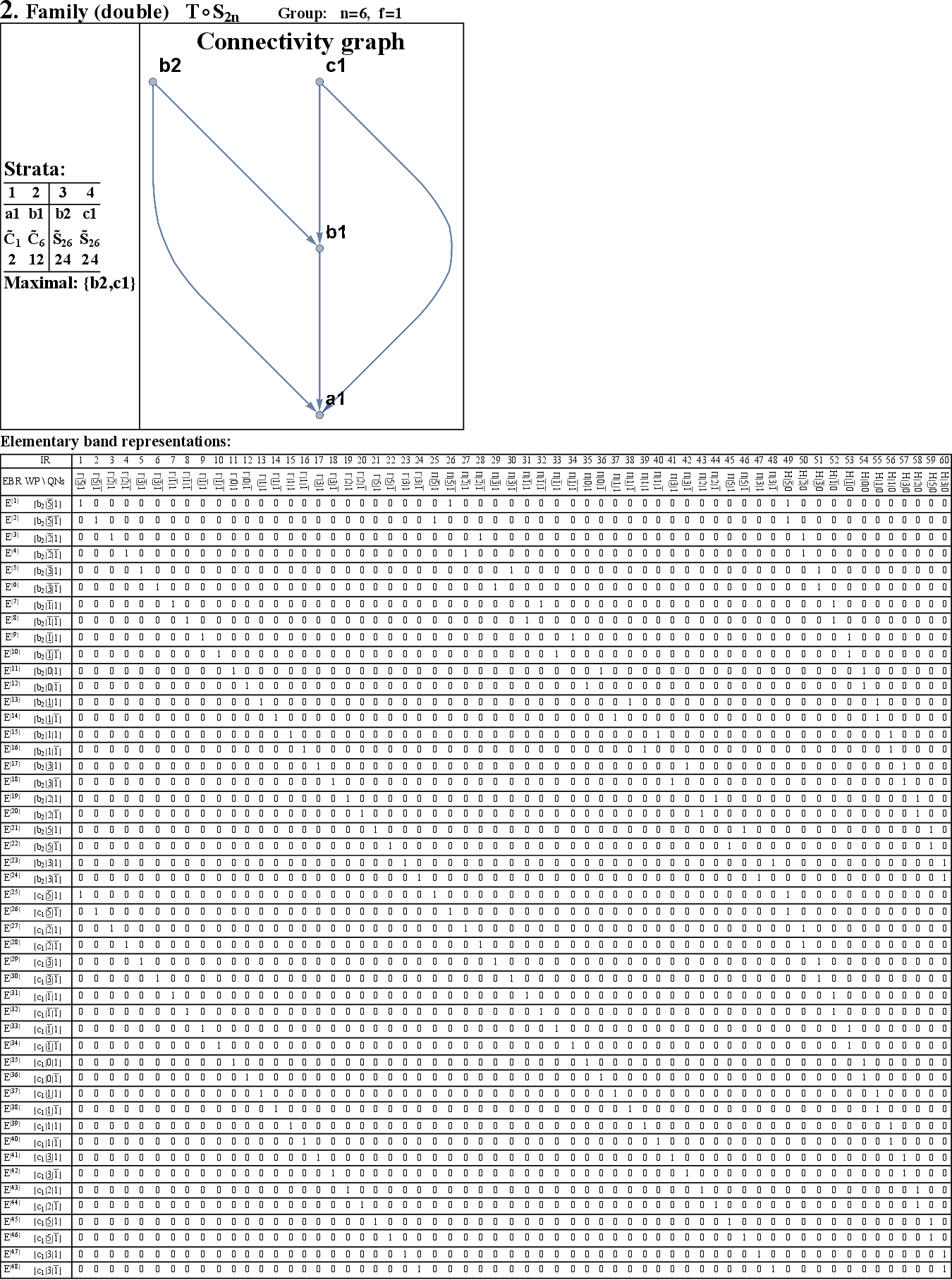}\end{figure*}
\begin{figure*}[b]\includegraphics[width=0.75\textwidth]{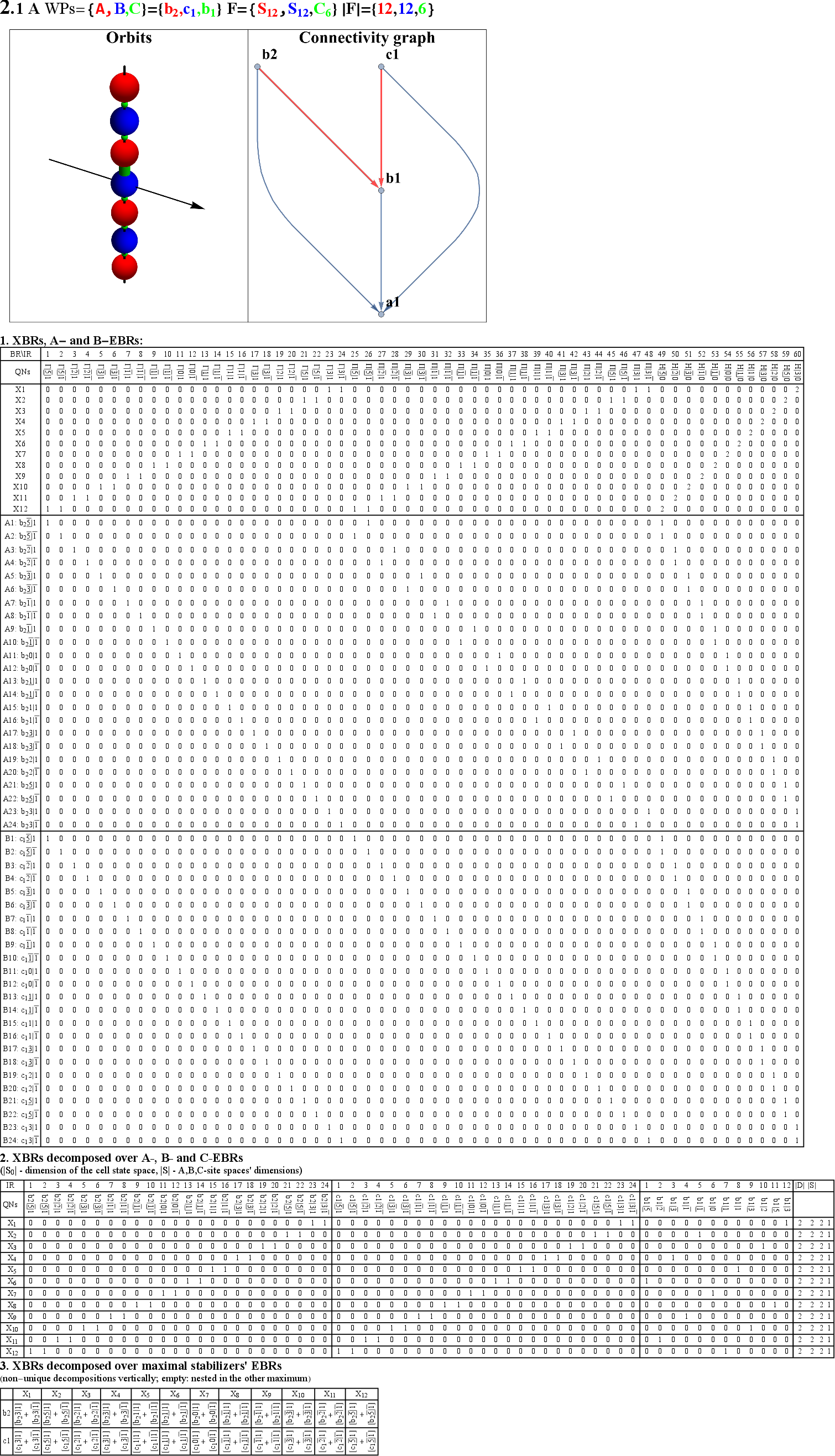}\end{figure*}
\begin{figure*}[b]\includegraphics[width=0.94\textwidth]{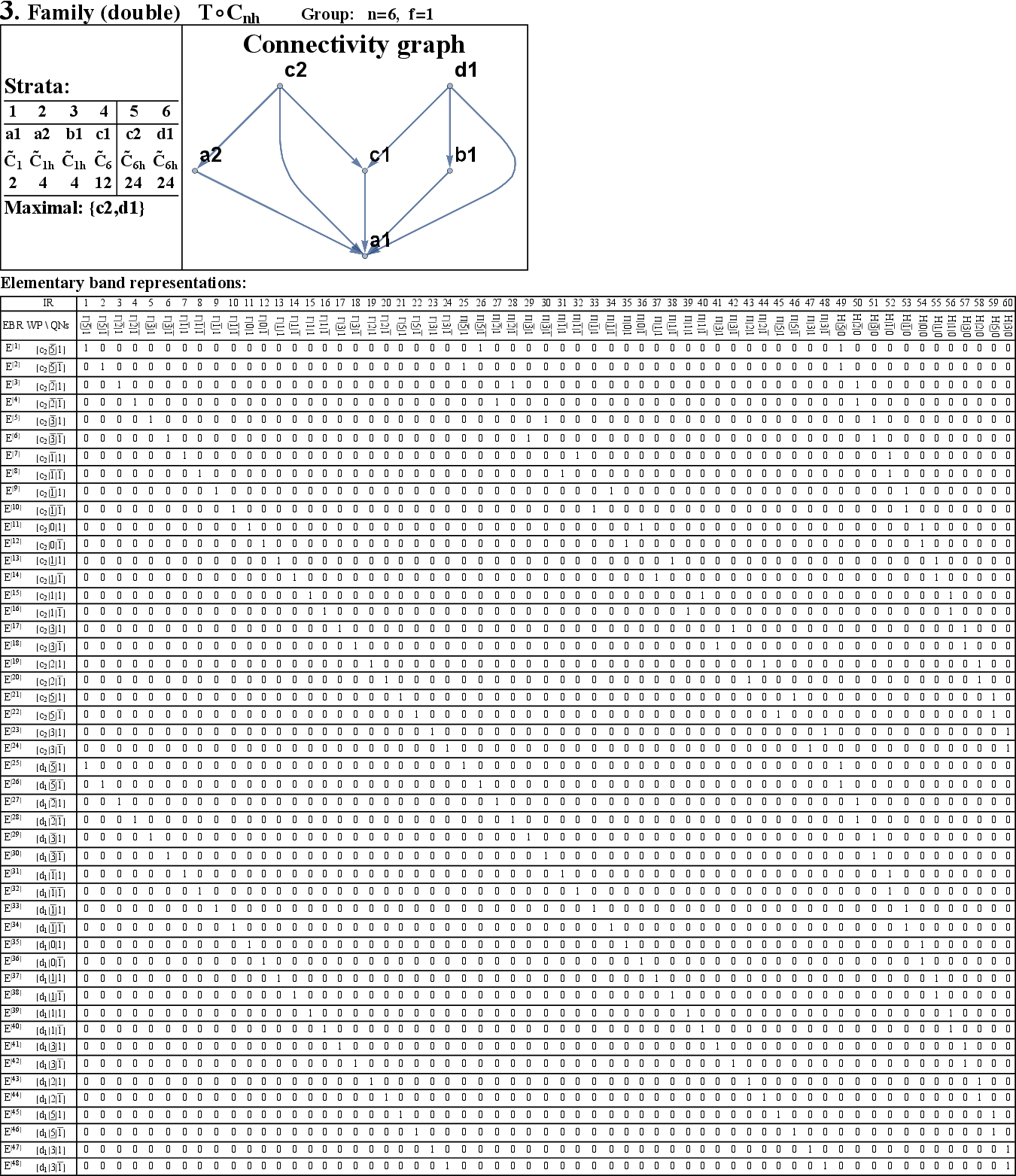}\end{figure*}
\begin{figure*}[b]\includegraphics[width=0.75\textwidth]{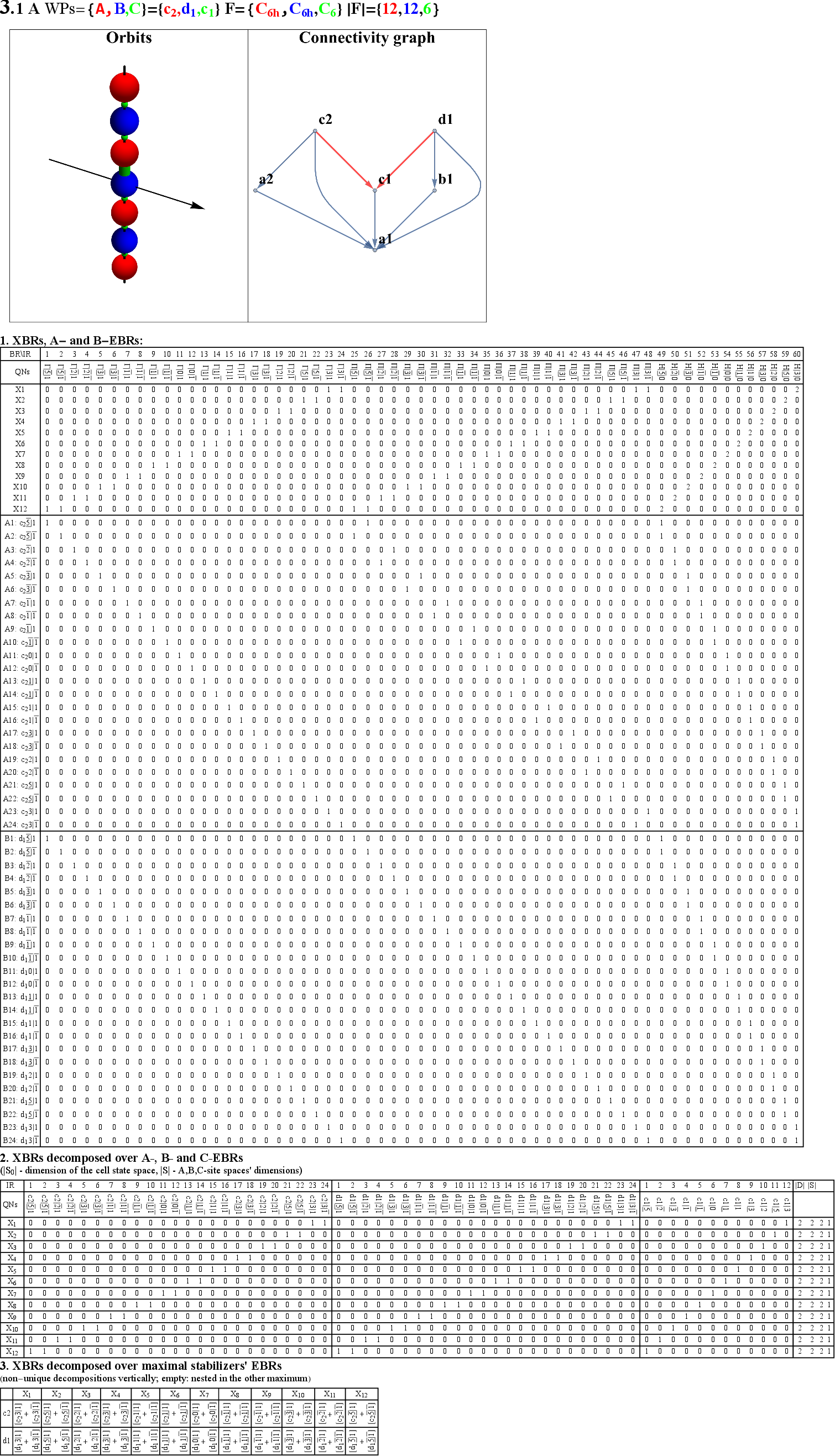}\end{figure*}
\begin{figure*}[b]\includegraphics[width=0.94\textwidth]{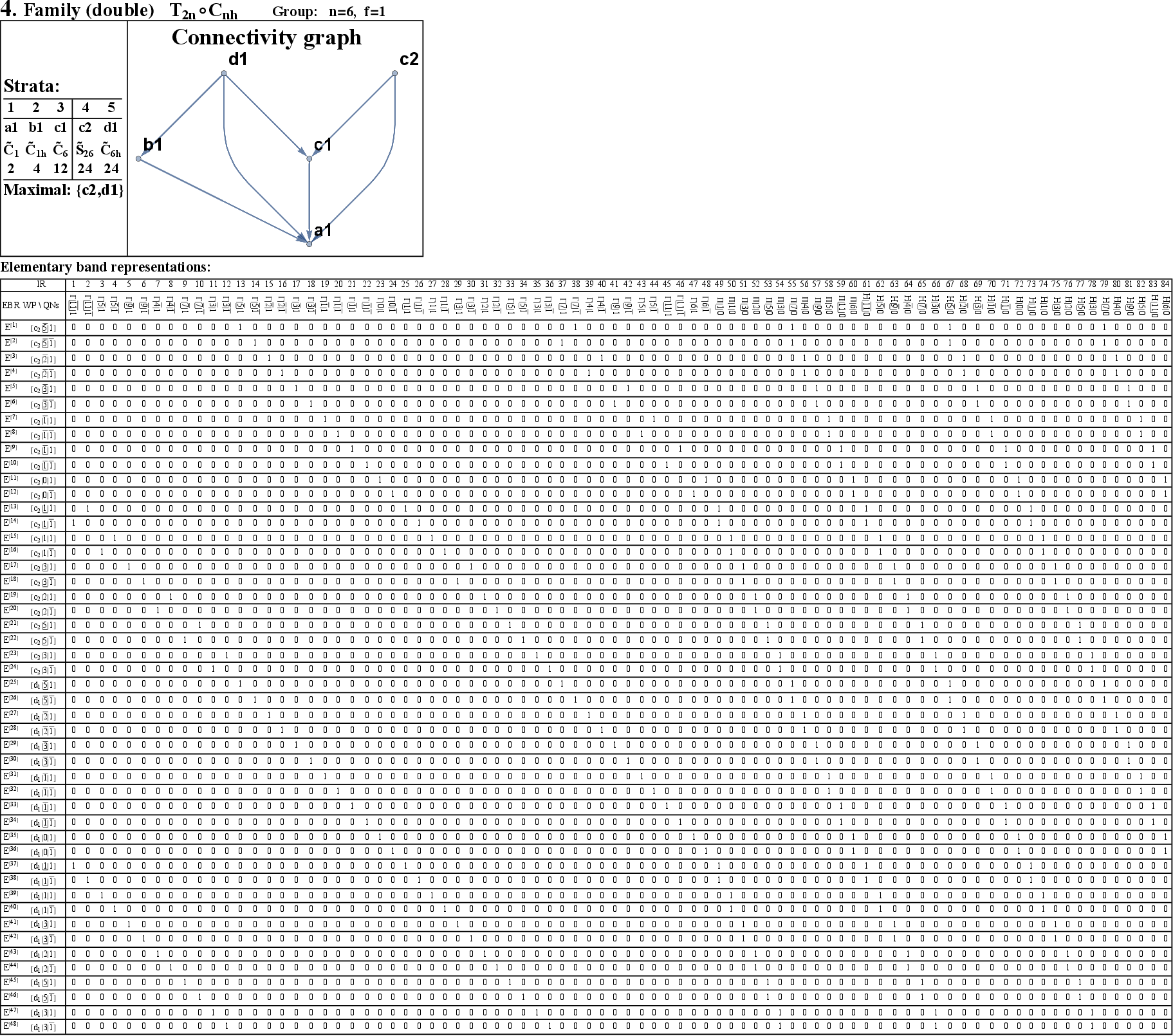}\end{figure*}
\begin{figure*}[b]\includegraphics[width=0.85\textwidth]{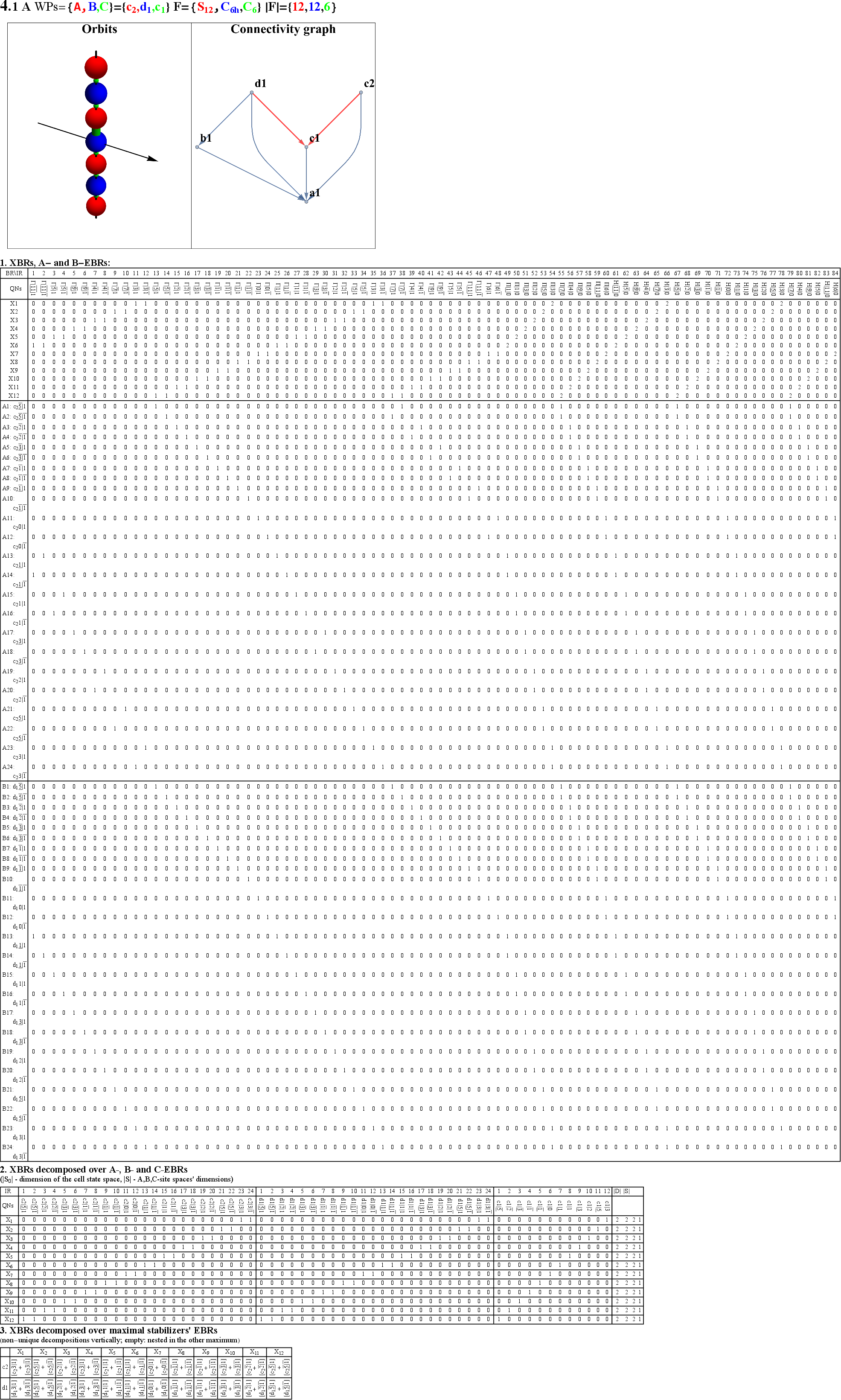}\end{figure*}
\begin{figure*}[b]\includegraphics[width=0.94\textwidth]{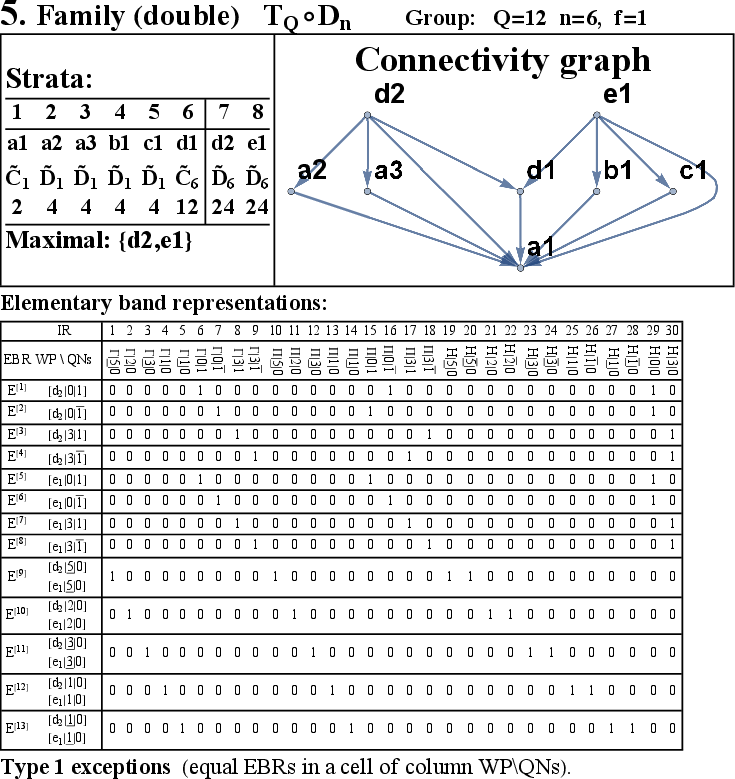}\end{figure*}
\begin{figure*}[b]\includegraphics[width=0.75\textwidth]{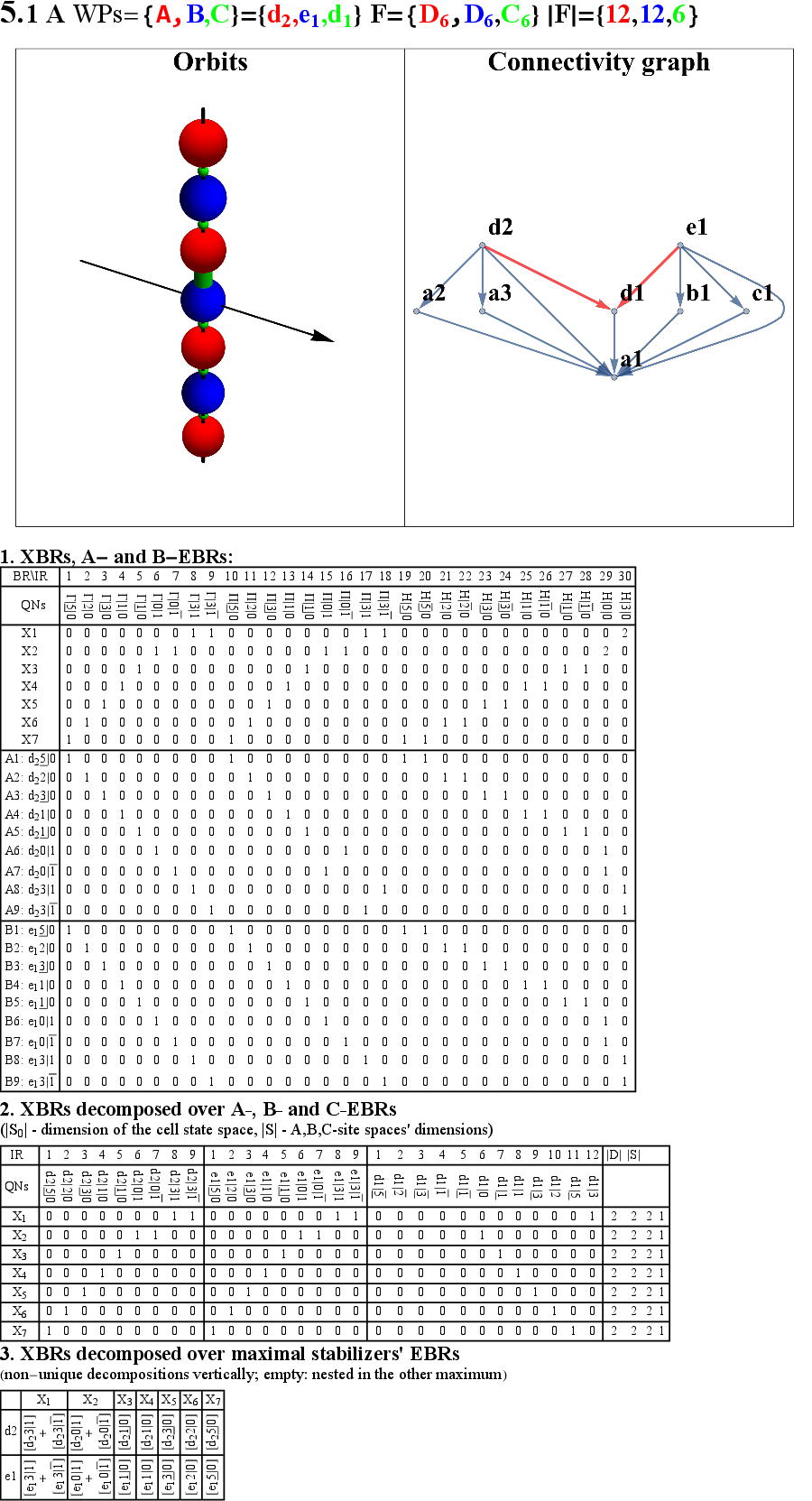}\end{figure*}
\clearpage
\begin{figure*}[b]\includegraphics[width=0.94\textwidth]{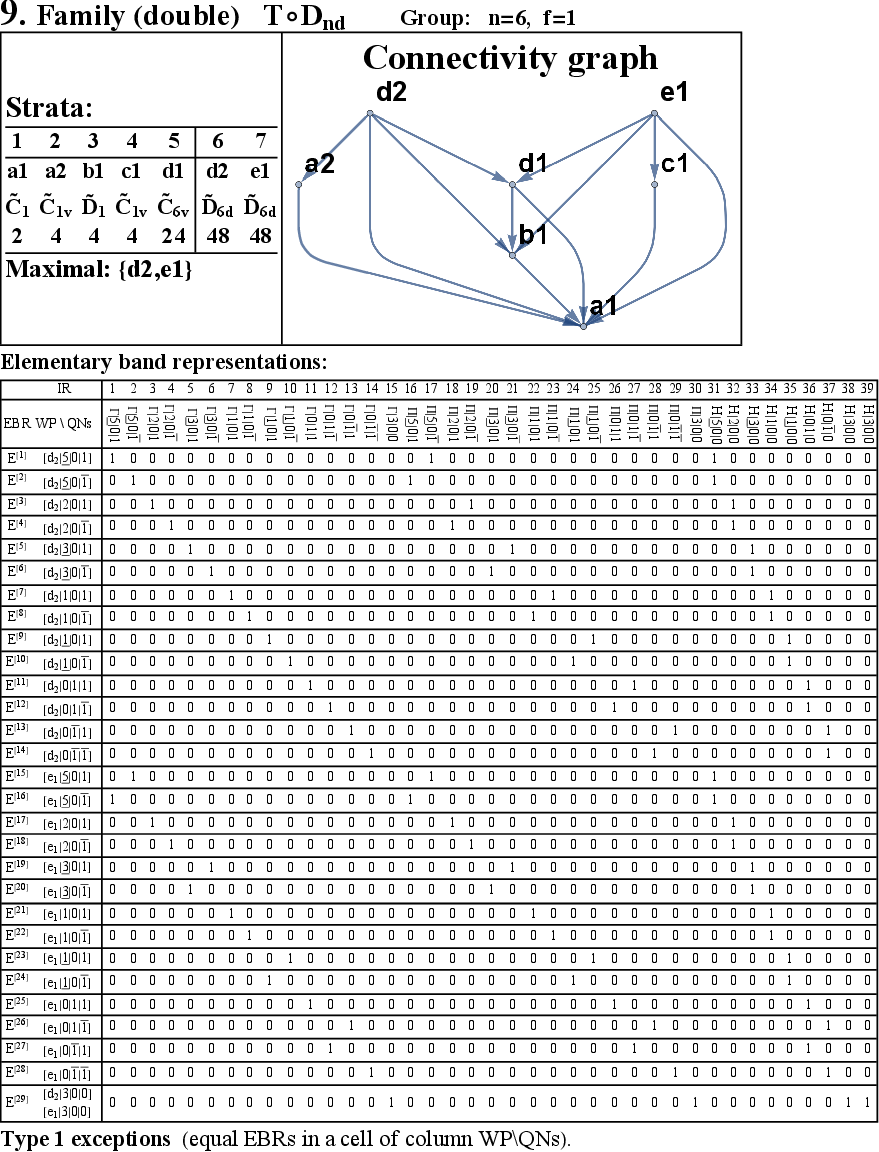}\end{figure*}
\begin{figure*}[b]\includegraphics[width=0.80\textwidth]{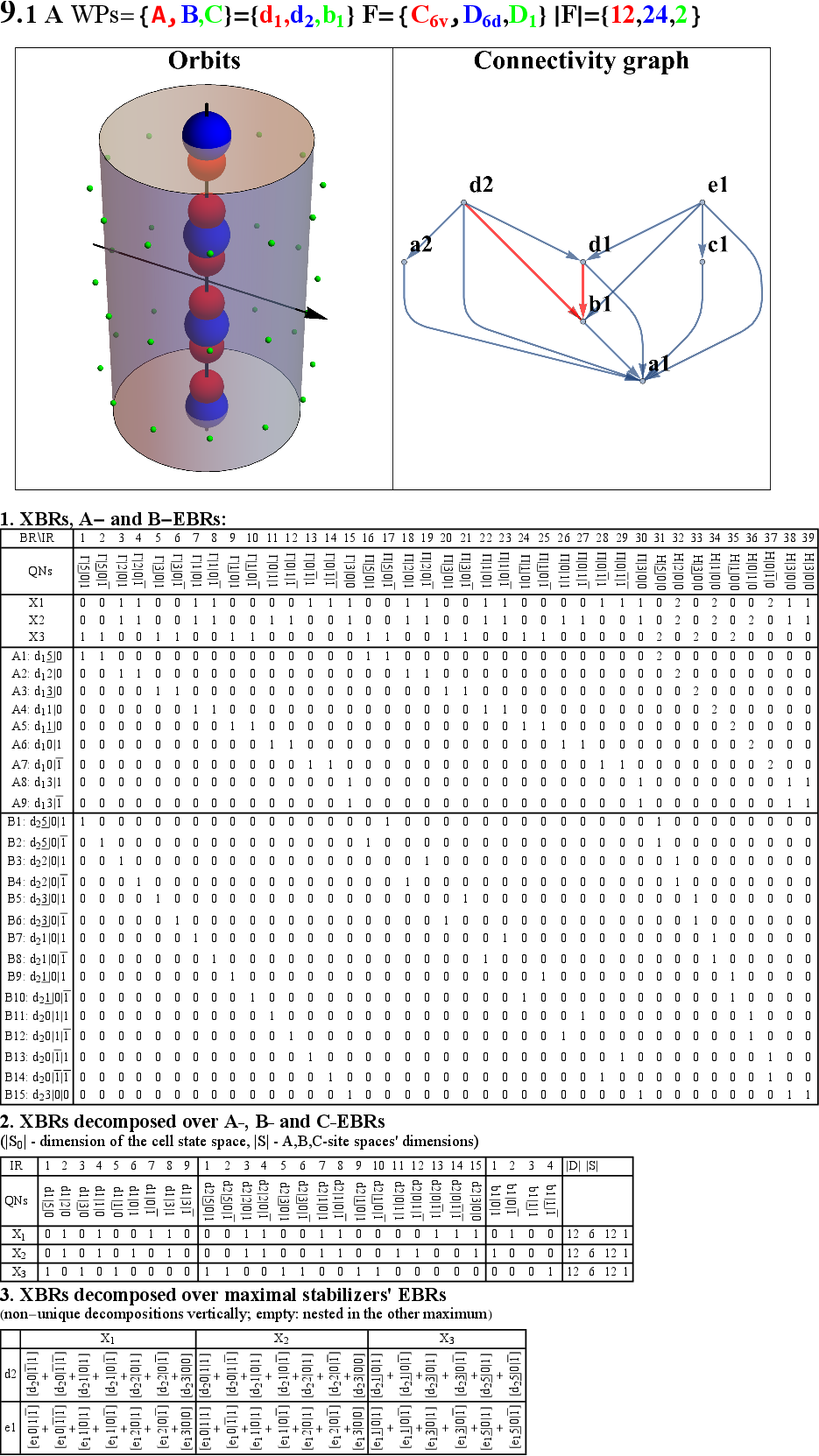}\end{figure*}
\begin{figure*}[b]\includegraphics[width=0.80\textwidth]{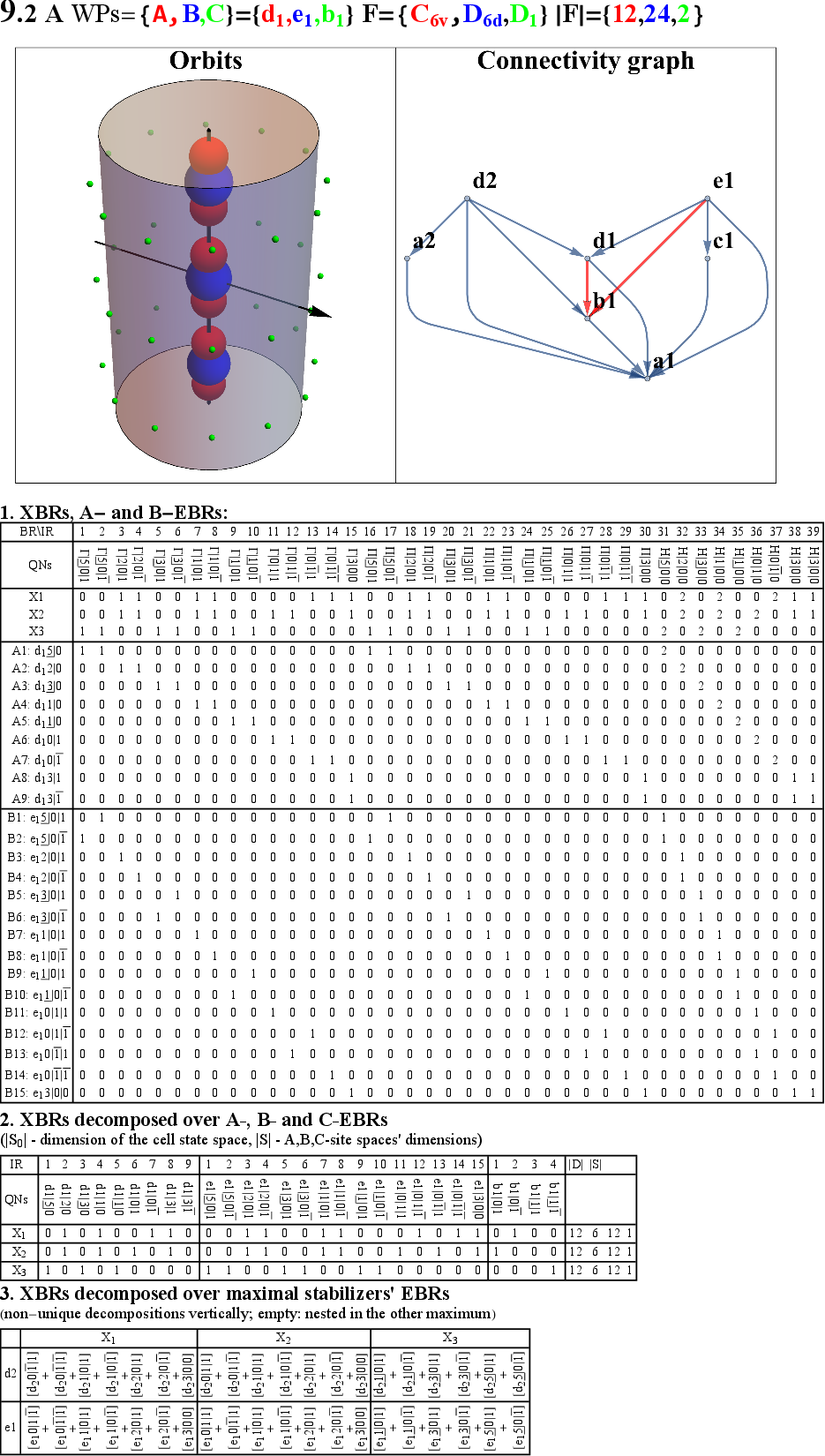}\end{figure*}
\begin{figure*}[b]\includegraphics[width=0.66\textwidth]{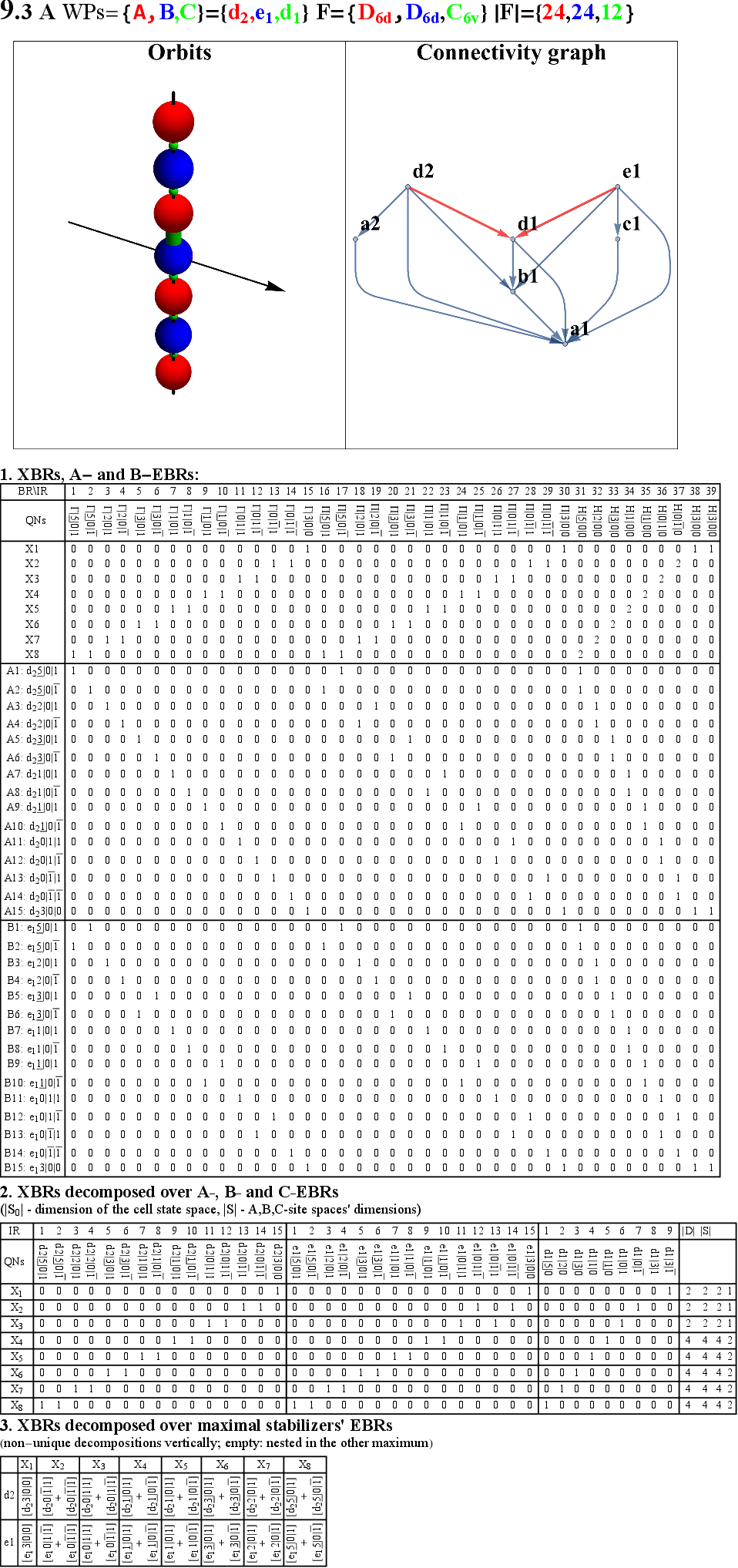}\end{figure*}
\begin{figure*}[b]\includegraphics[width=0.85\textwidth]{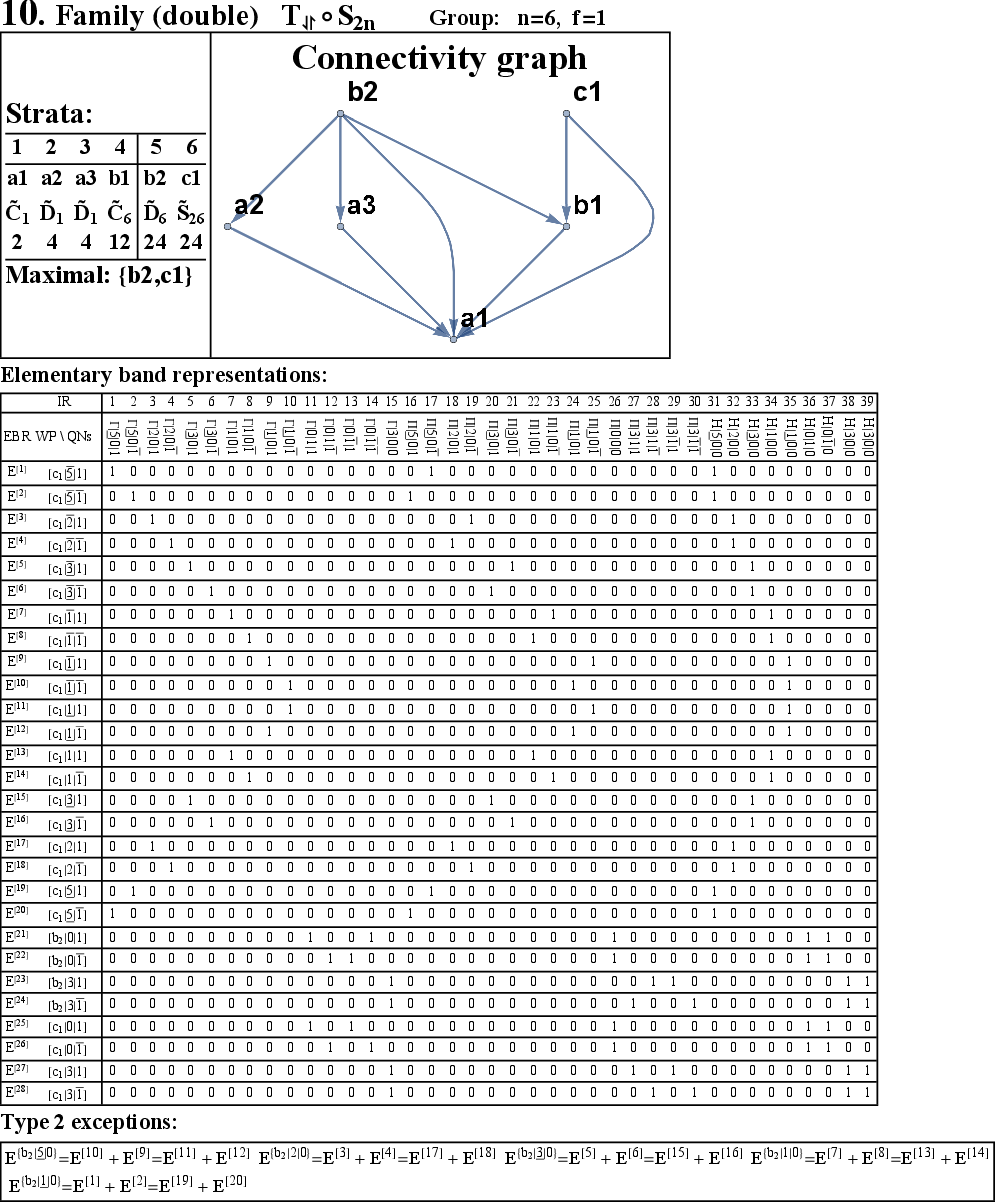}\end{figure*}
\begin{figure*}[b]\includegraphics[width=0.75\textwidth]{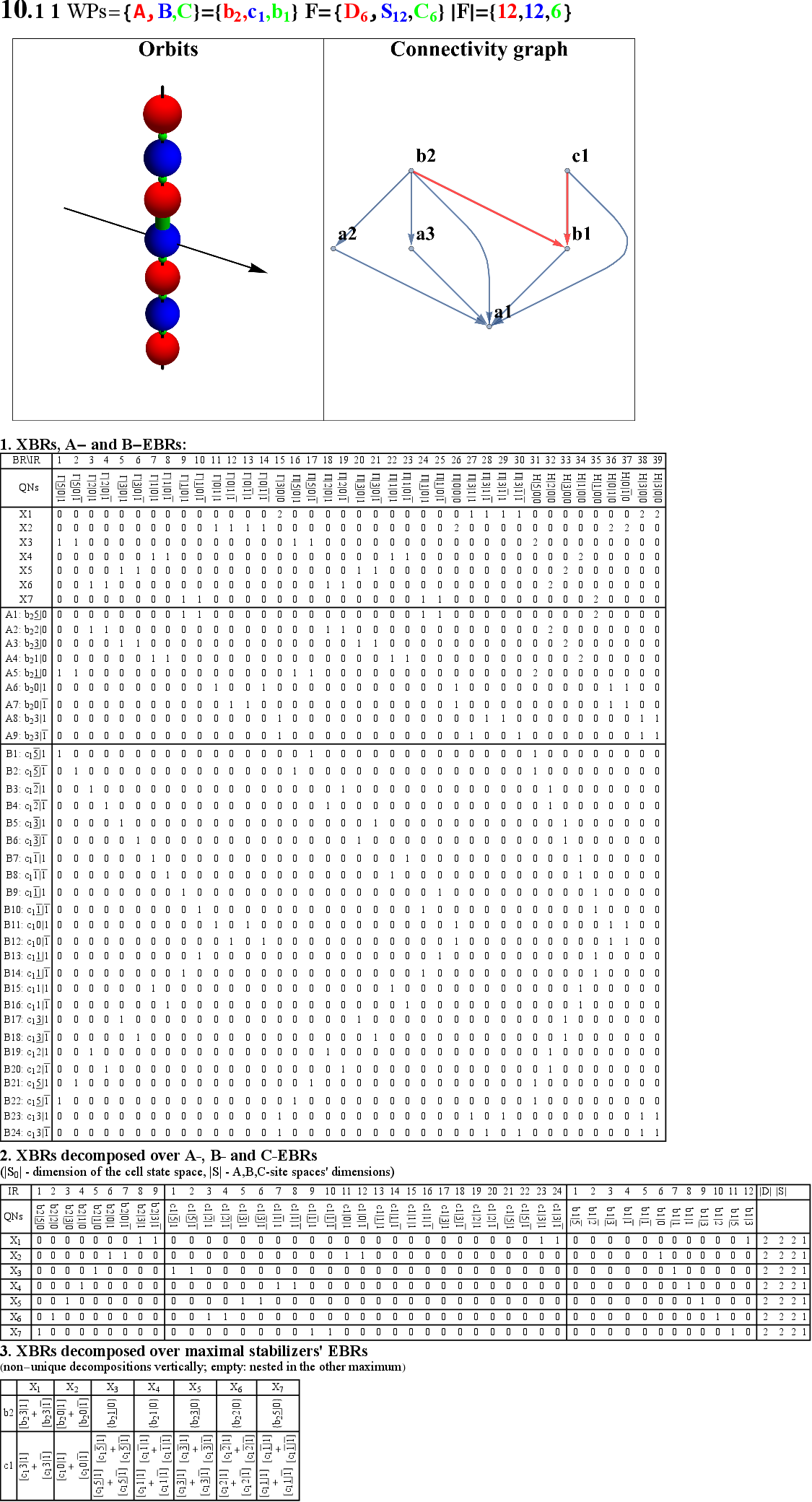}\end{figure*}
\clearpage
\begin{figure*}[b]\includegraphics[width=0.94\textwidth]{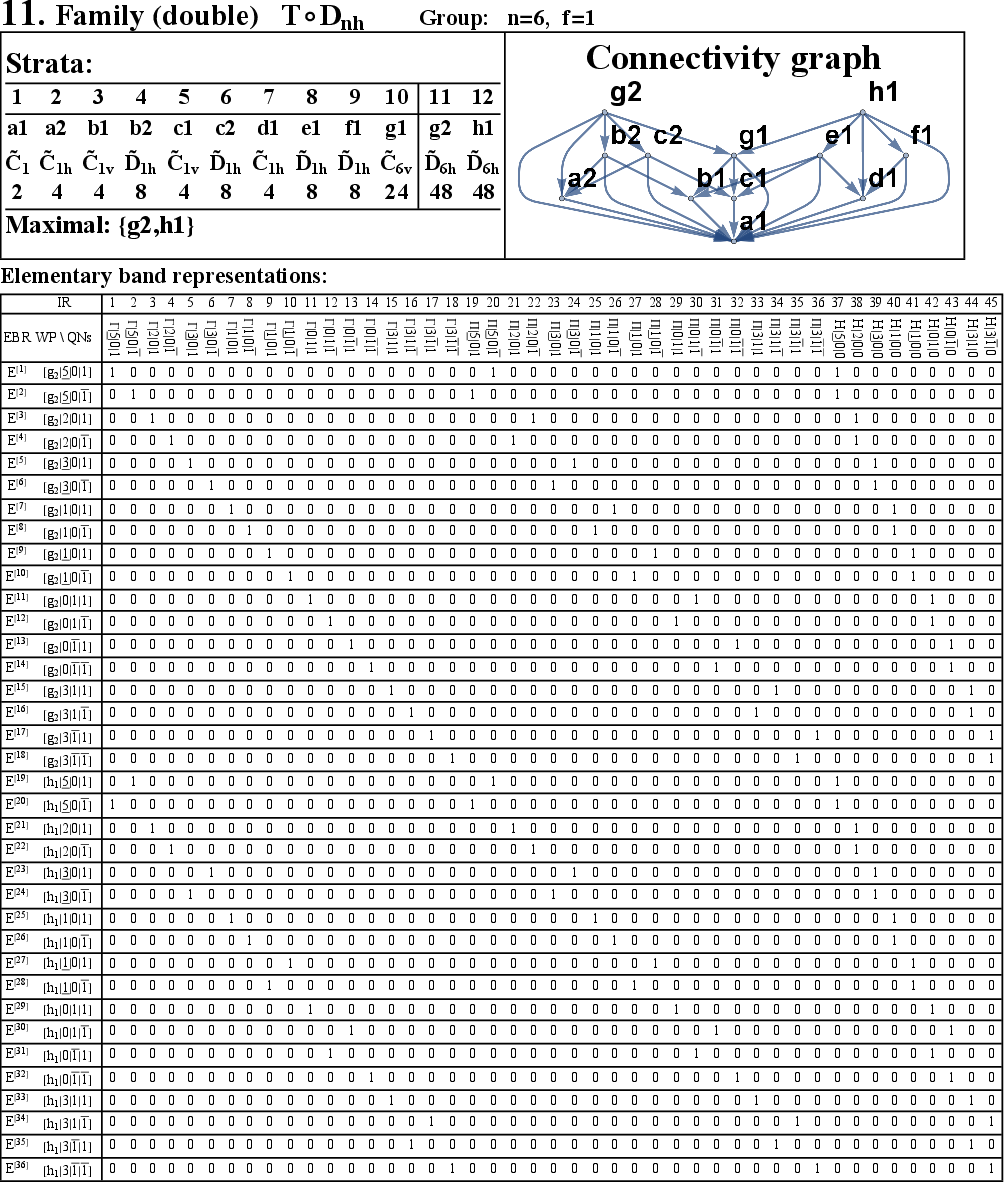}\end{figure*}
\begin{figure*}[b]\includegraphics[width=0.94\textwidth]{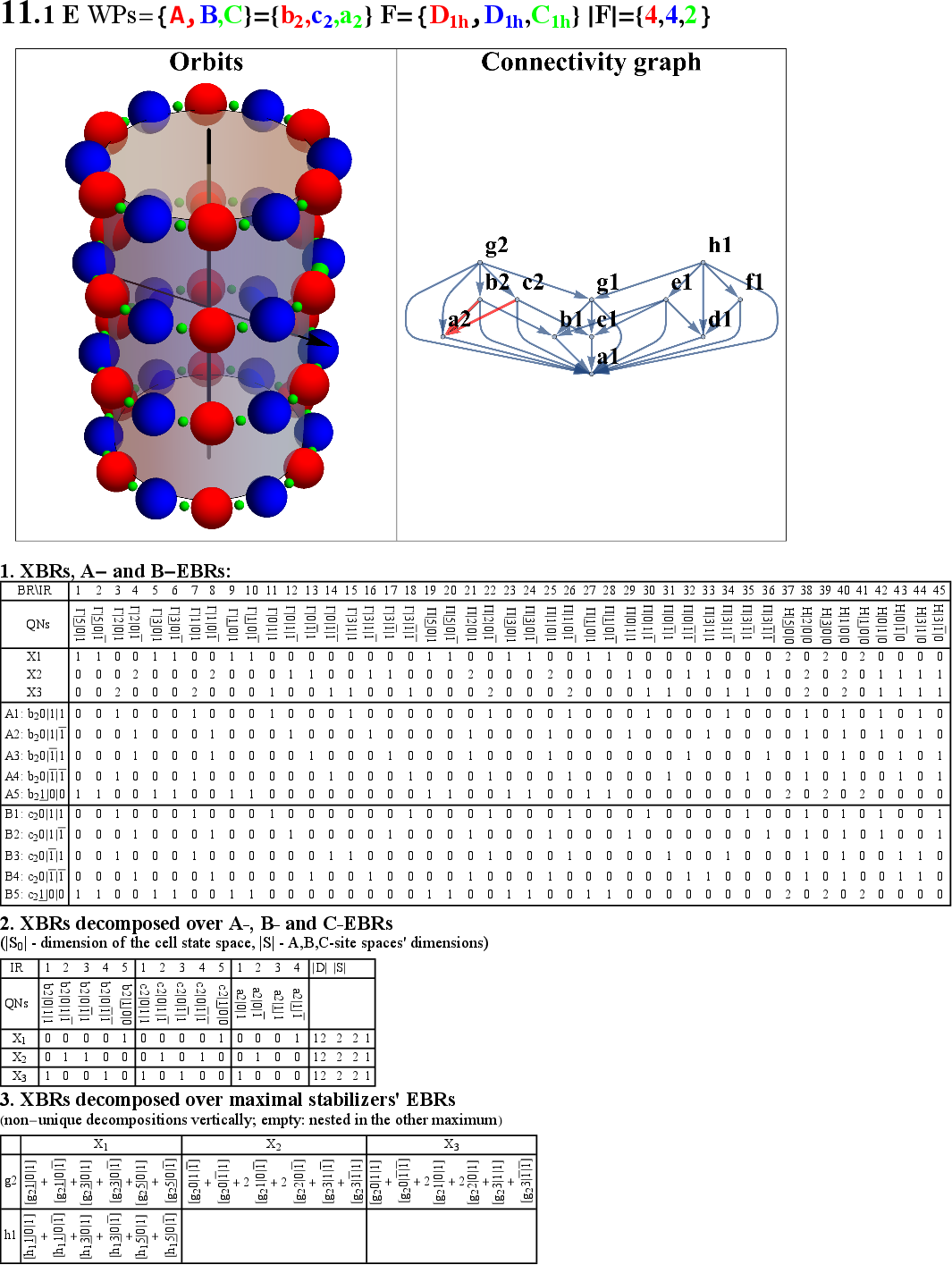}\end{figure*}
\begin{figure*}[b]\includegraphics[width=0.94\textwidth]{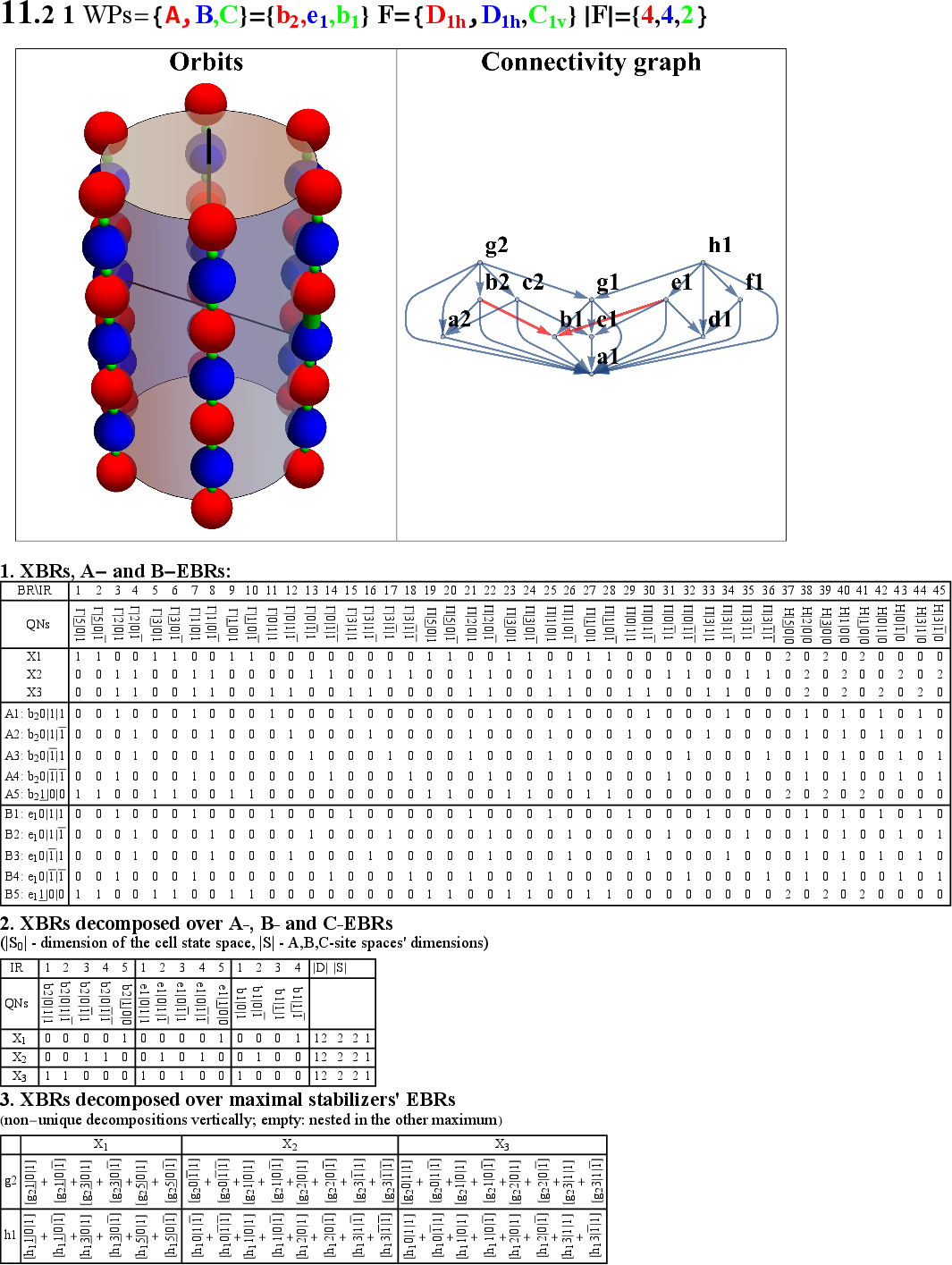}\end{figure*}
\begin{figure*}[b]\includegraphics[width=0.94\textwidth]{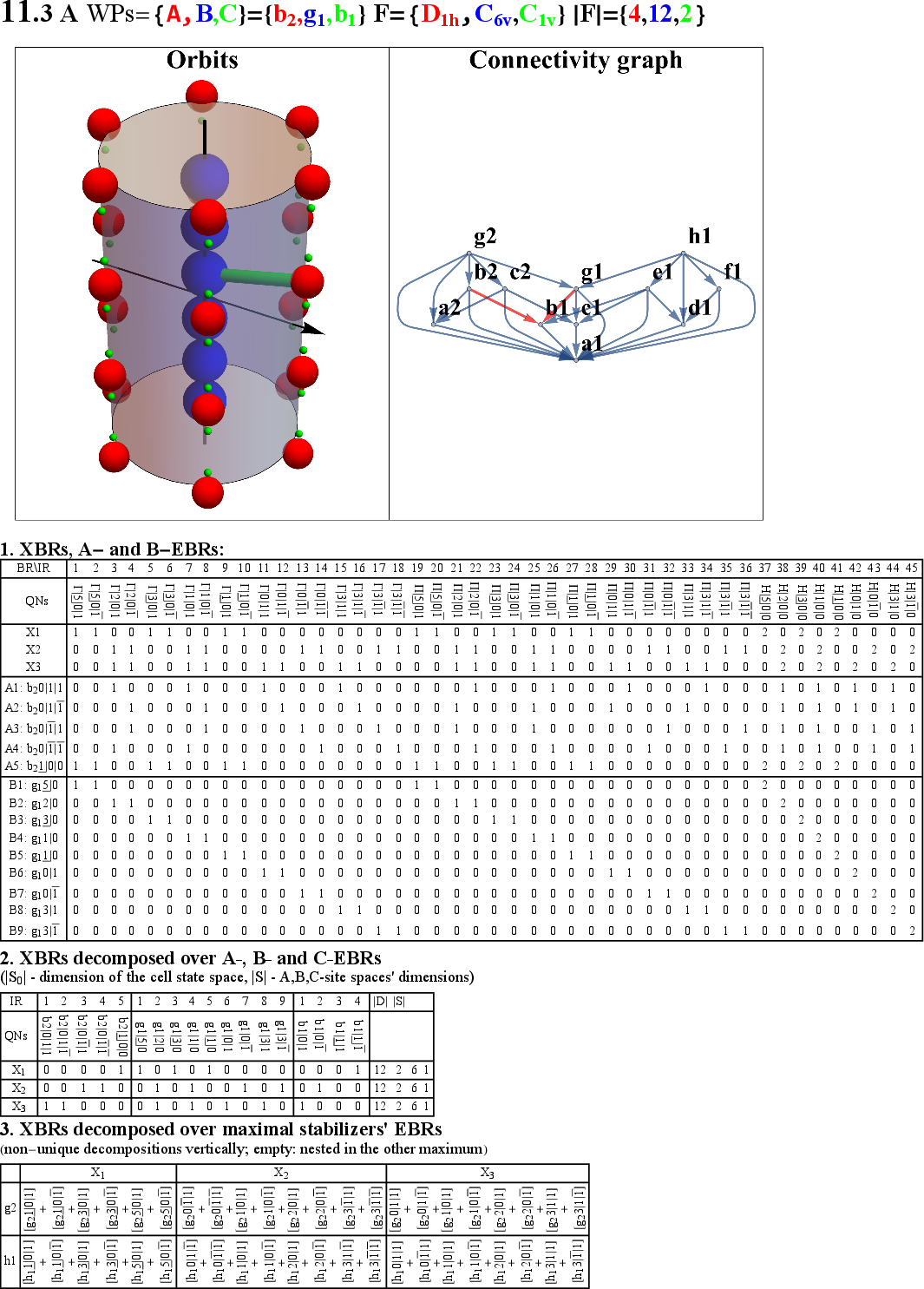}\end{figure*}
\begin{figure*}[b]\includegraphics[width=0.85\textwidth]{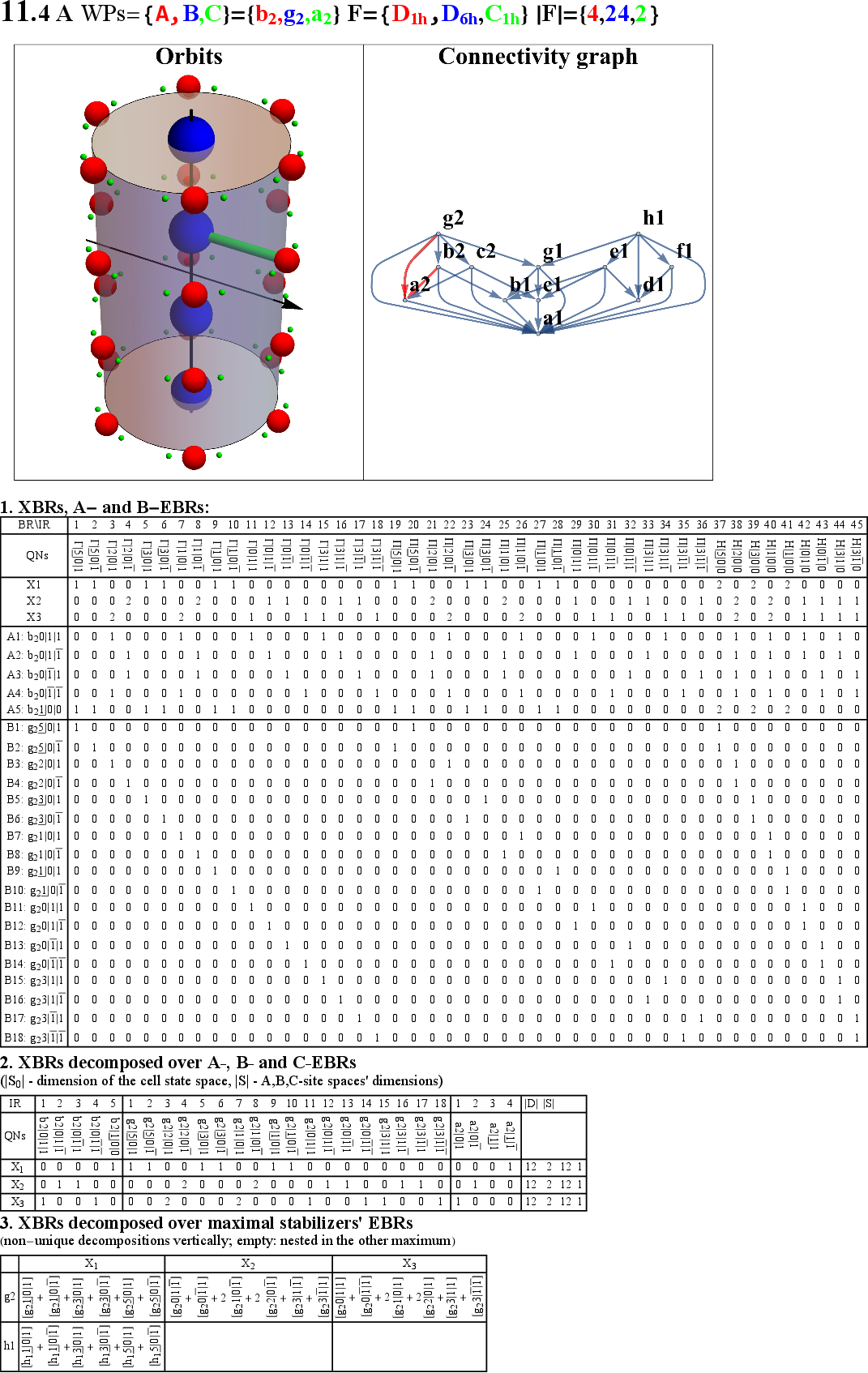}\end{figure*}
\begin{figure*}[b]\includegraphics[width=0.94\textwidth]{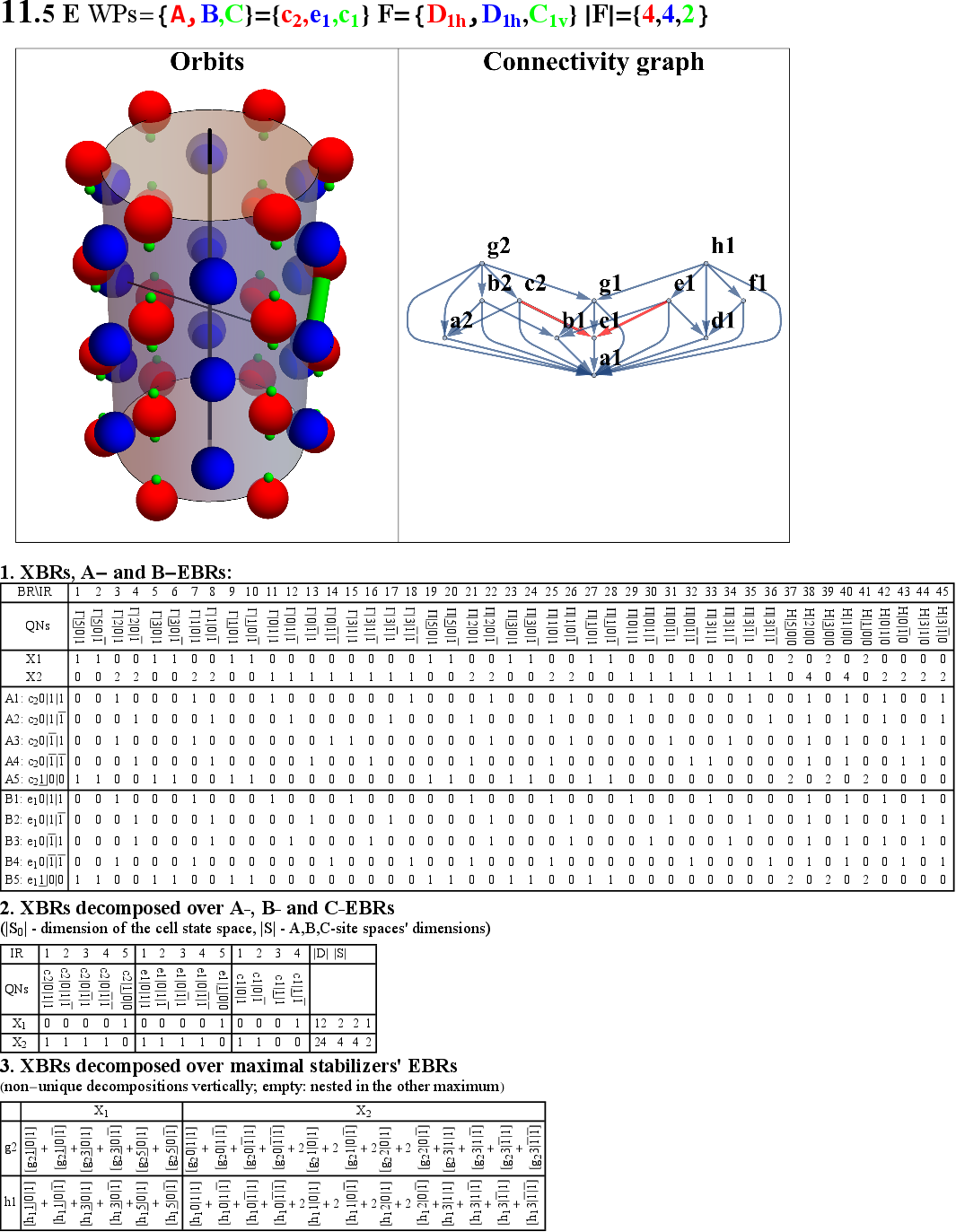}\end{figure*}
\begin{figure*}[b]\includegraphics[width=0.94\textwidth]{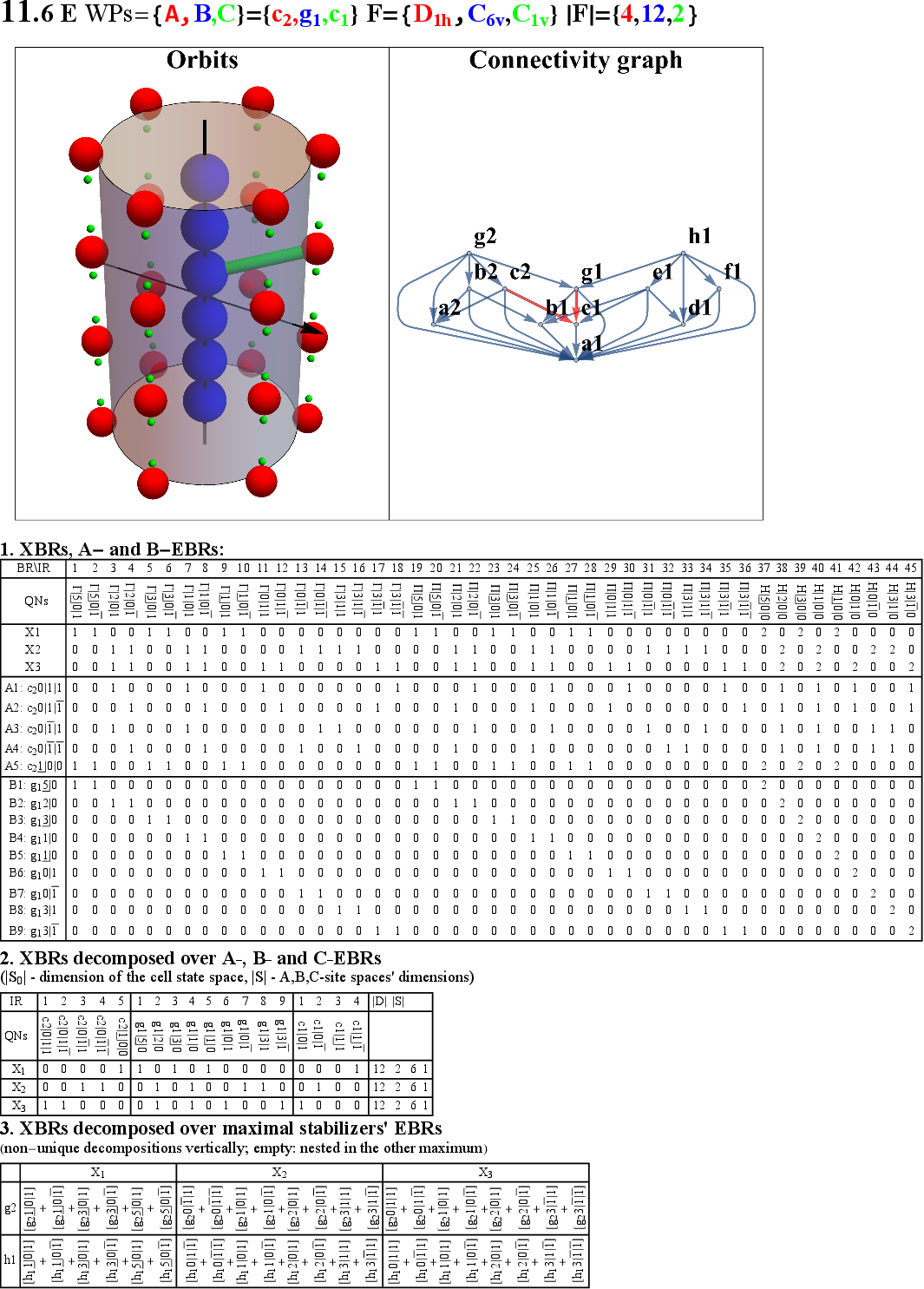}\end{figure*}
\begin{figure*}[b]\includegraphics[width=0.85\textwidth]{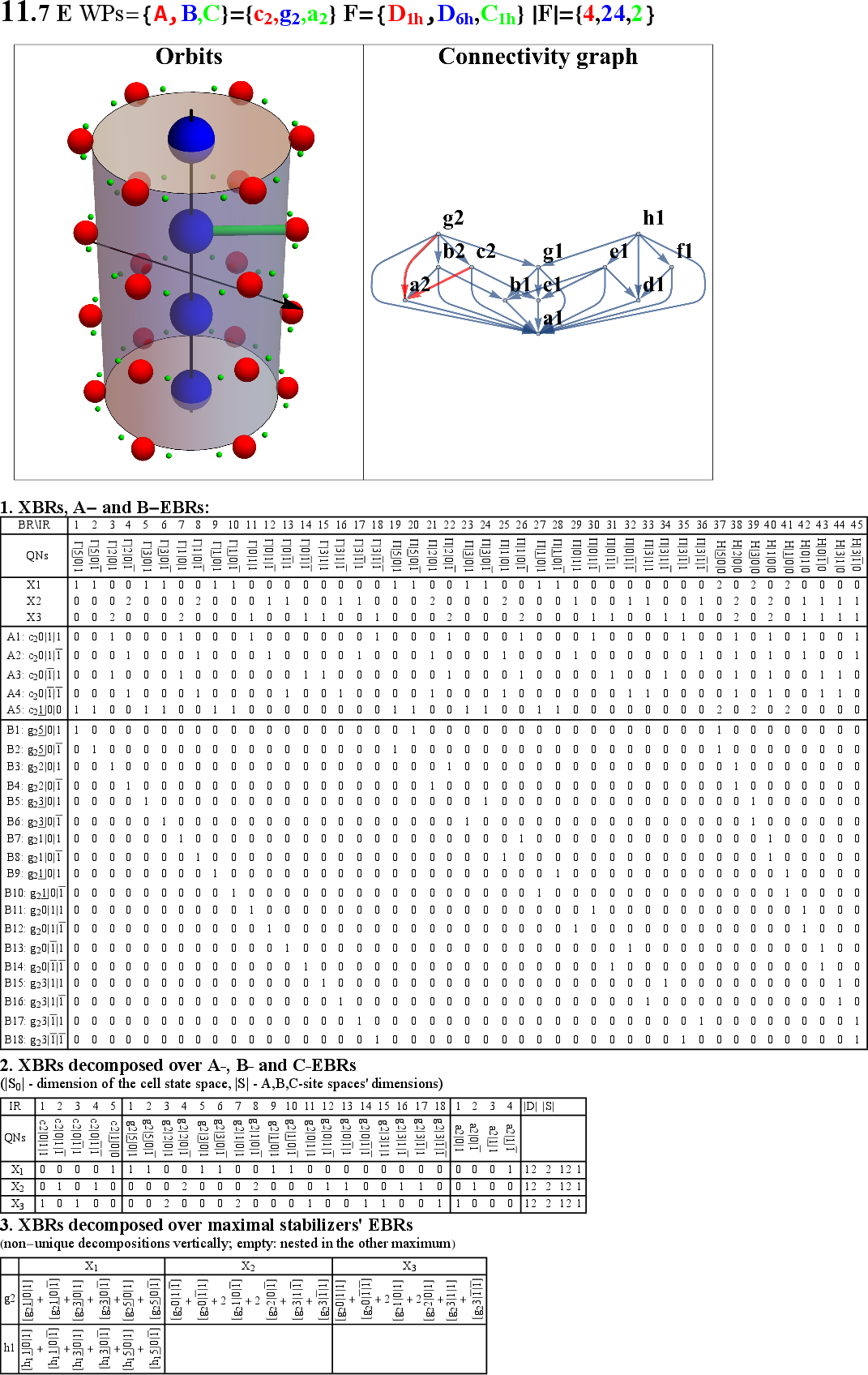}\end{figure*}
\begin{figure*}[b]\includegraphics[width=0.94\textwidth]{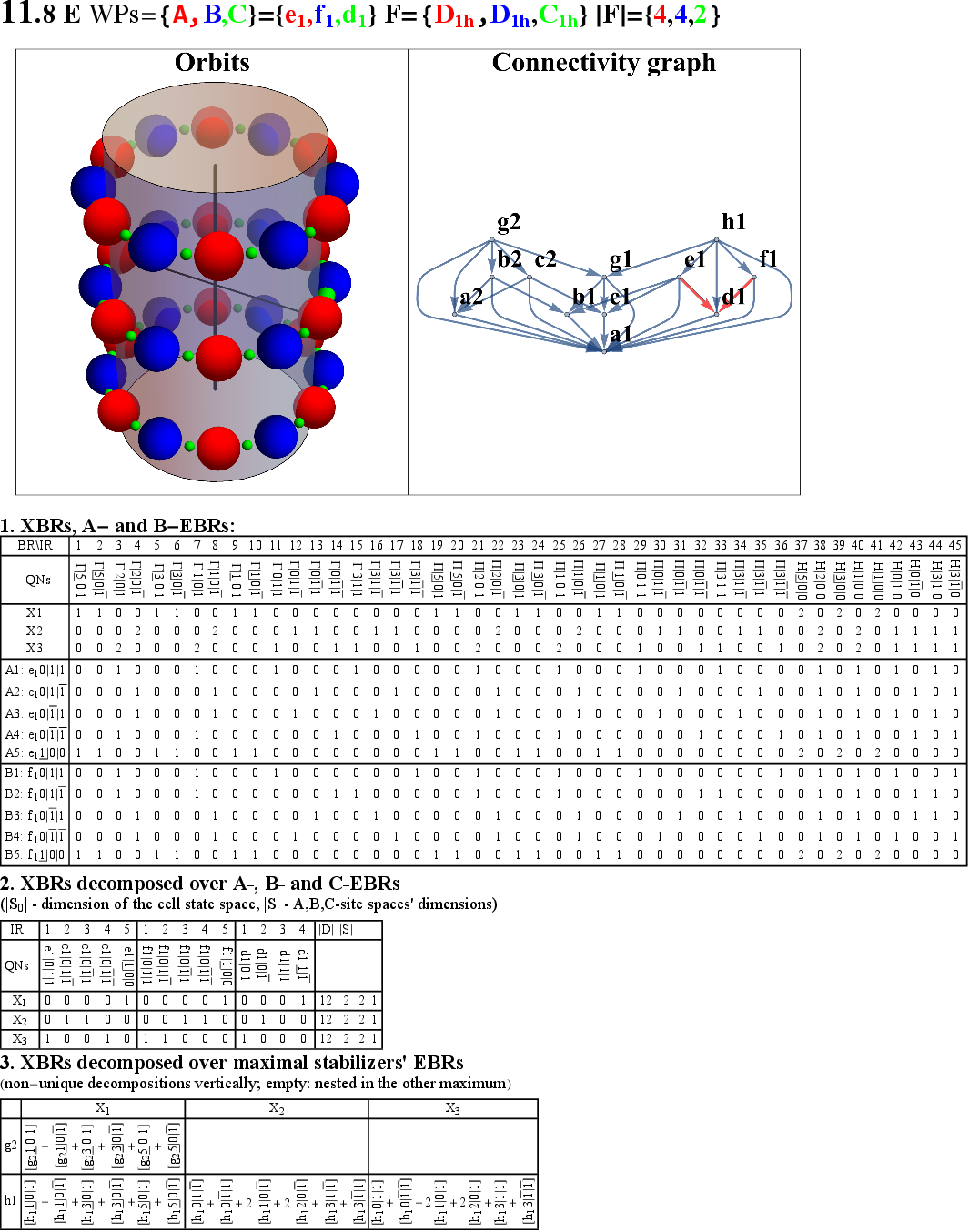}\end{figure*}
\begin{figure*}[b]\includegraphics[width=0.94\textwidth]{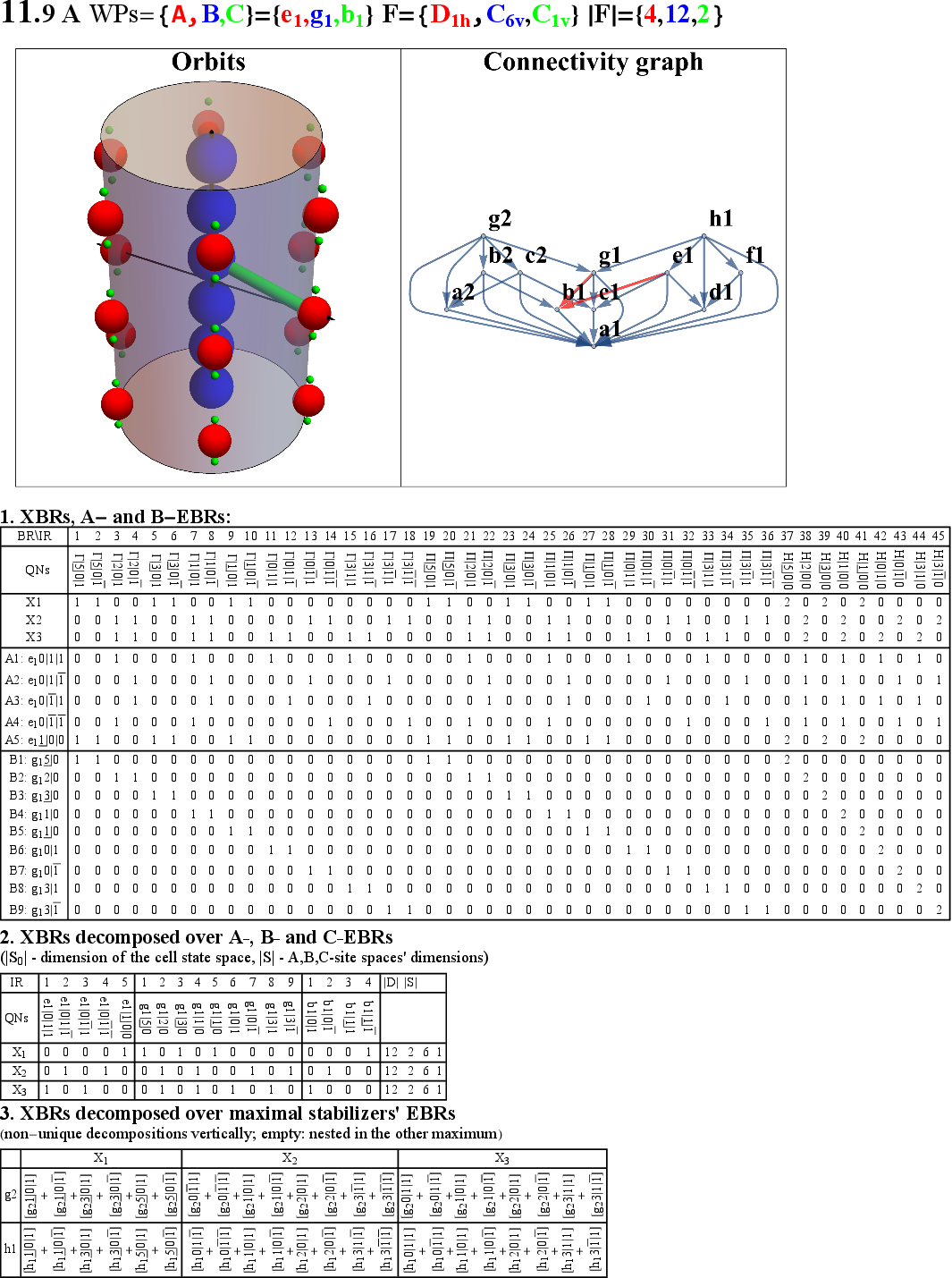}\end{figure*}
\begin{figure*}[b]\includegraphics[width=0.94\textwidth]{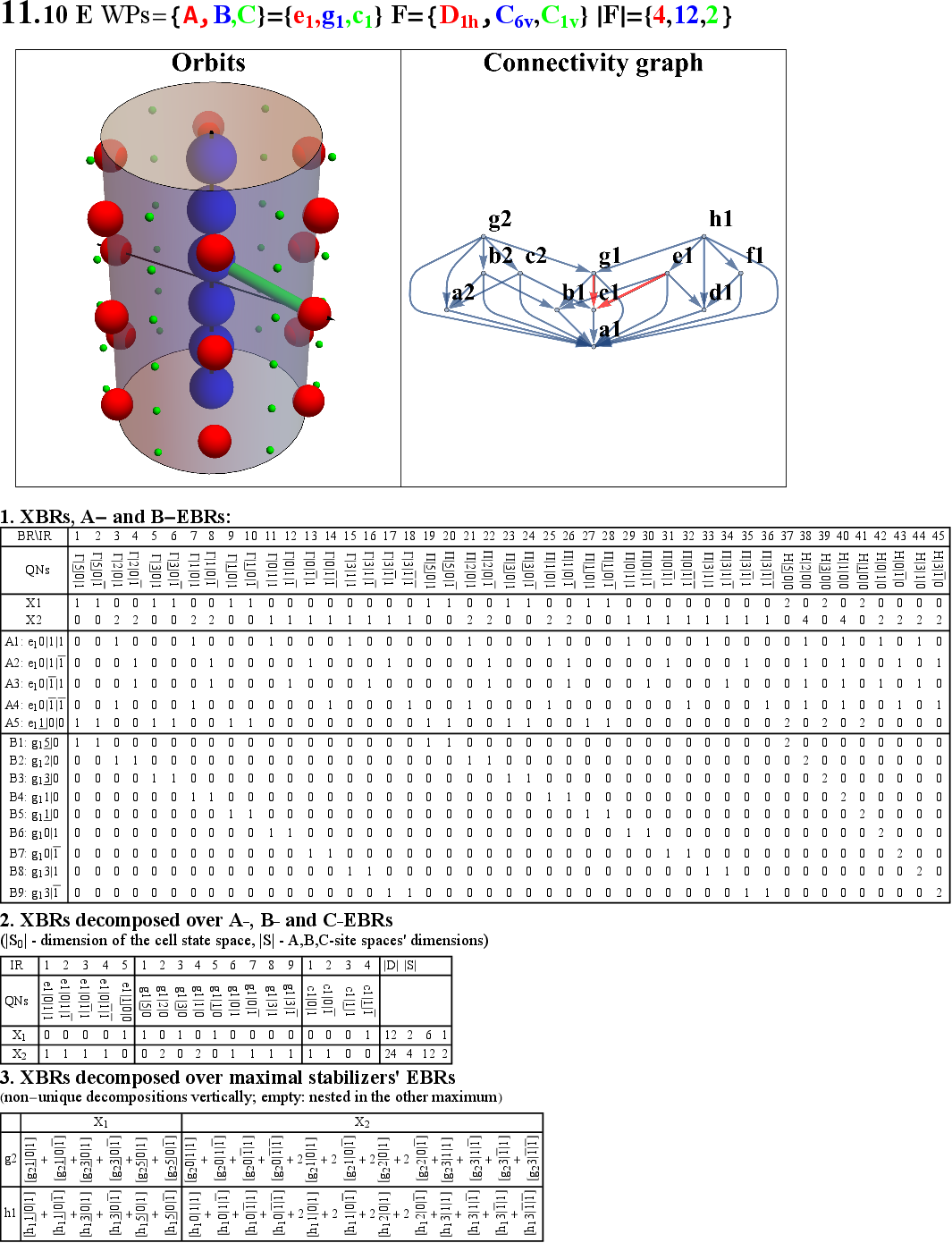}\end{figure*}
\begin{figure*}[b]\includegraphics[width=0.90\textwidth]{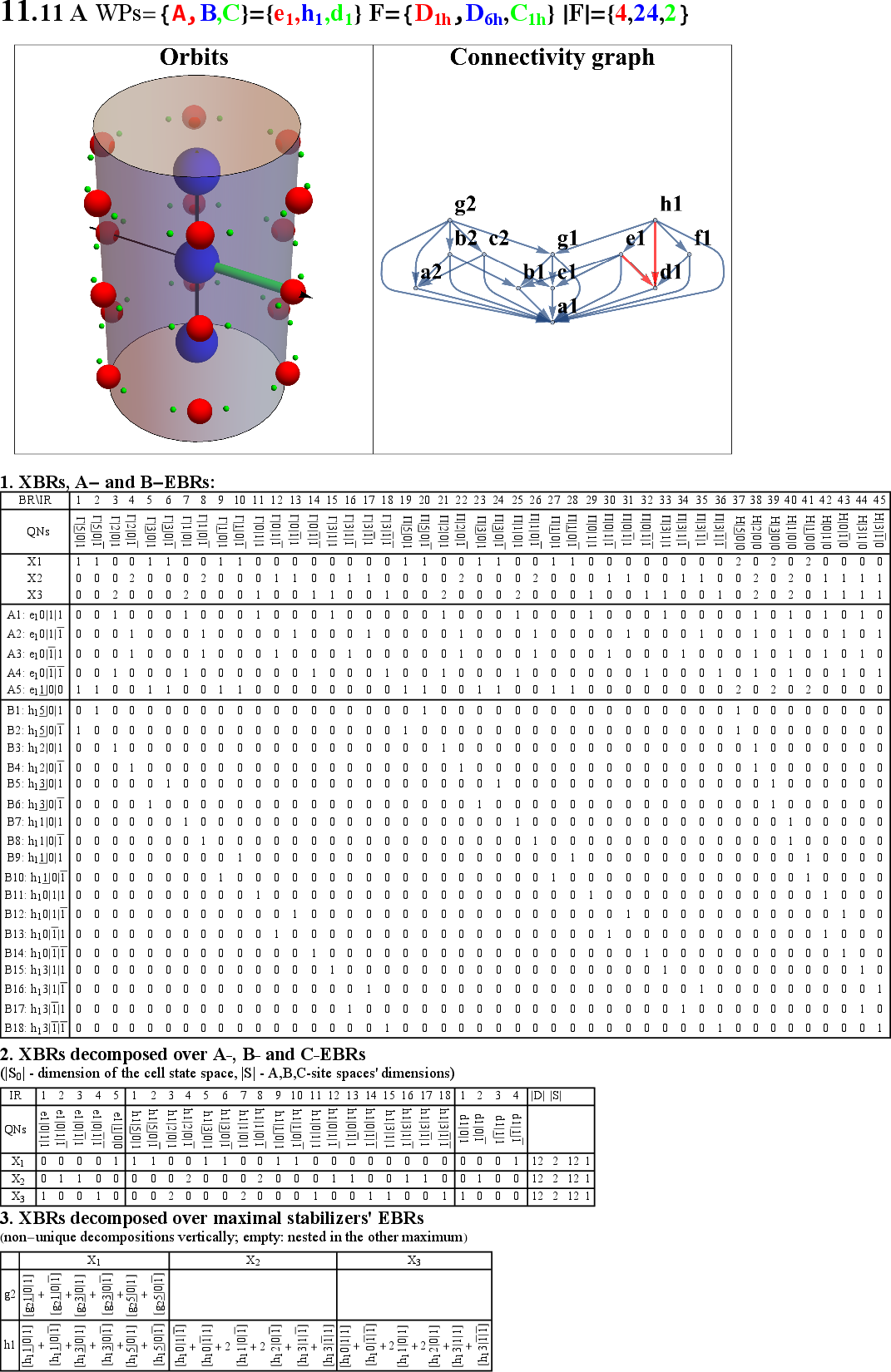}\end{figure*}
\begin{figure*}[b]\includegraphics[width=0.90\textwidth]{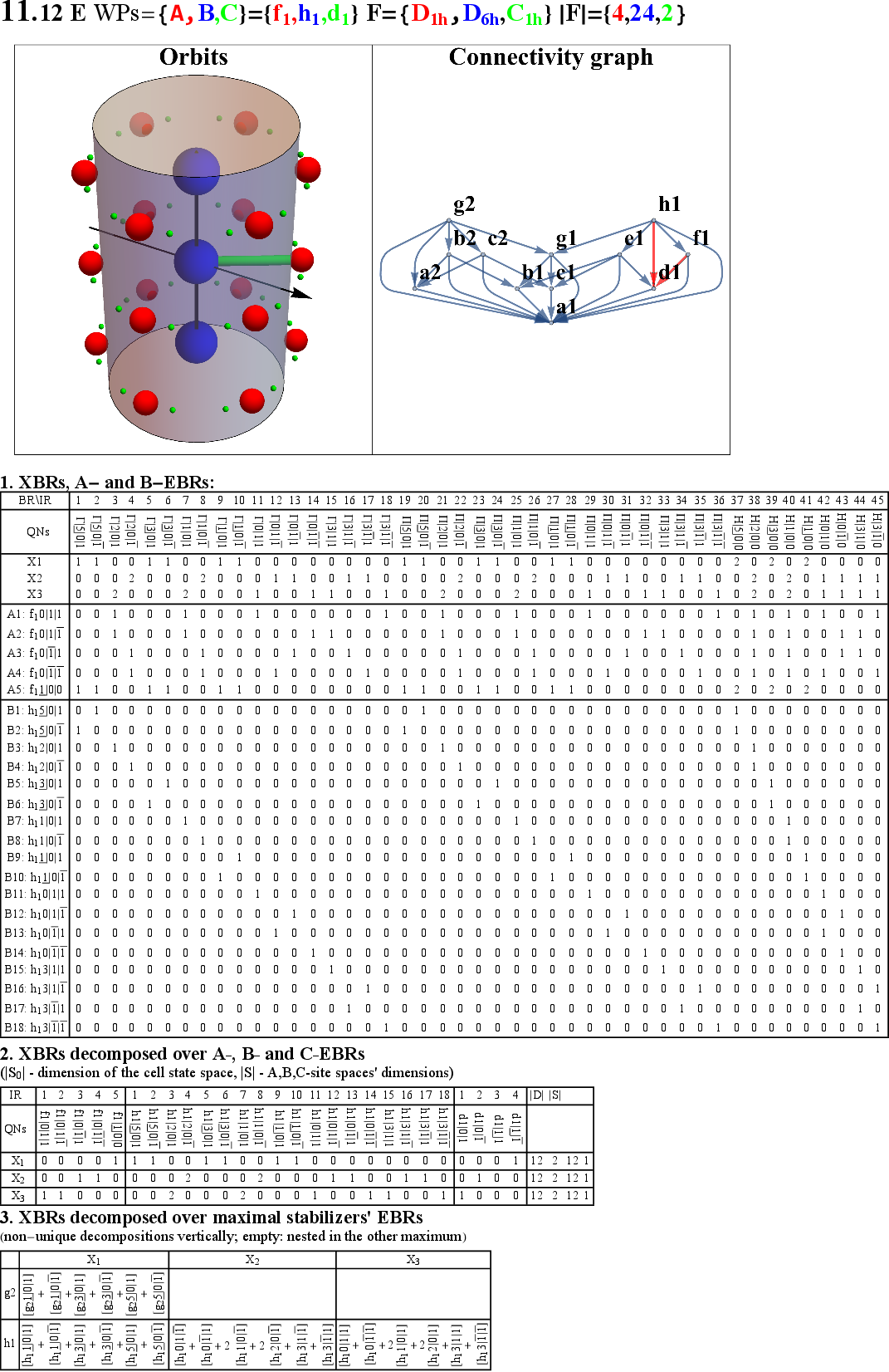}\end{figure*}
\begin{figure*}[b]\includegraphics[width=0.70\textwidth]{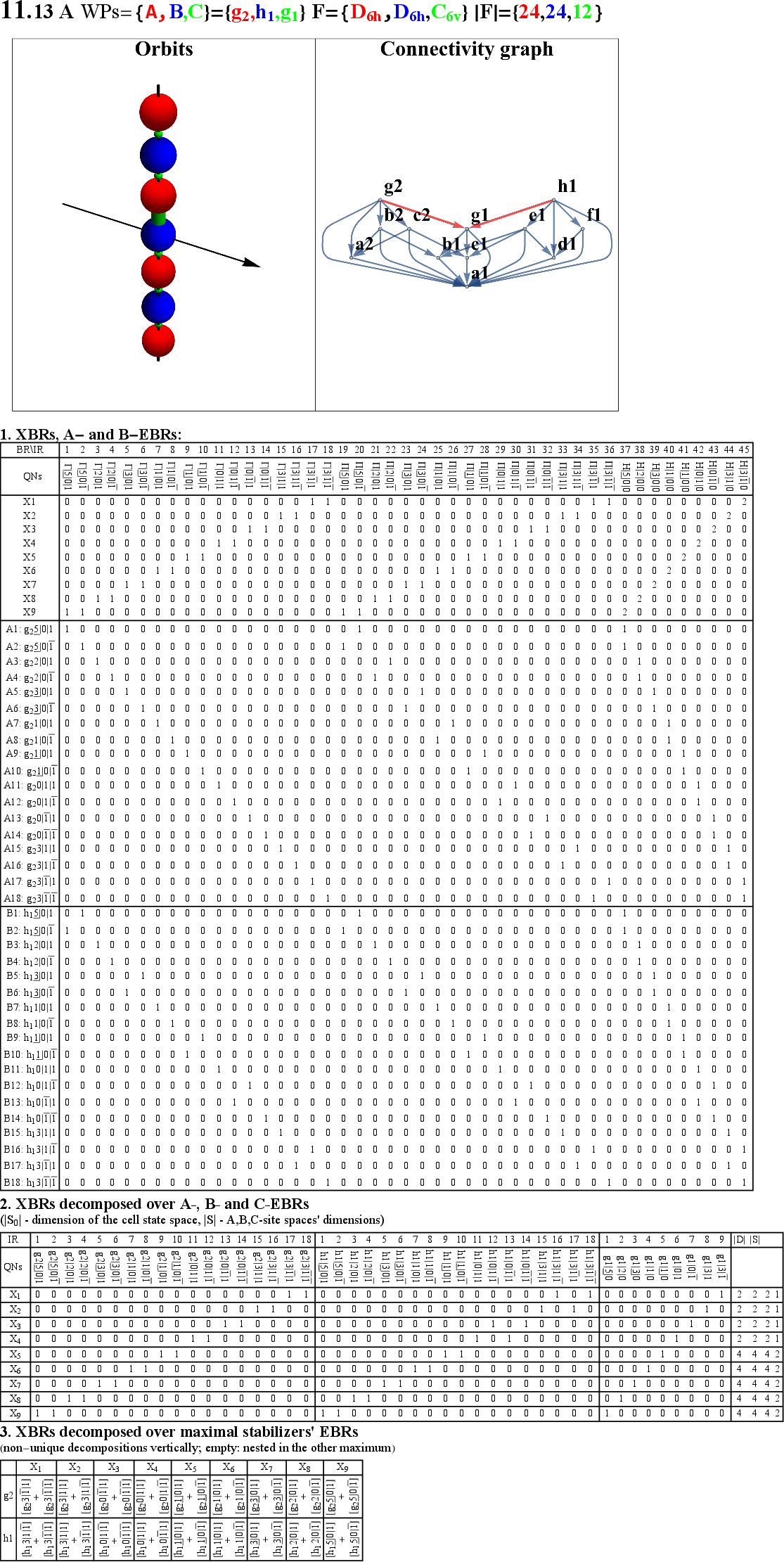}\end{figure*}
\clearpage
\begin{figure*}[b]\includegraphics[width=0.94\textwidth]{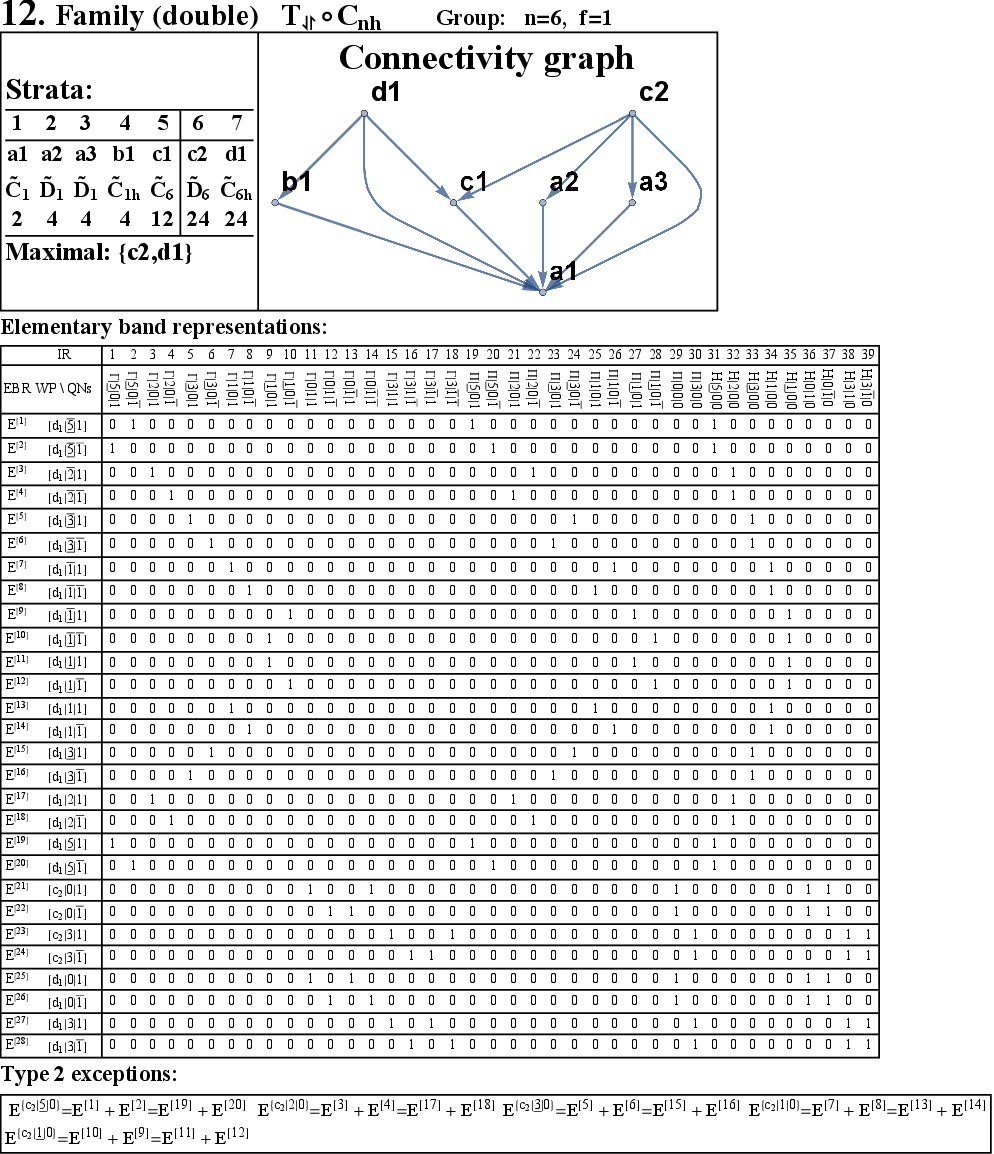}\end{figure*}
\begin{figure*}[b]\includegraphics[width=0.75\textwidth]{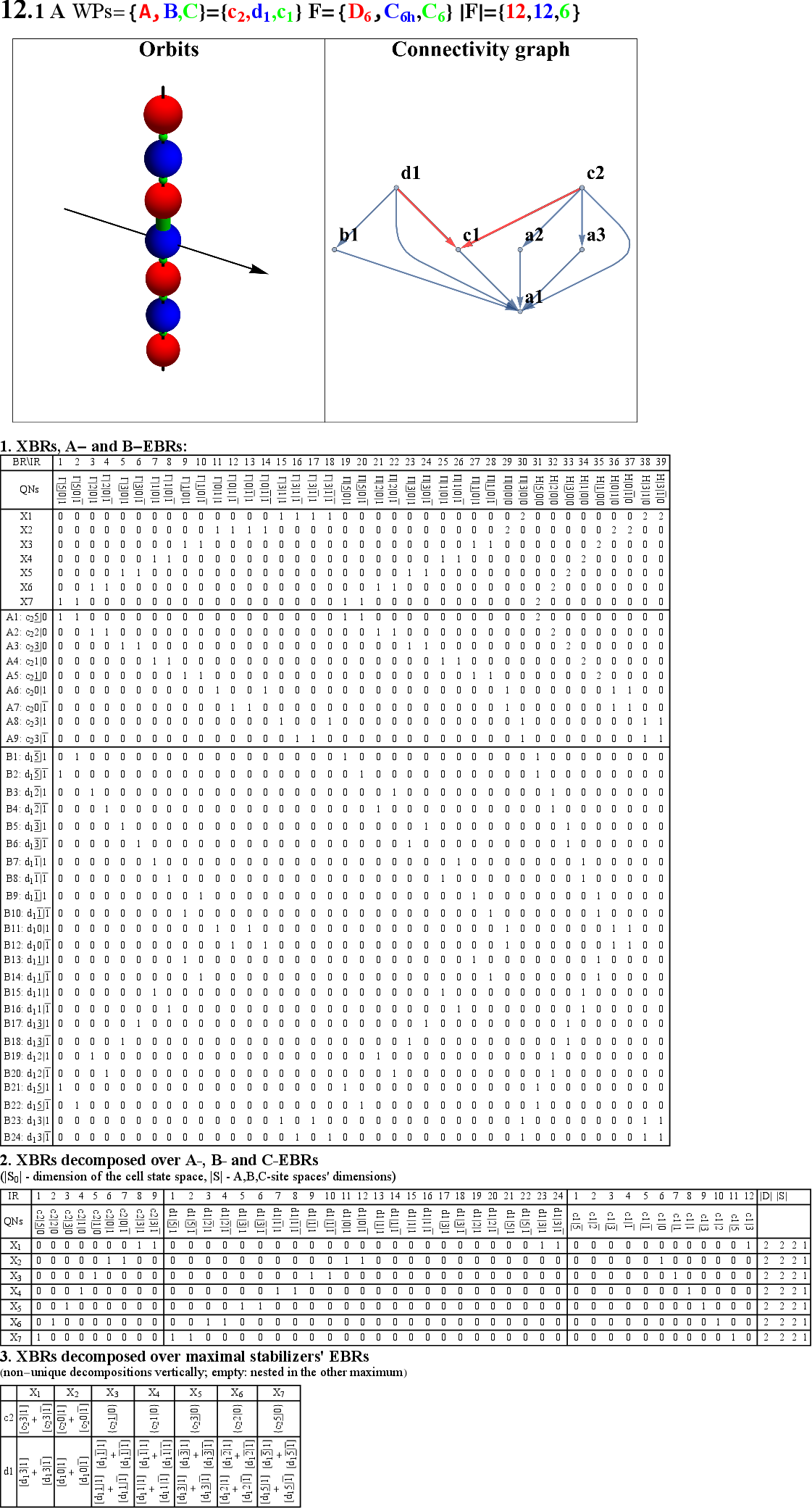}\end{figure*}
\begin{figure*}[b]\includegraphics[width=0.94\textwidth]{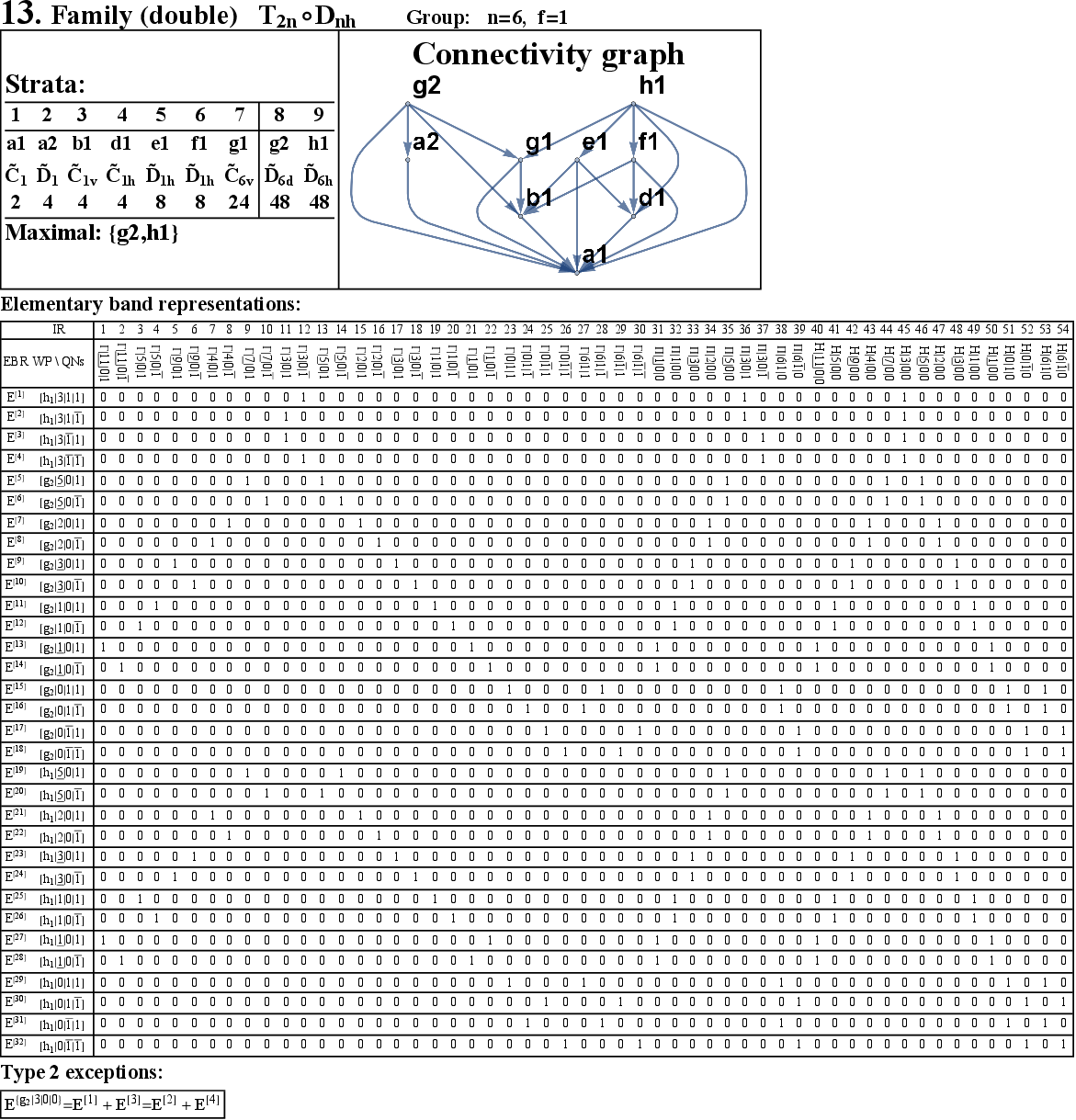}\end{figure*}
\begin{figure*}[b]\includegraphics[width=0.94\textwidth]{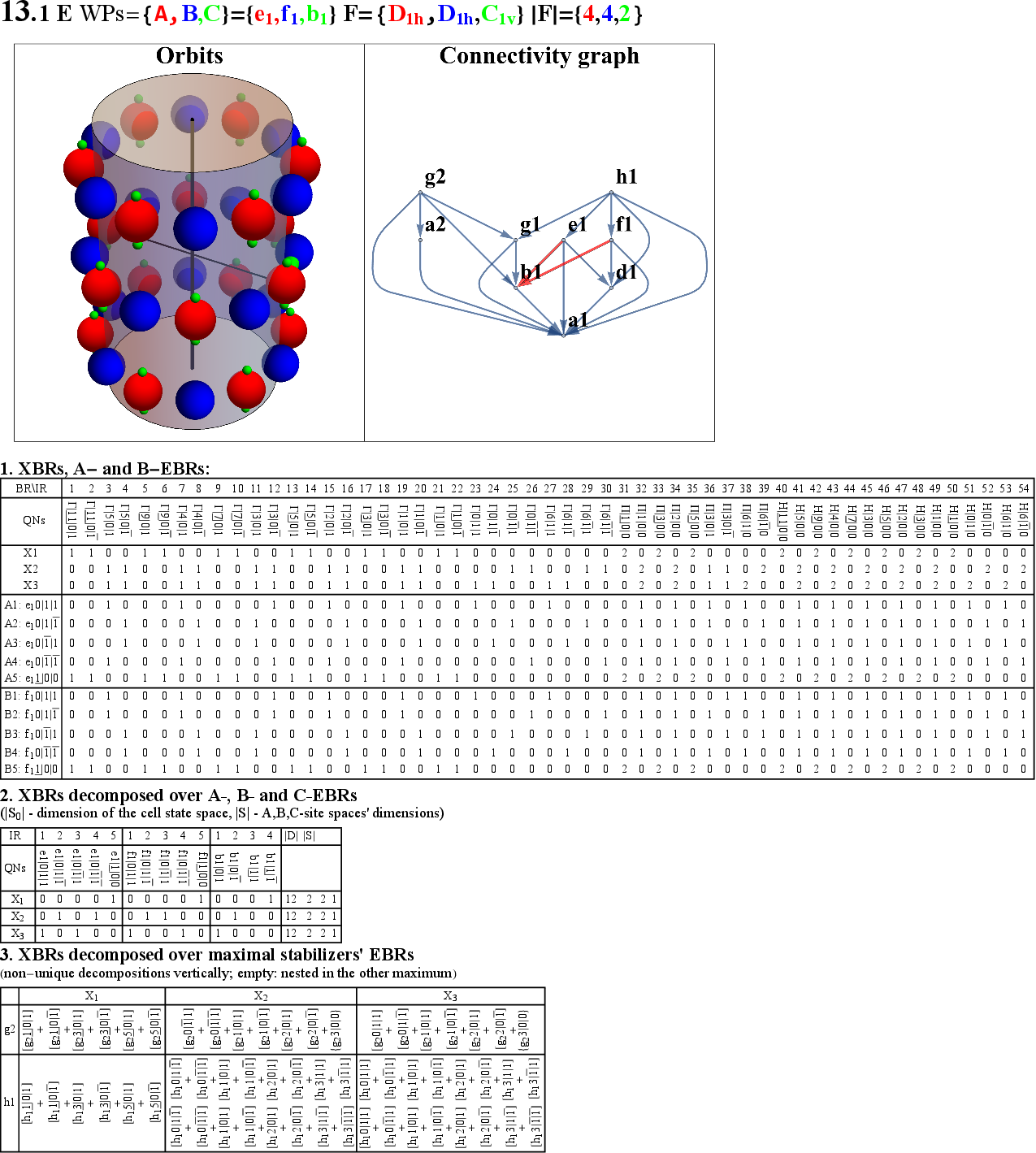}\end{figure*}
\begin{figure*}[b]\includegraphics[width=0.94\textwidth]{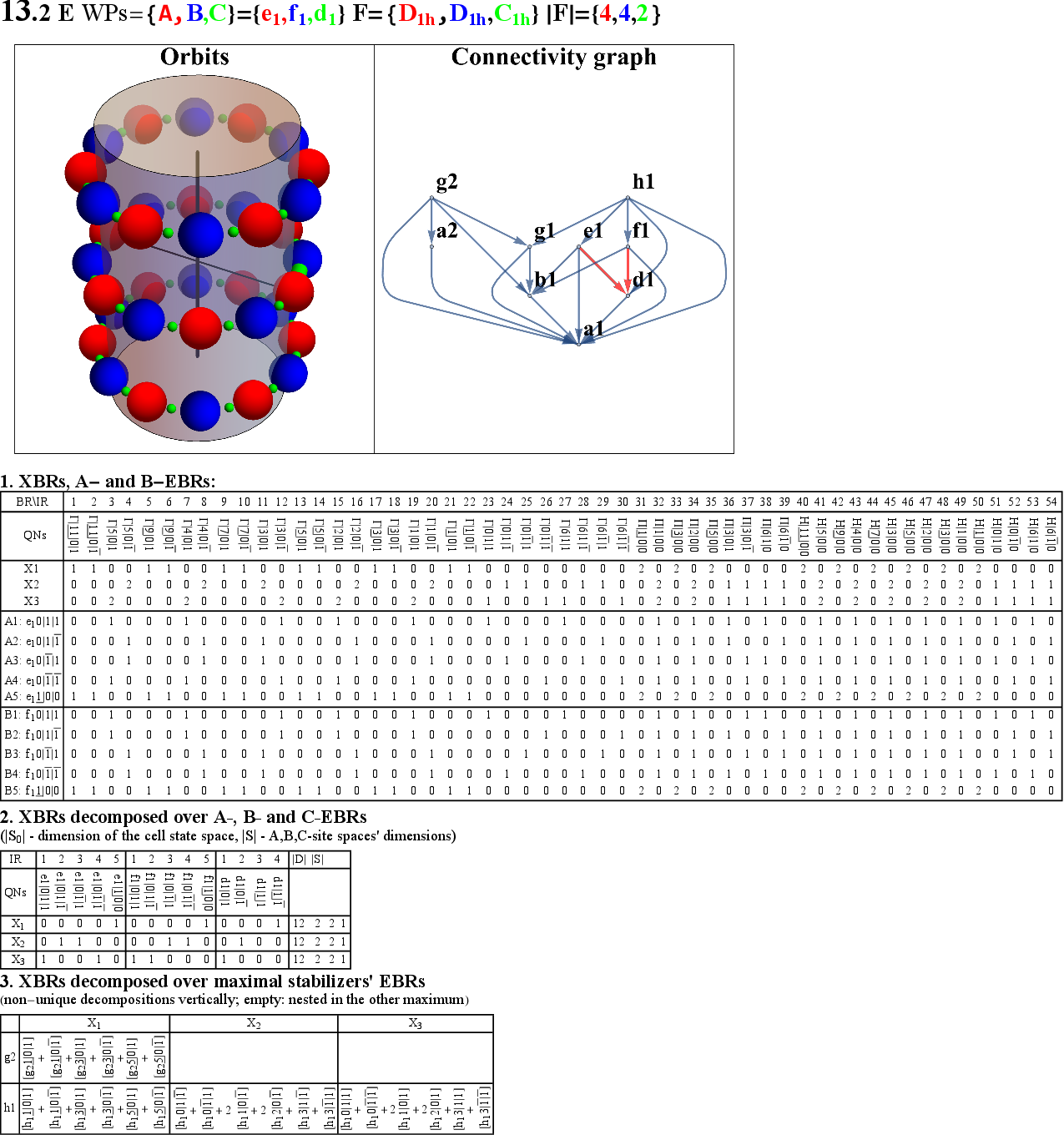}\end{figure*}
\begin{figure*}[b]\includegraphics[width=0.94\textwidth]{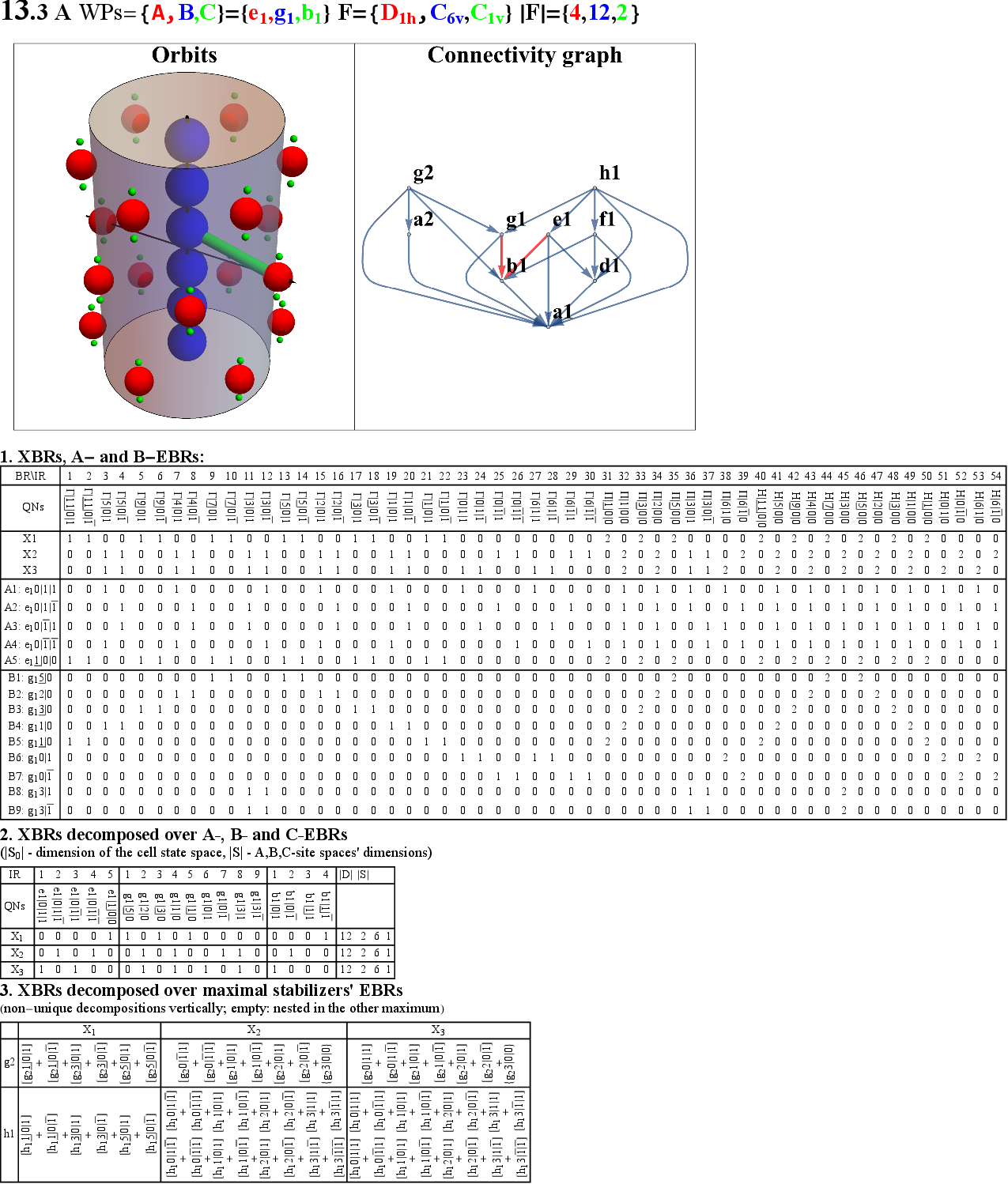}\end{figure*}
\begin{figure*}[b]\includegraphics[width=0.94\textwidth]{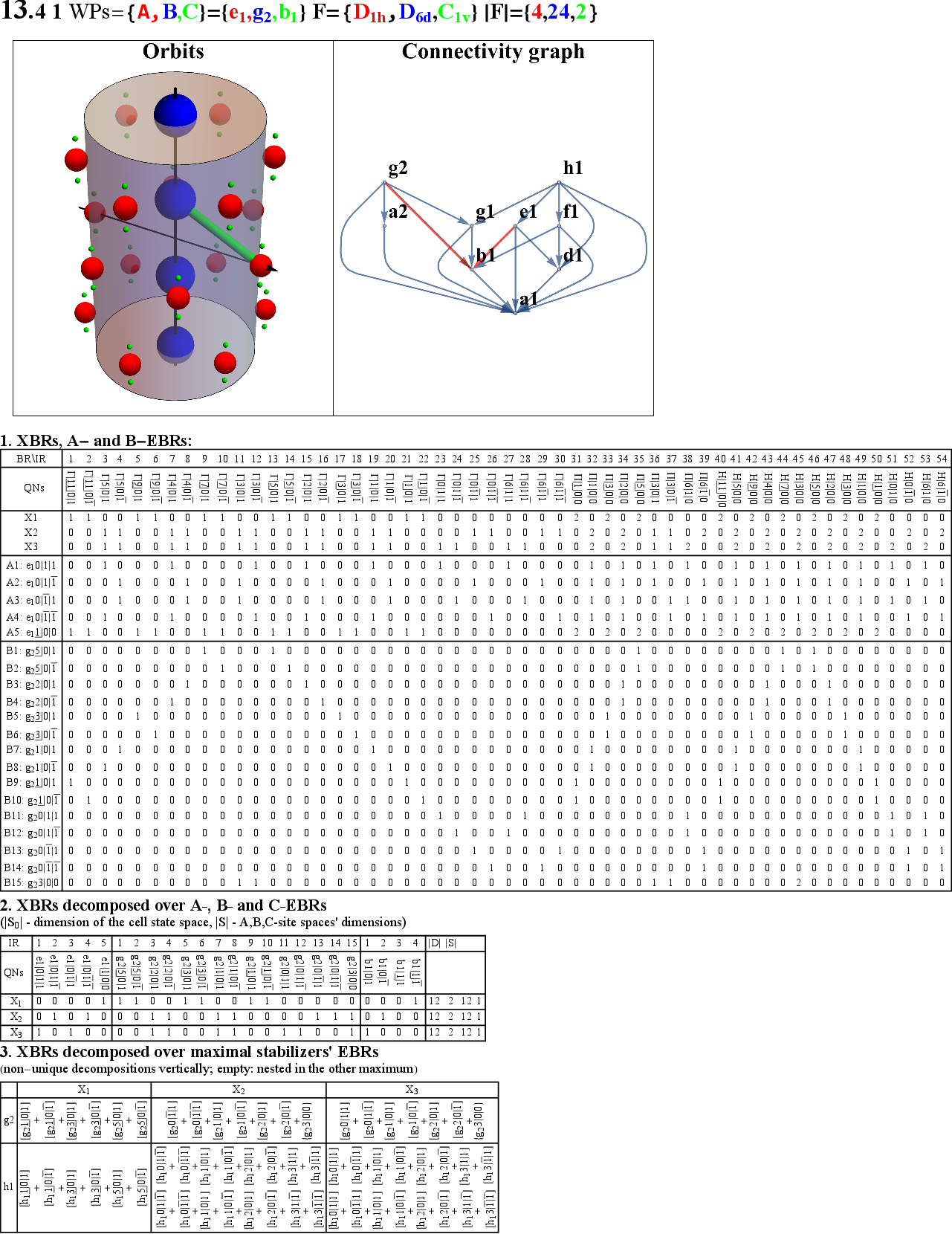}\end{figure*}
\begin{figure*}[b]\includegraphics[width=0.94\textwidth]{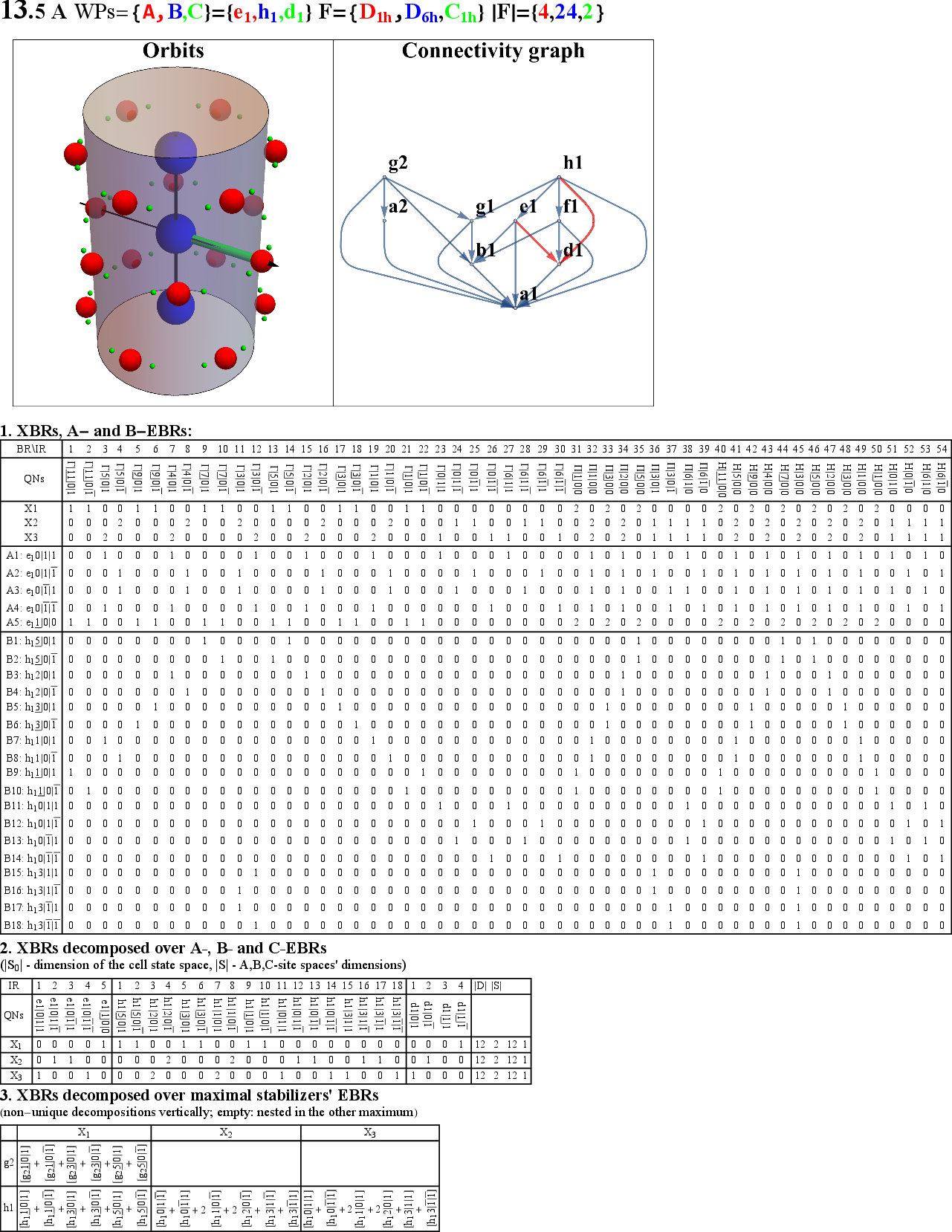}\end{figure*}
\begin{figure*}[b]\includegraphics[width=0.94\textwidth]{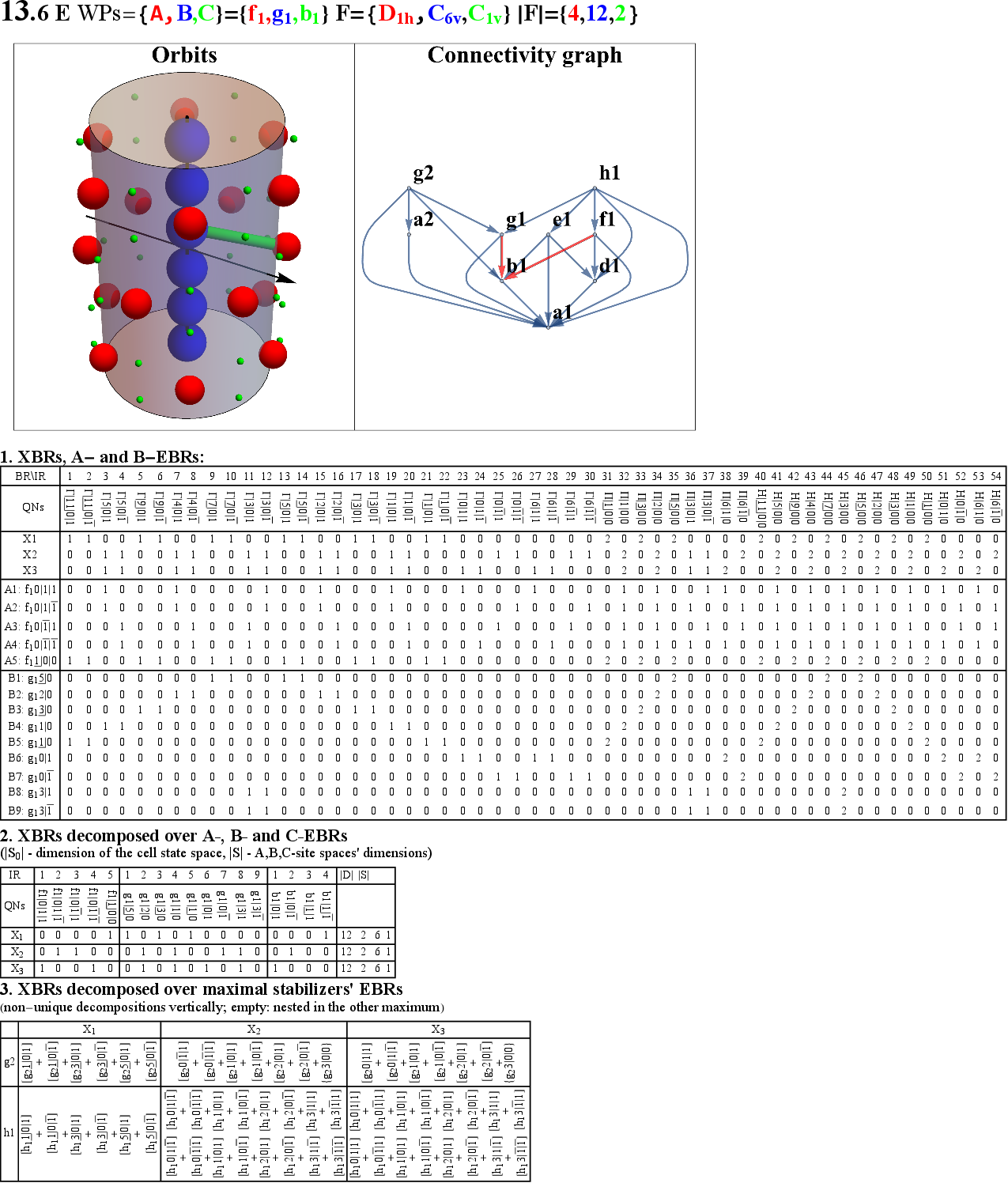}\end{figure*}
\begin{figure*}[b]\includegraphics[width=0.94\textwidth]{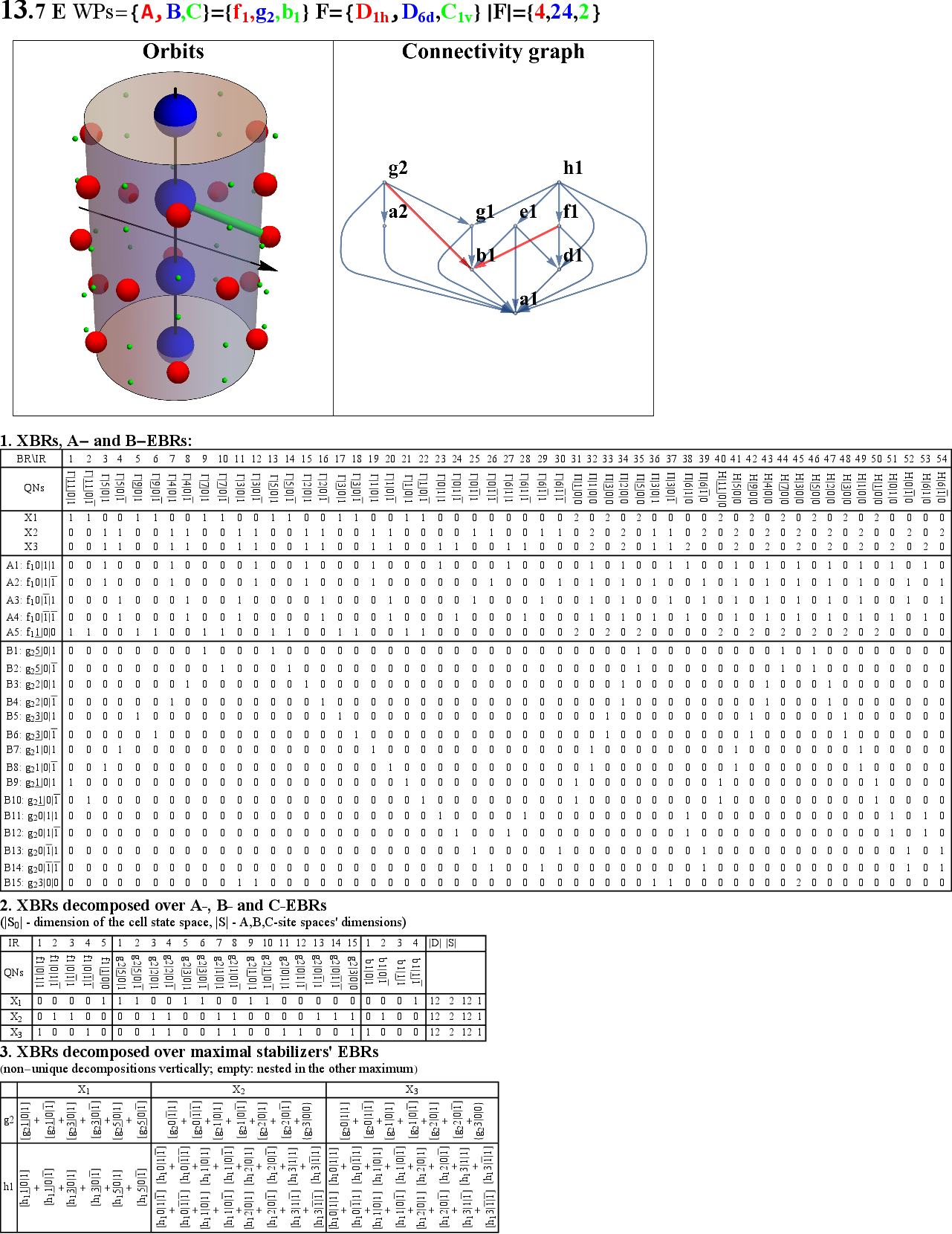}\end{figure*}
\clearpage
\begin{figure*}[b]\includegraphics[width=0.94\textwidth]{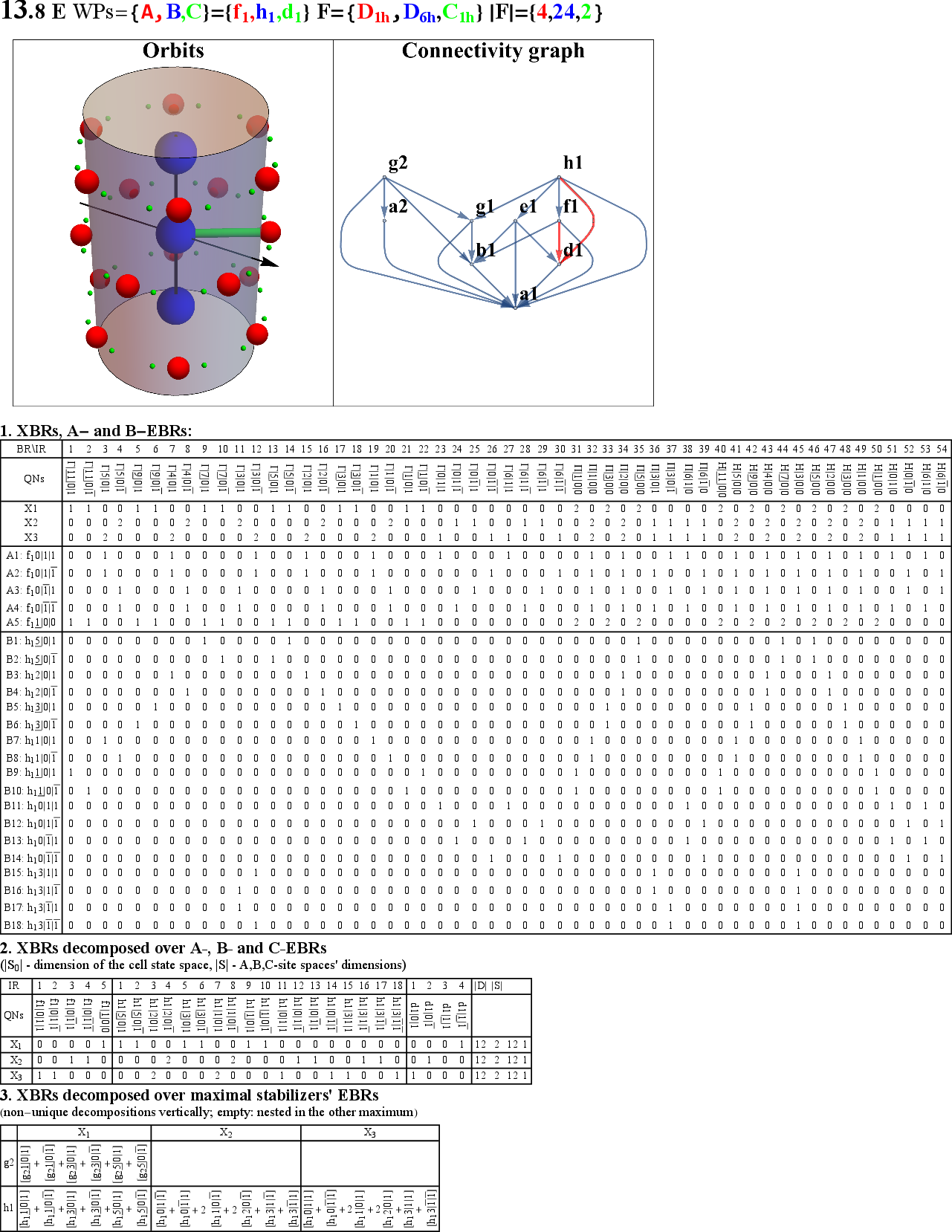}\end{figure*}
\begin{figure*}[b]\includegraphics[width=0.94\textwidth]{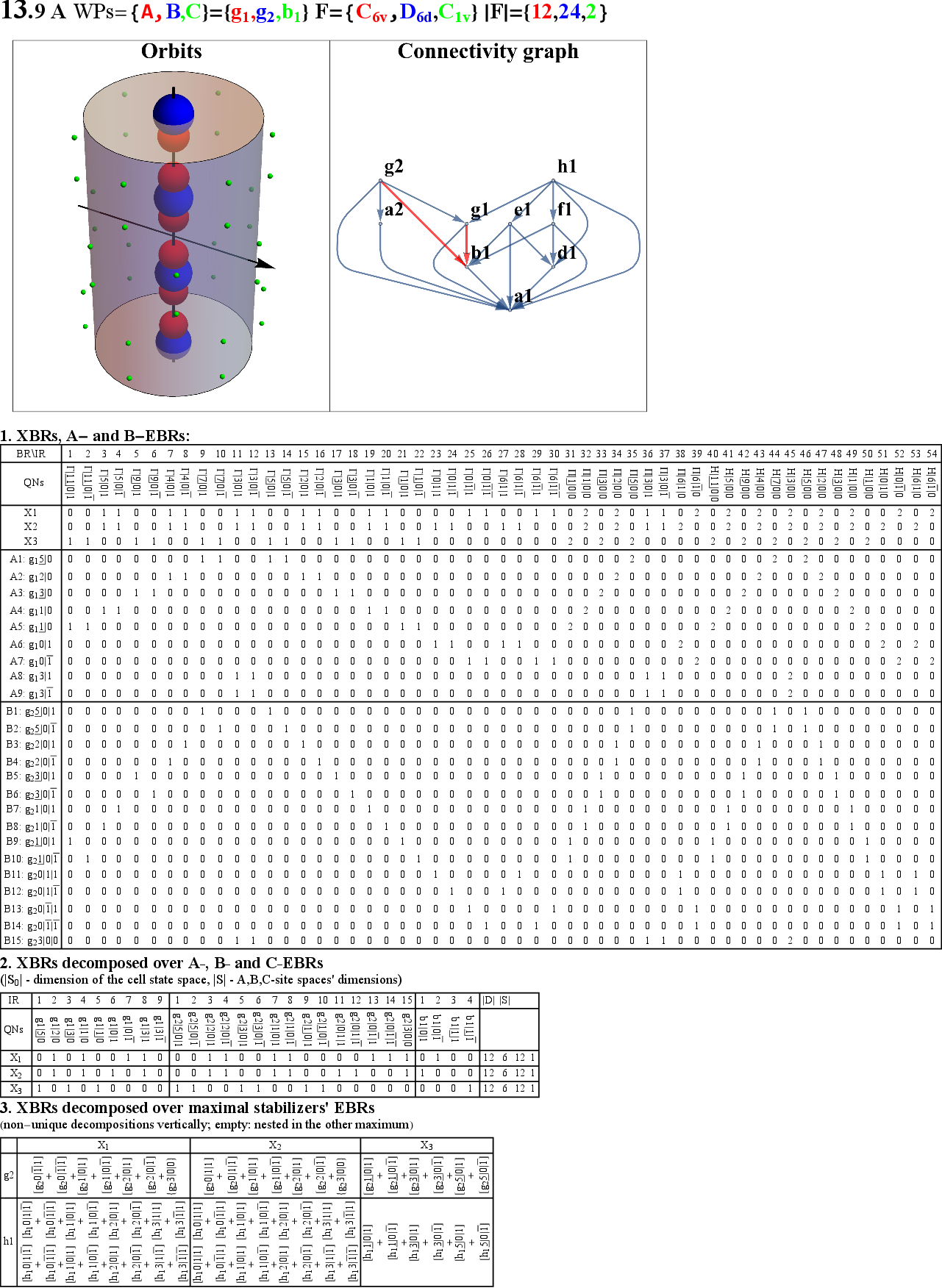}\end{figure*}
\begin{figure*}[b]\includegraphics[width=0.85\textwidth]{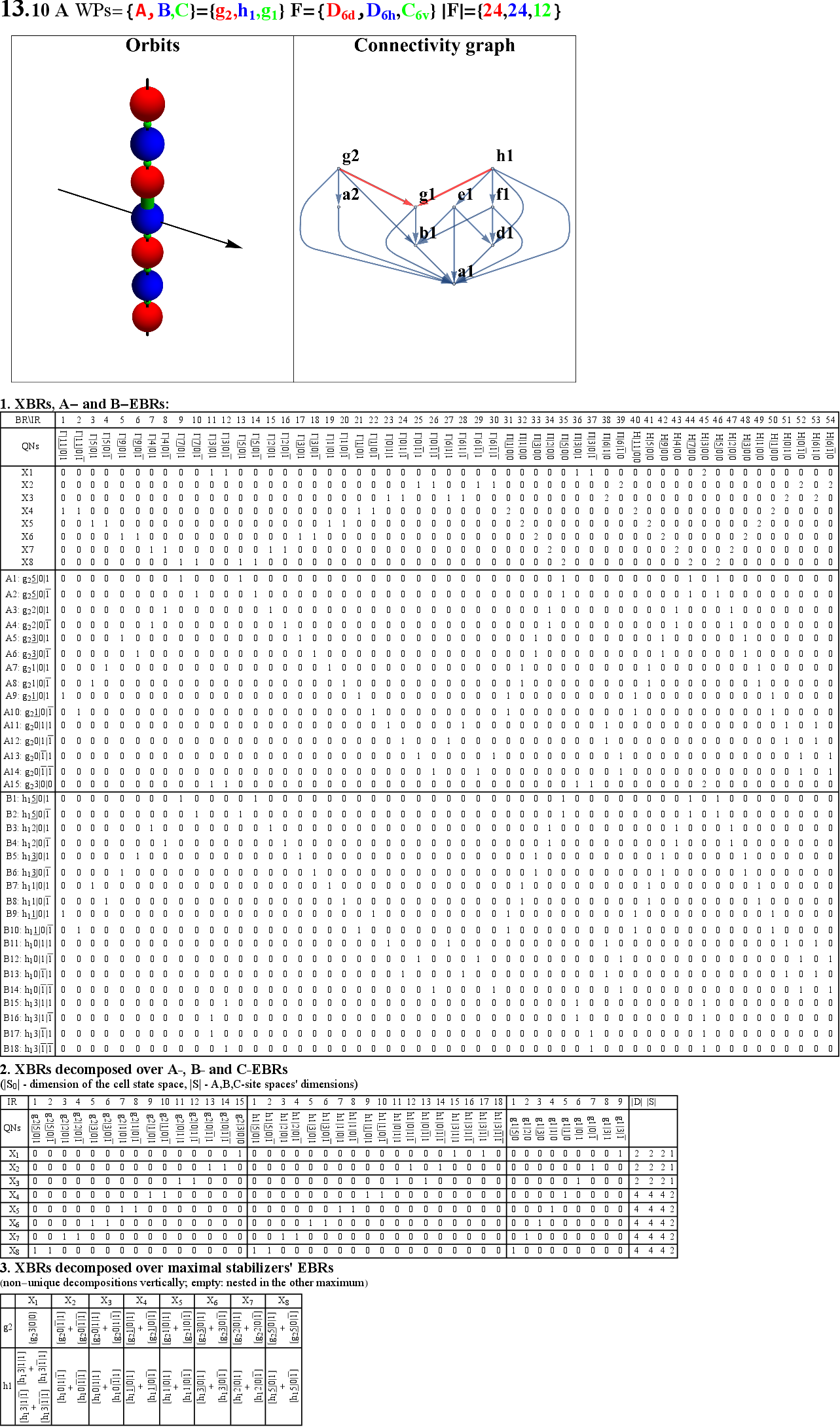}\end{figure*}
\clearpage
\end{widetext}
\end{document}